\documentstyle[10pt,epsf,epsfig,hangcaption,xspace,amssymb,amsfonts,amsmath,amsthm,cite,
dp_delphititle,lineno]{dp_delphi}
\makeindex
\def\DpPaperGroup{EP}
\def\DpPaperRef{2003-057}
\def\DpDate{18 June 2003}
\def\DpAuthors{DELPHI Collaboration}
\def\DpSubmit{(Accepted by  Eur. Phys. J.)}
\def\DpTitle{{ Measurement of $\Vcb$ using the semileptonic decay
$\Bdb \rightarrow \Dstarp \ell^- \overline{\nu}_{\ell}$}}
\def\DpComment{ }
\def\DpEMail{ }

\newcommand{\fd}{f_{\Bdb}}

\newcommand{\ps}{{\rm ps}}

\newcommand{\Dsm}{{\rm D}_s^-}

\newcommand{\Dm}{{\rm D}^-}
\newcommand{\Do}{{\rm D}^0}
\newcommand{\Dob}{\overline{{\rm D}}^0}

\newcommand{\Dstar}{{\rm D}^{\ast}}

\newcommand{\Dstarp}{{\rm D}^{\ast +}}
\newcommand{\Dstarstar}{{\rm D}^{\ast \ast}}

\newcommand{\Vcb}{\left | {\rm V}_{cb} \right |}
\newcommand{\Vub}{\left | {\rm V}_{ub} \right |}

\newcommand{\Bdb}{\overline{\rm{B^{0}_{d}}}}

\newcommand{\Dstarm}{\mbox{D}^{\ast -}}

\newcommand{\Km}{\mbox{K}^-}
\newcommand{\Kp}{\mbox{K}^+}

\newcommand{\GeV}{\rm{GeV}}
\newcommand{\GeVc}{\rm{GeV/c}}
\newcommand{\GeVcd}{\rm{GeV/c^2}}

\newcommand{\MeVc}{\rm{MeV/c}}
\newcommand{\MeVcd}{\rm{MeV/c^2}}

\newcommand{\mumu}{\ifmmode {\mu^+\mu^-} \else ${\mu^+\mu^-} $ \fi}

\newcommand{\dgsl}{{\rm d}\Gamma (\Bdb \rightarrow \Dstarp \ell^-
\overline{\nu}_{\ell})}

\newcommand{\ba}{\begin{array}}
\newcommand{\ea}{\end{array}}
\newcommand{\bc}{\begin{center}}
\newcommand{\ec}{\end{center}}
\newcommand{\be}{\begin{eqnarray}}
\newcommand{\eeq}{\end{eqnarray}}
\newcommand{\bes}{\begin{eqnarray*}}
\newcommand{\ees}{\end{eqnarray*}}
\newcommand{\Kz}{\ifmmode {\rm K^0_s} \else ${\rm K^0_s} $ \fi}
\newcommand{\Zz}{\ifmmode {\rm Z} \else ${\rm Z } $ \fi}
\newcommand{\qqbar}{\ifmmode {\rm q\bar{q}} \else ${\rm q\bar{q}} $ \fi}
\newcommand{\ccbar}{\ifmmode {\rm c\bar{c}} \else ${\rm c\bar{c}} $ \fi}
\newcommand{\bbbar}{\ifmmode {\rm b\bar{b}} \else ${\rm b\bar{b}} $ \fi}
\newcommand{\xxbar}{\ifmmode {\rm x\bar{x}} \else ${\rm x\bar{x}} $ \fi}
\newcommand{\rphi}{\ifmmode {\rm R\phi} \else ${\rm R\phi} $ \fi}
\begin{document}
%%%%%%%%%%%%%%%%%%%%%%%%%% They are a problem with Coll.Sty ?
\makeatletter
%\input{dp_system:coll.sty}
% Collapse citation numbers to ranges.  Non-numeric and undefined labels
% are handled.  No sorting is done.  E.g., 1,3,2,3,4,5,foo,1,2,3,?,4,5
% gives 1,3,2-5,foo,1-3,?,4,5
\newcount\@tempcntc
\def\@citex[#1]#2{\if@filesw\immediate\write\@auxout{\string\citation{#2}}\fi
  \@tempcnta\z@\@tempcntb\m@ne\def\@citea{}\@cite{\@for\@citeb:=#2\do
    {\@ifundefined
       {b@\@citeb}{\@citeo\@tempcntb\m@ne\@citea\def\@citea{,}{\bf ?}\@warning
       {Citation `\@citeb' on page \thepage \space undefined}}%
    {\setbox\z@\hbox{\global\@tempcntc0\csname b@\@citeb\endcsname\relax}%
     \ifnum\@tempcntc=\z@ \@citeo\@tempcntb\m@ne
       \@citea\def\@citea{,}\hbox{\csname b@\@citeb\endcsname}%
     \else
      \advance\@tempcntb\@ne
      \ifnum\@tempcntb=\@tempcntc
      \else\advance\@tempcntb\m@ne\@citeo
      \@tempcnta\@tempcntc\@tempcntb\@tempcntc\fi\fi}}\@citeo}{#1}}
\def\@citeo{\ifnum\@tempcnta>\@tempcntb\else\@citea\def\@citea{,}%
  \ifnum\@tempcnta=\@tempcntb\the\@tempcnta\else
   {\advance\@tempcnta\@ne\ifnum\@tempcnta=\@tempcntb \else \def\@citea{--}\fi
    \advance\@tempcnta\m@ne\the\@tempcnta\@citea\the\@tempcntb}\fi\fi}
 
\makeatother
%%%%%%%%%%%%%%%%%%%%%%%%%%%%
% Generate the title page
\begin{titlepage}
\pagenumbering{roman}
\CERNpreprint{\DpPaperGroup}{\DpPaperRef} % Reference of the paper
\date{{\small\DpDate}} % Date of the paper
\title{\DpTitle} % Title of the paper
\address{\DpAuthors} % General name of the author(s)
\begin{shortabs} % Start the abstract
\noindent
% Changes from David 21/5/03
% Hopefully Final version: 19/5/03
%   abstract.tex
%
\noindent

% Added DELPHI average 
%SA0 29-4-03
% Last modifications: BRs from PDG '02 values
%EAO 29-4-03
%
%SAO 13-4-03
% Added DELPHI average 
%EA0 13-4-03
%SAO 28-1-03
% Changed systematics -> final numbers 
%EA0 28-1-03
% SAO 5-2-03
% Changed systematics again (to symmetrize them) 
% EAO 5-2-03
% SAO 18-2-03
% Added sentence and Vcb value from F(1)Vcb 
% EAO 18-2-03
%=========================================================================%
%===================> DELPHI note abstract =====> To be filled <=====%
Data from $\Zz$ decays in DELPHI have been searched for
$\Bdb \rightarrow \Dstarp \ell^- \overline{\nu}_{\ell}$ with the $\Dstarp$ 
decaying to
$\Do \pi^+$ and $\Do \rightarrow \Km \pi^+$, $\Km \pi^+ \pi^+ \pi^-$
or $\Km \pi^+ (\pi^0)$. These events are used to
measure the CKM matrix element $\Vcb$ and the form factor slope, $\rho_{A_1}^2$:
\bc
${\cal F}_{D^*}(1) \Vcb= 0.0392 \pm 0.0018 \pm 0.0023; ~\rho_{A_1}^2= 1.32 \pm 
0.15 \pm 0.33$
\ec
corresponding to a branching fraction: 
\bc
${\rm BR}(\Bdb \rightarrow \Dstarp \ell^- \overline{\nu}_{\ell})
=(5.90 \pm 0.22 \pm 0.50)\%$.
\ec

Combining these and previous DELPHI measurements gives:
\bc
 ${\cal F}_{D^*}(1)\Vcb = 0.0377 \pm 0.0011 \pm 0.0019,~
 \rho_{A_1}^2 =  1.39 \pm 0.10 \pm 0.33$ and 
 ${\rm BR}(\Bdb \rightarrow \Dstarp \ell^- \overline{\nu}_{\ell})=(5.39 \pm 
0.11 \pm 0.34)\%$.
\ec
Using ${\cal F}_{D^*}(1)=0.91 \pm 0.04$, yields: 
\bc
$\Vcb= 0.0414 \pm 0.0012(stat.) \pm 0.0021(syst.) \pm 0.0018(theory).$
\ec

The $b$-quark semileptonic branching fraction into a $\Dstarp$ emitted
from higher mass charmed excited states has also been measured to be:
\bc

 ${\rm BR}(b \rightarrow \Dstarp {\rm X} \ell^- \overline{\nu}_{\ell})
= (0.67\pm 0.08\pm0.10)\%$.
\ec

%=========================================================================%

\end{shortabs}
\vfill
\begin{center}
\DpSubmit \ \\ % Horrible hack to allow to have DpSubmit empty
\DpComment \ \\
\DpEMail \ \\
\end{center}
\vfill
\clearpage
\headsep 10.0pt
\addtolength{\textheight}{10mm}
\addtolength{\footskip}{-5mm}
\begingroup
% Commands to process the author names
%
\newcommand{\DpName}[2]{\hbox{#1$^{\ref{#2}}$},\hfill}
\newcommand{\DpNameTwo}[3]{\hbox{#1$^{\ref{#2},\ref{#3}}$},\hfill}
\newcommand{\DpNameThree}[4]{\hbox{#1$^{\ref{#2},\ref{#3},\ref{#4}}$},\hfill}
\newskip\Bigfill \Bigfill = 0pt plus 1000fill
\newcommand{\DpNameLast}[2]{\hbox{#1$^{\ref{#2}}$}\hspace{\Bigfill}}
%
%\small
\footnotesize
\noindent
\DpName{J.Abdallah}{LPNHE}
\DpName{P.Abreu}{LIP}
\DpName{W.Adam}{VIENNA}
\DpName{P.Adzic}{DEMOKRITOS}
\DpName{T.Albrecht}{KARLSRUHE}
\DpName{T.Alderweireld}{AIM}
\DpName{R.Alemany-Fernandez}{CERN}
\DpName{T.Allmendinger}{KARLSRUHE}
\DpName{P.P.Allport}{LIVERPOOL}
\DpName{U.Amaldi}{MILANO2}
% ...
\normalsize
\endgroup
\titlefoot{Department of Physics and Astronomy, Iowa State
     University, Ames IA 50011-3160, USA
    \label{AMES}}
\titlefoot{Physics Department, Universiteit Antwerpen,
     Universiteitsplein 1, B-2610 Antwerpen, Belgium \\
     \indent~~and IIHE, ULB-VUB,
     Pleinlaan 2, B-1050 Brussels, Belgium \\
     \indent~~and Facult\'e des Sciences,
     Univ. de l'Etat Mons, Av. Maistriau 19, B-7000 Mons, Belgium
    \label{AIM}}
\titlefoot{Physics Laboratory, University of Athens, Solonos Str.
     104, GR-10680 Athens, Greece
    \label{ATHENS}}
\titlefoot{Department of Physics, University of Bergen,
     All\'egaten 55, NO-5007 Bergen, Norway
    \label{BERGEN}}
\titlefoot{Dipartimento di Fisica, Universit\`a di Bologna and INFN,
     Via Irnerio 46, IT-40126 Bologna, Italy
    \label{BOLOGNA}}
\titlefoot{Centro Brasileiro de Pesquisas F\'{\i}sicas, rua Xavier Sigaud 150,
     BR-22290 Rio de Janeiro, Brazil \\
     \indent~~and Depto. de F\'{\i}sica, Pont. Univ. Cat\'olica,
     C.P. 38071 BR-22453 Rio de Janeiro, Brazil \\
     \indent~~and Inst. de F\'{\i}sica, Univ. Estadual do Rio de Janeiro,
     rua S\~{a}o Francisco Xavier 524, Rio de Janeiro, Brazil
    \label{BRASIL}}
\titlefoot{Coll\`ege de France, Lab. de Physique Corpusculaire, IN2P3-CNRS,
     FR-75231 Paris Cedex 05, France
    \label{CDF}}
\titlefoot{CERN, CH-1211 Geneva 23, Switzerland
    \label{CERN}}
\titlefoot{Institut de Recherches Subatomiques, IN2P3 - CNRS/ULP - BP20,
     FR-67037 Strasbourg Cedex, France
    \label{CRN}}
\titlefoot{Now at DESY-Zeuthen, Platanenallee 6, D-15735 Zeuthen, Germany
    \label{DESY}}
\titlefoot{Institute of Nuclear Physics, N.C.S.R. Demokritos,
     P.O. Box 60228, GR-15310 Athens, Greece
    \label{DEMOKRITOS}}
\titlefoot{FZU, Inst. of Phys. of the C.A.S. High Energy Physics Division,
     Na Slovance 2, CZ-180 40, Praha 8, Czech Republic
    \label{FZU}}
\titlefoot{Dipartimento di Fisica, Universit\`a di Genova and INFN,
     Via Dodecaneso 33, IT-16146 Genova, Italy
    \label{GENOVA}}
\titlefoot{Institut des Sciences Nucl\'eaires, IN2P3-CNRS, Universit\'e
     de Grenoble 1, FR-38026 Grenoble Cedex, France
    \label{GRENOBLE}}
\titlefoot{Helsinki Institute of Physics, P.O. Box 64,
     FIN-00014 University of Helsinki, Finland
    \label{HELSINKI}}
\titlefoot{Joint Institute for Nuclear Research, Dubna, Head Post
     Office, P.O. Box 79, RU-101 000 Moscow, Russian Federation
    \label{JINR}}
\titlefoot{Institut f\"ur Experimentelle Kernphysik,
     Universit\"at Karlsruhe, Postfach 6980, DE-76128 Karlsruhe,
     Germany
    \label{KARLSRUHE}}
\titlefoot{Institute of Nuclear Physics,Ul. Kawiory 26a,
     PL-30055 Krakow, Poland
    \label{KRAKOW1}}
\titlefoot{Faculty of Physics and Nuclear Techniques, University of Mining
     and Metallurgy, PL-30055 Krakow, Poland
    \label{KRAKOW2}}
\titlefoot{Universit\'e de Paris-Sud, Lab. de l'Acc\'el\'erateur
     Lin\'eaire, IN2P3-CNRS, B\^{a}t. 200, FR-91405 Orsay Cedex, France
    \label{LAL}}
\titlefoot{School of Physics and Chemistry, University of Lancaster,
     Lancaster LA1 4YB, UK
    \label{LANCASTER}}
\titlefoot{LIP, IST, FCUL - Av. Elias Garcia, 14-$1^{o}$,
     PT-1000 Lisboa Codex, Portugal
    \label{LIP}}
\titlefoot{Department of Physics, University of Liverpool, P.O.
     Box 147, Liverpool L69 3BX, UK
    \label{LIVERPOOL}}
\titlefoot{Dept. of Physics and Astronomy, Kelvin Building,
     University of Glasgow, Glasgow G12 8QQ
    \label{GLASGOW}}
\titlefoot{LPNHE, IN2P3-CNRS, Univ.~Paris VI et VII, Tour 33 (RdC),
     4 place Jussieu, FR-75252 Paris Cedex 05, France
    \label{LPNHE}}
\titlefoot{Department of Physics, University of Lund,
     S\"olvegatan 14, SE-223 63 Lund, Sweden
    \label{LUND}}
\titlefoot{Universit\'e Claude Bernard de Lyon, IPNL, IN2P3-CNRS,
     FR-69622 Villeurbanne Cedex, France
    \label{LYON}}
\titlefoot{Dipartimento di Fisica, Universit\`a di Milano and INFN-MILANO,
     Via Celoria 16, IT-20133 Milan, Italy
    \label{MILANO}}
\titlefoot{Dipartimento di Fisica, Univ. di Milano-Bicocca and
     INFN-MILANO, Piazza della Scienza 2, IT-20126 Milan, Italy
    \label{MILANO2}}
\titlefoot{IPNP of MFF, Charles Univ., Areal MFF,
     V Holesovickach 2, CZ-180 00, Praha 8, Czech Republic
    \label{NC}}
\titlefoot{NIKHEF, Postbus 41882, NL-1009 DB
     Amsterdam, The Netherlands
    \label{NIKHEF}}
\titlefoot{National Technical University, Physics Department,
     Zografou Campus, GR-15773 Athens, Greece
    \label{NTU-ATHENS}}
\titlefoot{Physics Department, University of Oslo, Blindern,
     NO-0316 Oslo, Norway
    \label{OSLO}}
\titlefoot{Dpto. Fisica, Univ. Oviedo, Avda. Calvo Sotelo
     s/n, ES-33007 Oviedo, Spain
    \label{OVIEDO}}
\titlefoot{Department of Physics, University of Oxford,
     Keble Road, Oxford OX1 3RH, UK
    \label{OXFORD}}
\titlefoot{Dipartimento di Fisica, Universit\`a di Padova and
     INFN, Via Marzolo 8, IT-35131 Padua, Italy
    \label{PADOVA}}
\titlefoot{Rutherford Appleton Laboratory, Chilton, Didcot
     OX11 OQX, UK
    \label{RAL}}
\titlefoot{Dipartimento di Fisica, Universit\`a di Roma II and
     INFN, Tor Vergata, IT-00173 Rome, Italy
    \label{ROMA2}}
\titlefoot{Dipartimento di Fisica, Universit\`a di Roma III and
     INFN, Via della Vasca Navale 84, IT-00146 Rome, Italy
    \label{ROMA3}}
\titlefoot{DAPNIA/Service de Physique des Particules,
     CEA-Saclay, FR-91191 Gif-sur-Yvette Cedex, France
    \label{SACLAY}}
\titlefoot{Instituto de Fisica de Cantabria (CSIC-UC), Avda.
     los Castros s/n, ES-39006 Santander, Spain
    \label{SANTANDER}}
\titlefoot{Inst. for High Energy Physics, Serpukov
     P.O. Box 35, Protvino, (Moscow Region), Russian Federation
    \label{SERPUKHOV}}
\titlefoot{J. Stefan Institute, Jamova 39, SI-1000 Ljubljana, Slovenia
     and Laboratory for Astroparticle Physics,\\
     \indent~~Nova Gorica Polytechnic, Kostanjeviska 16a, SI-5000 Nova Gorica, Slovenia, \\
     \indent~~and Department of Physics, University of Ljubljana,
     SI-1000 Ljubljana, Slovenia
    \label{SLOVENIJA}}
\titlefoot{Fysikum, Stockholm University,
     Box 6730, SE-113 85 Stockholm, Sweden
    \label{STOCKHOLM}}
\titlefoot{Dipartimento di Fisica Sperimentale, Universit\`a di
     Torino and INFN, Via P. Giuria 1, IT-10125 Turin, Italy
    \label{TORINO}}
\titlefoot{INFN,Sezione di Torino, and Dipartimento di Fisica Teorica,
     Universit\`a di Torino, Via P. Giuria 1,\\
     \indent~~IT-10125 Turin, Italy
    \label{TORINOTH}}
\titlefoot{Dipartimento di Fisica, Universit\`a di Trieste and
     INFN, Via A. Valerio 2, IT-34127 Trieste, Italy \\
     \indent~~and Istituto di Fisica, Universit\`a di Udine,
     IT-33100 Udine, Italy
    \label{TU}}
\titlefoot{Univ. Federal do Rio de Janeiro, C.P. 68528
     Cidade Univ., Ilha do Fund\~ao
     BR-21945-970 Rio de Janeiro, Brazil
    \label{UFRJ}}
\titlefoot{Department of Radiation Sciences, University of
     Uppsala, P.O. Box 535, SE-751 21 Uppsala, Sweden
    \label{UPPSALA}}
\titlefoot{IFIC, Valencia-CSIC, and D.F.A.M.N., U. de Valencia,
     Avda. Dr. Moliner 50, ES-46100 Burjassot (Valencia), Spain
    \label{VALENCIA}}
\titlefoot{Institut f\"ur Hochenergiephysik, \"Osterr. Akad.
     d. Wissensch., Nikolsdorfergasse 18, AT-1050 Vienna, Austria
    \label{VIENNA}}
\titlefoot{Inst. Nuclear Studies and University of Warsaw, Ul.
     Hoza 69, PL-00681 Warsaw, Poland
    \label{WARSZAWA}}
\titlefoot{Fachbereich Physik, University of Wuppertal, Postfach
     100 127, DE-42097 Wuppertal, Germany
    \label{WUPPERTAL}}
\addtolength{\textheight}{-10mm}
\addtolength{\footskip}{5mm}
\clearpage
\headsep 30.0pt
\end{titlepage}
%%%%%%%%%%%%%%%%%%%%%%%%%
%
% Change for the document body
%%\pagestyle{heading} % for page numbering
\pagenumbering{arabic} % page numbering in number
\setcounter{footnote}{0} %
\large
%\linenumbers %%%CD
% Last modification (table 7) 13/12/2003 
% Last modification (delphi average) 4/6/2003 
% Changed some plots and some David comments 21/5/03 
% Changes from David 21/5/03
% Hopefully Final version: 19/5/03
% *****************************************************************
% Corrections done by P. Roudeau 13-11-2002 are indicated as
% PR 13-11-02
% End PR 13-11-02
% *****************************************************************
% Corrections done by P. Roudeau 9-1-2003 are indicated as
% SPR 9-1-03
% EPR 9-1-03
% *****************************************************************
% Corrections done by A. Oyanguren 15,27,28-1-03 
% and 5,7-2-03 are indicated as
% SA0 15-1-03, 27-1-03, 28-1-03, 5-2-03, 7-2-03, 14-2-03, 18-2-03, 13-3-03
% EA0 15-1-03, 27-1-03, 28-1-03, 5-2-03, 7-2-03, 14-2-03, 18-2-03, 13-3-03
% All figures have been named old_name2.eps to distinguish this version. 
% ***********************************************************************
% Corrections done by P. Roudeau 28-1-2003 are indicated as
% SPR 28-1-03
% EPR 28-1-03
% *********************************************************************
% Corrections done by P. Roudeau 4-2-2003 are indicated as
% SPR 4-2-03
% EPR 4-2-03
% ********************************************************************
% Added DELPHI average A. Oyanguren 13-4-2003
% SAO 13-4-03
% EAO 13-4-03
%***********************************************************************
%  A. Oyanguren 29-4-2003 -> Change all numbers due to BRs PDG'02 values
%***********************************************************************

\section {Introduction}
\label{sec:intro}
The Cabibbo, Kobayashi and Maskawa (CKM) matrix 
element ${\rm V}_{cb}$ is a parameter of the Standard Model and its value
needs to be fixed by experiments. This parameter determines the decay rate of 
$b$-hadrons because $\Vub$ governs the other possible
% SPR 9-1-03
% ``by only'' --> ``only''
% EPR 9-1-03
charged weak decay of $b$-quarks and contributes  only 1-2$\%$ to the total
decay rate. The value of $\Vcb$ cannot be measured directly and there are two
decay processes for which theoretical uncertainties are expected to be 
% SPR 9-1-03
% Reference \cite{ref:f1vcb} added in the sentence below
% EPR 9-1-03
under control \cite{ref:f1vcb}: the inclusive semileptonic decay of 
$b$-hadrons corresponding
to the $b \rightarrow c \ell^- \overline{\nu}_{\ell}$ transition and the
exclusive decay channel 
$\Bdb \rightarrow \Dstarp \ell^- \overline{\nu}_{\ell}$.
% SPR 9-1-03
% footnote displaced from page 3 to the introduction
% EPR 9-1-03
The latter is used in the following analysis, the $\Dstarp$ 
\footnote{ Throughout this paper charge-conjugate states are 
implicitly included.}
is reconstructed 
through its decay to $\Do \pi^+$ and the $\Do$ meson is isolated
using three decay channels: $\Km \pi^+$, $\Km \pi^+ \pi^+ \pi^-$
and $\Km \pi^+ (\pi^0)$.

This study benefits from the reprocessing of DELPHI data taken
between 1992 and 1995 through improved versions of the event reconstruction
algorithms. As a consequence, the number of signal events has increased
by more than a factor two over those reported in 
\cite{ref:firstvcb} using the same decay final states.
% SPR 9-1-03
% $\rightarrow$ added below
% EPR 9-1-03
An additional decay channel of the $\Do$ ($\rightarrow \Km \pi^+ (\pi^0)$)
has been analysed which provides 
another factor of two increase. There are also improvements
on the $\Dstar$ mass 
and $b$-meson energy reconstruction. 
% SPR 9-1-03
% sentence below modified
% EPR 9-1-03
The $\Dstarp$ signal is now
narrower which gives a better isolation of the signal over
the combinatorial background. As the main source of 
experimental systematic uncertainty originates from the contribution
of $\Dstarp$ mesons emitted in the decay of excited charmed states
produced in $b$-hadron semileptonic decays, additional observables have been
defined to control the level of this contamination in a better way.
The evaluation of the remaining contamination from the 
% SPR 4-2-03
% ``double charm'' added below
% EPR 4-2-03
double charm cascade decays
($b \rightarrow \Dstarp \overline{{\rm D}} {\rm X},~
\overline{{\rm D}} \rightarrow \ell^- \overline{\nu}_{\ell} {\rm Y}$)
benefits from recent measurements of their rates.

\section{Measurement of $\Vcb$ from the decay $\Bdb \rightarrow \Dstarp \ell^- 
\overline{\nu}_{\ell}$}
\label{sec:principle}
The value of $\Vcb$ is extracted by studying
the decay partial width for the process $\Bdb \rightarrow \Dstarp \ell^- 
\overline{\nu}_{\ell}$
 as a function of the recoil kinematics
% SPR 9-1-03
% Reference \cite{ref:f1vcb} added in the sentence below
% EPR 9-1-03
of the $\Dstarp$ meson \cite{ref:f1vcb}. The decay rate is parameterized as 
a function
of the variable $w$, defined as the product of the four-velocities of the 
$\Dstarp$ and $\Bdb$ mesons.
This variable is related to the square of
the four-momentum transfer from the $\Bdb$ to the $\ell^-{\overline \nu}_\ell$
system, $q^2$, by:
\begin{equation}
w = \frac{m_{\rm D^{*+}}^2+m_{\rm \overline B^0_d}^2-q^2}{2m_{\rm \overline 
B^0_d}
%\cdot 
m_{\rm D^{*+}}}
\end{equation}
and its value ranges from $1.0$, when the $\Dstarp$ is produced at rest in the 
$\Bdb$ rest 
% SPR 9-1-03
% HQET in made explicit in the sentence below
% EPR 9-1-03
frame, to about $1.5$.  Using the Heavy Quark Effective Theory 
(HQET)\cite{hqet}, 
the 
differential partial width 
for this
decay is given by: 
\begin{eqnarray}
\label{eq:decayw}
{\frac{{\rm d} \Gamma}{{\rm d} w}}&=
&\frac{G_F^2 \Vcb^2}{48 \pi^3}{\cal K}(w){\cal F}_{D^*}^2(w),
\label{eq:dgdw}
\end{eqnarray}
where ${\cal K}(w)$ contains kinematic factors:
\begin{equation}
{\cal K}(w)=m_{D^*}^3 (m_B-m_{D^*})^2 \sqrt{w^2-1}(w+1)^2
\left( 1 + \frac{4w}{w+1} \frac{1-2wr+r^2}{(1-r)^2} \right)
\end{equation}
with $r=m_{D^*}/m_B$
and ${\cal F}_{D^*}(w)$ is a hadronic form factor.

 Although the shape of the
form factor, ${\cal F}_{D^*}(w)$, is not known, its magnitude
at zero recoil, corresponding to $w=1$, can be estimated using HQET.  
It is convenient to express ${\cal F}_{D^*}(w)$ in terms 
of the axial form factor
$h_{A_1}(w)$ and of the reduced helicity form factors $\tilde{H}_0$ 
and $\tilde{H}_{\pm}$:
\begin{equation}
{\cal F}_{D^*}(w)= h_{A_1}(w)\sqrt{\frac
{\tilde{H}_0^2+\tilde{H}_{+}^2+\tilde{H}_{-}^2}
{1 + \frac{4w}{w+1} \frac{1-2wr+r^2}{(1-r)^2}}}.
\end{equation}
The reduced helicity form factors are themselves expressed in terms of the 
ratios between the other HQET form factors 
($h_{V}(w),~h_{A_2}(w),~h_{A_3}(w)$) and $h_{A_1}(w)$:
\begin{eqnarray}
\tilde{H}_0(w)=1+\frac{w-1}{1-r}\left [ 1- R_2(w)\right] \\
\tilde{H}_{\pm}(w)=\frac{\sqrt{1-2wr+r^2}}{1-r}
\left[1 \mp \sqrt{\frac{w-1}{w+1} R_1(w)} \right]
\end{eqnarray}
with
\begin{equation}
 R_1(w)=\frac{h_V(w)}{h_{A_1}(w)}~{\rm and}~
 R_2(w)=\frac{h_{A_3}(w)+rh_{A_2}(w) }{h_{A_1}(w)}.
\label{eq:ratios}
\end{equation}
Values for $R_1(w)$ and $R_2(w)$ have been measured by CLEO 
\cite{ref:cleoform} using different models.

The  unknown function $h_{A_1}(w)$ is approximated with an expansion
around $w=1$ \cite{CLN}:
\begin{equation}
h_{A_1}(w)= h_{A_1}(1) \times 
\left[ 1 - 8 \rho_{A_1}^2 z + (53 \rho_{A_1}^2 - 15)z^2
- (231 \rho_{A_1}^2 -91) z^3 \right],
\label{eq:clnextr}
\end{equation}
where $\rho_{A_1}^2$ is the slope parameter at zero recoil and 
$\large z= \frac{\sqrt{w+1} - \sqrt{2}}{\sqrt{w+1} + \sqrt{2}}$.
An alternative parametrization, obtained earlier, can be found in
\cite{ref:grinstein}.
%The ratio between the axial and vector form factors is included in
%${\cal K}(w)$.
%Theoretical predictions restrict 
%values of $\rho^2$ to be in the range: $-0.14<\rho^2<1.54$.

In the heavy quark limit ($m_{\rm b}\rightarrow \infty$), 
${\cal F}_{D^*}(1)= h_{A_1}(1)$
coincides with the Isgur-Wise function \cite{neub,neuc} which is
normalized to unity at the point of zero recoil.
Corrections to ${\cal F}_{D^*}(1)$ have been calculated to take into account
the effects of finite quark masses and QCD corrections 
%\cite{luke}. %AO 11/11/03
They yield ${\cal F}_{D^*}(1)$ $=0.91 \pm 0.04$ \cite{ref:PDG02}.
%(Appendix \ref{appendixC}).
%\cite{patricknote}. 

Experiments determine the product ${\cal F}_{D^*}(1) \Vcb$
by fitting this quantity and the slope $\rho_{A_1}^2$,
using the expression (\ref{eq:dgdw}), convoluted with the experimental
resolution on the $w$ variable.
%the measured $\frac{{\rm d} \Gamma}{{\rm d} w}$ distribution.
Since the phase space factor ${\cal K}(w)$ tends to zero
as $w\rightarrow 1$, the decay rate vanishes 
in this limit and the 
accuracy of the extrapolation relies on achieving a reasonably
constant reconstruction efficiency in the region close to $w=1$.  

Results of the following analysis are expressed in terms of the 
 $q^2$ variable.

\section{The DELPHI detector}

% SPR 9-1-03
% sentence below modified
% EPR 9-1-03
The DELPHI detector and its performance have been described in detail 
elsewhere \cite{ref:delphi}. 

The tracking system consisted of the 
Vertex Detector (VD), the Inner Detector (ID), the Time Projection Chamber
(TPC) and the Outer Detector (OD) in the barrel region while the Forward 
Chambers (FCA, FCB) covered the end-cap regions. The 
average momentum resolution for high momentum charged particles traversing 
the DELPHI magnetic field of $1.2$ T was 
% SPR 28-1-03
% Correct expression for sigmap/p**2
% EPR 28-1-03
$\sigma(p)/p^2 = 0.0006$ 
$(\GeVc)^{-1}$ in
the polar angle region between $30^\circ$ and $150^\circ$.  

The VD surrounded the beam pipe and consisted of three concentric layers of
silicon micro-strip detectors at radii $6$, $9$ and $11$ cm.
Until 1994 the VD layers were single-sided and provided only information 
in the $R\phi$ plane\footnote{ In the DELPHI coordinate system, $z$ is 
along the electron beam direction, $\phi$ and R are the azimuthal angle and 
radius
in the $xy$ plane, and $\theta$ is the polar angle with respect to 
the $z$ axis.}. In 1994 the innermost and outermost layers were 
replaced by double-sided silicon micro-strip modules 
providing 
% SPR 9-1-03
% sentence modified
% EPR 9-1-03
%full three-dimensional point information was provided.  
both $R\phi$ and $Rz$ measurements.
The ID was placed outside the VD and consisted of a jet chamber
providing $R\phi$ information, and a trigger chamber providing a measurement 
of the $z$ coordinate. 
% SPR 9-1-03
% sentence added
% EPR 9-1-03
In 1995 a new ID was installed with
a longer jet chamber and straw tubes replacing the trigger chambers.
The VD and ID were surrounded by the TPC, the main 
DELPHI tracking device. The TPC provided up to 16 space points per particle 
trajectory between radii of $40$ and $110$ cm. The OD consisted of 5 
layers of drift tubes and complemented the TPC by improving the momentum
resolution of charged particles. 

Hadrons were identified using the specific ionization ($dE/dx$) in the TPC
and the Cherenkov radiation in the barrel Ring Imaging CHerenkov detector 
(RICH) placed between the TPC and OD detectors. 
The muon identification relied mainly on the muon chambers, a set of drift 
chambers giving three-dimensional information located at the periphery of 
DELPHI after approximately $1$ m of iron.
Electron identification relied mainly on the electromagnetic calorimeter 
in the barrel region (High density Projection Chamber, HPC) which was a 
sampling device having a relative energy resolution of $\pm 5.5 \%$ for 
electrons with $46$ $\GeVc$ momentum, and a spatial resolution 
of $\pm 2$~mm in $z$.  
   
\section{Hadronic Event Selection and Simulation}
\label{sec:A}

 Hadronic $\Zz$ decays collected by DELPHI between 1992 and 1995 have 
been analysed. Each event was divided into two hemispheres by a plane
orthogonal to the thrust axis. To ensure that the event was well contained 
inside the fiducial volume of the detector, the cosine of the 
polar angle of the thrust
axis of the event had to lie between -0.95 and +0.95.
Charged and neutral particles were
 clustered into jets by using the LUCLUS  algorithm \cite{ref:luc} with 
the resolution
parameter $d_{join}=5~ \GeV$.

 About 3.4 million events were selected from the full LEP1 data sets.
The JETSET 7.3 Parton Shower program \cite{ref:luc} was used to generate 
hadronic $\Zz$ decays, which were followed through the detailed detector 
simulation DELSIM \cite{ref:delphi} and finally processed by the same 
analysis chain as the real data. A sample of about nine million
$\Zz \rightarrow q \overline{q}$ events was used. To increase the statistical
significance of the simulation, an additional sample of about 3.6 million
$\Zz \rightarrow b \overline{b}$ events was analysed, equivalent to about
17 million hadronic $\Zz$ decays. Statistics for the hadronic samples 
 are given in Table \ref{tab:stat}.
\begin{table}[htb]
\begin{center}
  \begin{tabular}{|c|c|c|c|}
    \hline
Year &  Real data & Simulated & Simulated\\
  &      & $\Zz \rightarrow q \overline{q}$ 
& $\Zz \rightarrow b \overline{b}$\\
    \hline
 1992+1993 & 1355805  & 3916050 & 1096199    \\
 1994+1995 & 2012921 &  5012881 & 2495335   \\
\hline
 Total & 3368726  & 8928931  & 3591534    \\
    \hline
  \end{tabular}
  \caption[]{\it { Analysed number of events. In 1992 and 1993
only two-dimensional vertex reconstruction was available.}
  \label{tab:stat}}
\end{center}
\end{table}

%SAO 18-2-03
%changed sentence, removed pi* definition
%EAO 18-2-03

\section{Selection of events for analysis}
\label{sec:channel}

%SAO 18-2-03 
%Put back the hemisphere definition.
%EAO 18-2-03
Events corresponding to candidates for the decay
$\Bdb \rightarrow \Dstarp \ell^- \overline{\nu}_{\ell}$ are selected
by requiring the presence of an identified lepton and of a 
$\Dstarp$ candidate in the same event hemisphere
% PR 13-11-02
which is defined by the direction of the jet containing the lepton.
%as the jet containing the lepton.
% End PR 13-11-02
$\Dstarp$ are measured using the decays to $\Do \pi^+$. $\Do$ mesons 
are reconstructed using their decays into $\Km \pi^+$,
$\Km \pi^+ \pi^+ \pi^-$ and $\Km \pi^+ (\pi^0)$.

%and the charged pion is denoted as $\pistar$
%in the following.

To reduce the contribution 
from $\Zz$ decays into light flavours,
% SPR 9-1-03
% sentence below modified
% EPR 9-1-03
the standard DELPHI event $b$-tagging variable \cite{ref:delphi}
is used to enhance the sample 
in 
$\Zz \rightarrow b \overline{b}$ decays. The $b$-tagging variable is 
essentially
%a cut has been applied on a variable provided
% by the
%$b$-tagging algorithm for the whole event \cite{ref:btag}. 
%This variable depends on 
the probability that 
the analysed
event originates from light quarks and is required to be less than 0.5 
for $\Do \rightarrow \Km \pi^+
~{\rm and}~ \Km \pi^+ (\pi^0)$ decays and less than 0.1 for the 
$ \Do \rightarrow \Km \pi^+ \pi^+ \pi^-$ decay.
%These values correspond to efficiencies of xxx.x % (yyy.y%) for light-quark 
%events (including charm) and of xxx.x % (yyy.y%) for $b$-events.
In addition, the mass of the $\Dstarp-\ell^-$ system is restricted
to the range between 2.5 and 5.5 $\GeVcd$.

\subsection{Lepton identification}
% SPR 9-1-03
% `` a `` before momentum has been suppressed in the sentence below
% EPR 9-1-03
Muons and electrons with momentum larger than 2 $\GeVc$ and at least one
associated hit in the VD are selected.
 
Muons are identified using standard algorithms %\cite{ref:btag}
\cite{ref:delphi} based on the matching
of the track reconstructed in the tracking system to the 
track elements provided by the barrel and forward
muon chambers. Loose selection criteria
are applied and the efficiency is $\sim 80\%$ for $\sim 1\%$ 
probability of hadron misidentification. 

Electrons are identified using a neural
network algorithm 
%\cite{ref:electrons}
providing about 75$\%$ efficiency within the calorimeter
acceptance. The probability for a hadron to fake an electron was about 1$\%$.
Electrons from photon conversions are mainly produced in the outer ID wall 
and in the inner TPC frame. About 80$\%$ of them were removed, with negligible
loss of signal, by reconstructing  
% SPR 9-1-03
% ``their materialization'' --> ``the conversion''
% EPR 9-1-03
the conversion vertex.

\subsection{Isolation of the $\Do \rightarrow \Km \pi^+$ decay channel}
\label{sec:kpi}

The kaon candidate corresponds to a particle 
% SPR 9-1-03
% ``of same'' --> ``with the same''
% EPR 9-1-03
with the same charge as the lepton, 
with a momentum
 larger than 1 $\GeVc$ and not identified as a pion by the standard algorithms
%\cite{ref:btag}
\cite{ref:delphi}
% SPR 9-1-03
% s removed in informationS
% EPR 9-1-03
which combine information provided by the ionization deposited
in the gas volume of the TPC and by the RICH detectors.
The pion candidate must have a charge opposite to the kaon and a 
momentum larger than 0.5 $\GeVc$. The pion and kaon candidates must both 
have at least one 
associated VD hit in $\rphi$ 
and be situated in the same 
event hemisphere as 
% PR 13-11-02
the jet containing the lepton. 
% End PR 13-11-02
The two tracks must intersect in space
% SPR 9-1-03
% sentence modified below
% EPR 9-1-03
to form a $\Do$ decay vertex candidate 
and those with a $\chi^2$ probability lower
than 10$^{-3}$ are rejected. $\Km \pi^+$ systems with a mass
between 1.81 and 1.92 $\GeVcd$ and with a momentum larger than
6 $\GeVc$ are selected as signal candidates. Resolutions on the reconstructed
$\Do$ mass measured on real and simulated events are given in 
Table~\ref{tab:resolm}. The selected mass window for the signal corresponds
to about $\pm3\sigma$.
A $\Do$ track is then reconstructed using the parameters of the $\Km$ and 
$\pi^+$
tracks fitted at their common vertex and imposing the condition that
% SPR 9-1-03
% we have added:'', quoted as $m(\Do)$ in the following,'' in 
% the sentence below 
% EPR 9-1-03
the $\Do$ mass, quoted as $m(\Do)$ in the following,
is 1.8645 $\GeVcd$ \cite{ref:PDG02}. 

\begin{table}[htb]
\begin{center}
  \begin{tabular}{|c|c|c||c|c|}
    \hline
$\Do$ decay channel &  92-93 MC & 92-93 data &  94-95 MC & 94-95 data\\
                    &$\MeVcd$    &$\MeVcd$     &$\MeVcd$  &$\MeVcd$\\  
    \hline
 $\Km \pi^+$&$16.0\pm0.8$  &$19.8\pm2.0$  &$16.7\pm0.6$ &$20.3\pm1.4$ \\ 
    \hline
$\Km \pi^+ \pi^+ \pi^-$  &$10.4\pm0.7$  &$12.7\pm1.8$  &$10.5\pm0.5$ & 
$12.1\pm1.0$\\ 
\hline
  \end{tabular}
  \caption[]{\it { Mass resolution for $\Do$ signals measured on real 
and simulated events.}
  \label{tab:resolm}}
\end{center}
\end{table}
The B decay vertex is obtained from the intersection of
the $\Do$ and the lepton trajectories. This vertex must have
a $\chi^2$ probability larger than 10$^{-3}$ and only $\Do - \ell^-$ pairs of
total momentum larger than 10 $\GeVc$ are kept. The B decay vertex is then 
required to be at a minimum distance from the position of the beam
interaction point. 
% SPR 9-1-03
% Sentence below modified to give the definition of these 'distances''
% EPR 9-1-03
This algebraic distance is evaluated
along the direction of the $\Do-\ell^-$ momentum.
A minimum distance 
% SPR 9-1-03
% Sentence below modified to give the definition of these 'distances''
% EPR 9-1-03
between the D and B measured decay points,
evaluated along the 
$\Do$ momentum, is also required. These conditions 
%given in Table \ref{table:1},
%are different if the position of the vertices is well measured or not along
%the beam axis,
depend on the number of $z$ VD hits associated to the tracks.
If the 
decay distance is not measured along $z$ with the VD, only cuts on decay
distances transverse to the beam direction are applied.
% SPR 4-2-03
% Sentence below modified 
% EPR 4-2-03
These requirements, which are rather loose, are given in Table \ref{tab:cuts}.
%corresponding to a decay distance divided by its error larger than -2 to -1.

% EPR 28-1-03
% SPR 4-2-03
%  Table below modified
% EPR 4-2-03
% SA0 13-3-03
% columns 2 and 3 in table 3 aligned to the left. 
% EA0 13-3-03
%\begin{table}[htb]
%\begin{center}
%  \begin{tabular}{|c|c|c|}
%    \hline
%  &  B vert. $\Leftrightarrow$ main vert. & D vert. $\Leftrightarrow$ B vert\\
%    \hline
%  distance in space  & $\ell/\sigma_{\ell}>1 $ & $\ell/\sigma_{\ell}>-1 $ \\
%  transverse distance & $\ell/\sigma_{\ell}>2 $ & $\ell/\sigma_{\ell}>-0.5 $\\
%    \hline
%  \end{tabular}
%  \caption[]{\it { Minimum requirements on the decay distance $(\ell)$ between
%the B and the main vertices and also between the D and the B vertices.}
%  \label{tab:cuts}}
%\end{center}
%\end{table}
\begin{table}[htb]
\begin{center}
  \begin{tabular}{|c|l|l|}
    \hline
$\Do$ decay channel  &  B vert. $\Leftrightarrow$ main vert. & D vert. 
$\Leftrightarrow$ B vert.\\
    \hline
 $\Km \pi^+$ & $\ell/\sigma_{\ell}>-1,~-2 $ & $\ell/\sigma_{\ell}>-2,~-1 $ \\
$\Km \pi^+ \pi^+ \pi^-$ & $\ell/\sigma_{\ell}>1,~2 $ & 
$\ell/\sigma_{\ell}>-1,~-0.5 $  \\
$\Km \pi^+ (\pi^0)$   & $\ell/\sigma_{\ell}>2,~2 $ & $\ell/\sigma_{\ell}>-1,
~-1 $  \\
    \hline
  \end{tabular}
  \caption[]{\it { Minimum requirements on the decay distance $(\ell)$ between
the B decay and the main vertices and also between the D and the B decay
vertices. The first 
number corresponds to the cut in space whereas the second is the
transverse distance cut, which is applied only when the z coordinate
is not measured. Negative distances corespond to positions behind the beam 
interaction point.}
  \label{tab:cuts}}
\end{center}
\end{table}

The $\Dstarp$ signal is identified by its decay to $\Do \pi^+$. Each
particle of charge opposite to the lepton candidate and emitted in the same
event hemisphere as the jet containing the lepton is considered as a
candidate for the $\pi^+$.
% SPR 9-1-03
% Sentence below modified 
% EPR 9-1-03
The track of this particle must form a vertex
with the $\Do$ and the charged lepton trajectories and the vertex fit
probability has to be higher than 10$^{-3}$. Signals for the cascade decay
$\Dstarp \rightarrow \Do \pi^+$ correspond to a peak in 
the distribution of the mass difference
$\delta m~=~m(\Do \pi^+)-m(\Do)$. 

% SPR 9-1-03
% Sentence below modified 
% EPR 9-1-03
 The global efficiencies to select signal events 
have been estimated using the simulation
(see Table \ref{tab:effic}), accounting for
all analysis steps described above, apart from the branching fractions of 
the $\Dstarp$ and 
of the $\Do$ into the selected decay channels.

\begin{table}[htb]
\begin{center}
  \begin{tabular}{|c|c|c|}
    \hline
$\Do$ decay channel &  92-93 MC & 94-95 MC\\
    \hline
 $\Km \pi^+$& $(19.2 \pm 0.7)\%$ &  $(22.3 \pm 0.6)\%$    \\
    \hline
$\Km \pi^+ \pi^+ \pi^-$  &  $(8.6 \pm 0.3)\%$  &   $(10.8 \pm 0.3)\%$  \\
    \hline
$\Km \pi^+ (\pi^0)$  &  $(8.7 \pm 0.3)\%$  &   $(10.4 \pm 0.2 )\%$  \\
\hline
  \end{tabular}
  \caption[]{\it { Global efficiencies of the analysis chain to reconstruct 
and select simulated signal events. 
% SPR 9-1-03
% Sentence below modified 
% EPR 9-1-03
The quoted uncertainties are statistical.}
%Quoted uncertainties are only of
%statistical origin.}
  \label{tab:effic}}
\end{center}
\end{table}

%{\bf Give the plot of $\delta m$ for the Kpi channel, 
%separating the two periods,
%draw on the same plot the wrong-sign combinations.}

% SAO 15-1-03
% changed textwidth of figures, added negative \vspace
% EAO 15-1-03
\begin{figure}[h]
\vspace{-1.cm}
  \begin{center}
%    \mbox{\epsfig{file=masakpi2.eps,width=10cm,height=6.5cm}}
%    \mbox{\epsfig{file=masakpi2.eps,width=0.55\textwidth,
%height=0.4\textwidth}}
    \mbox{\epsfig{file=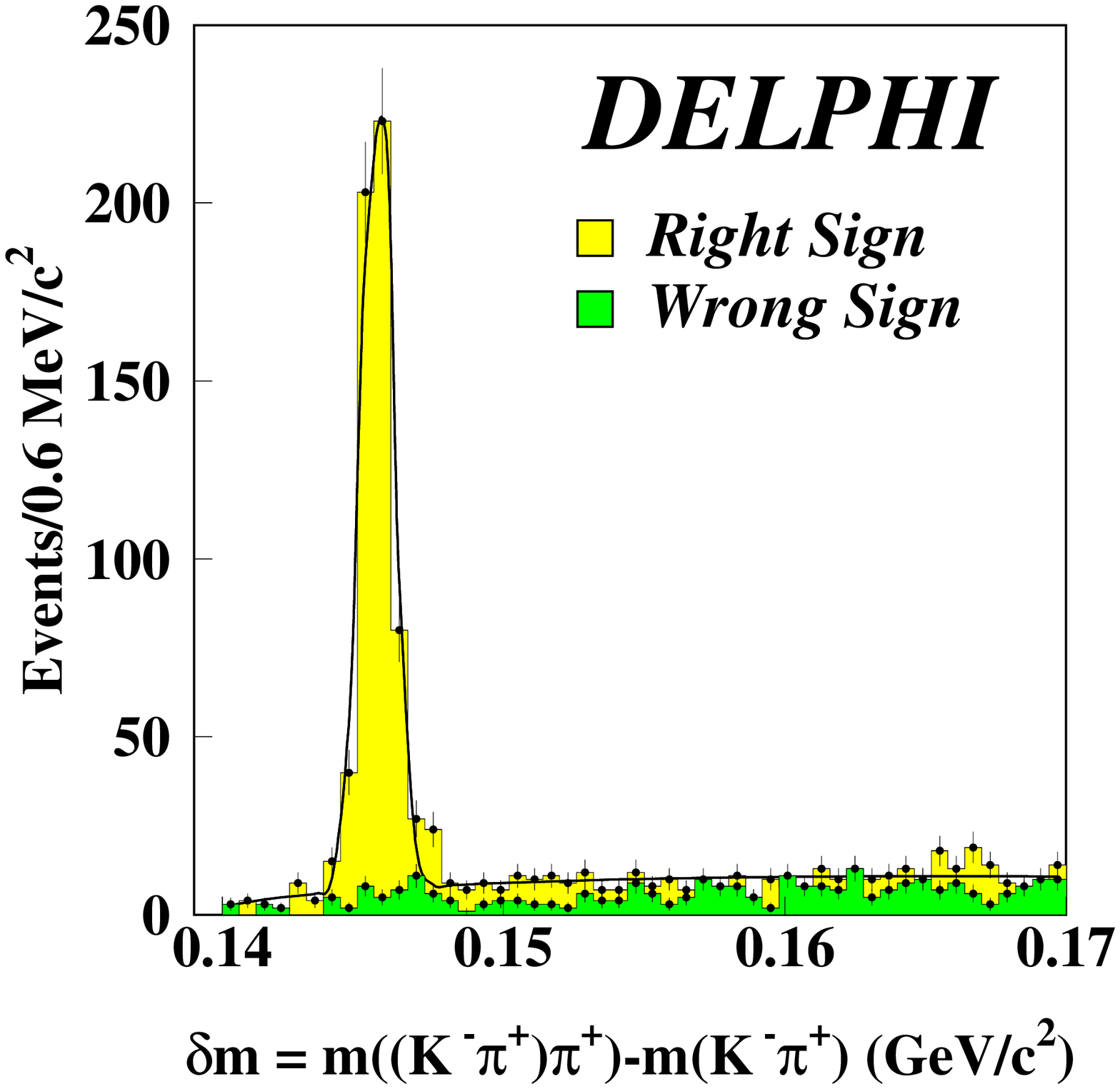,width=0.48\textwidth}}
  \end{center}
\vspace{-1.cm}
  \begin{center}
%    \mbox{\epsfig{file=masak3pi2.eps,width=10cm,height=6.5cm}}
%    
%\mbox{\epsfig{file=masak3pi2.eps,width=0.55\textwidth,height=0.4\textwidth}}
    \mbox{\epsfig{file=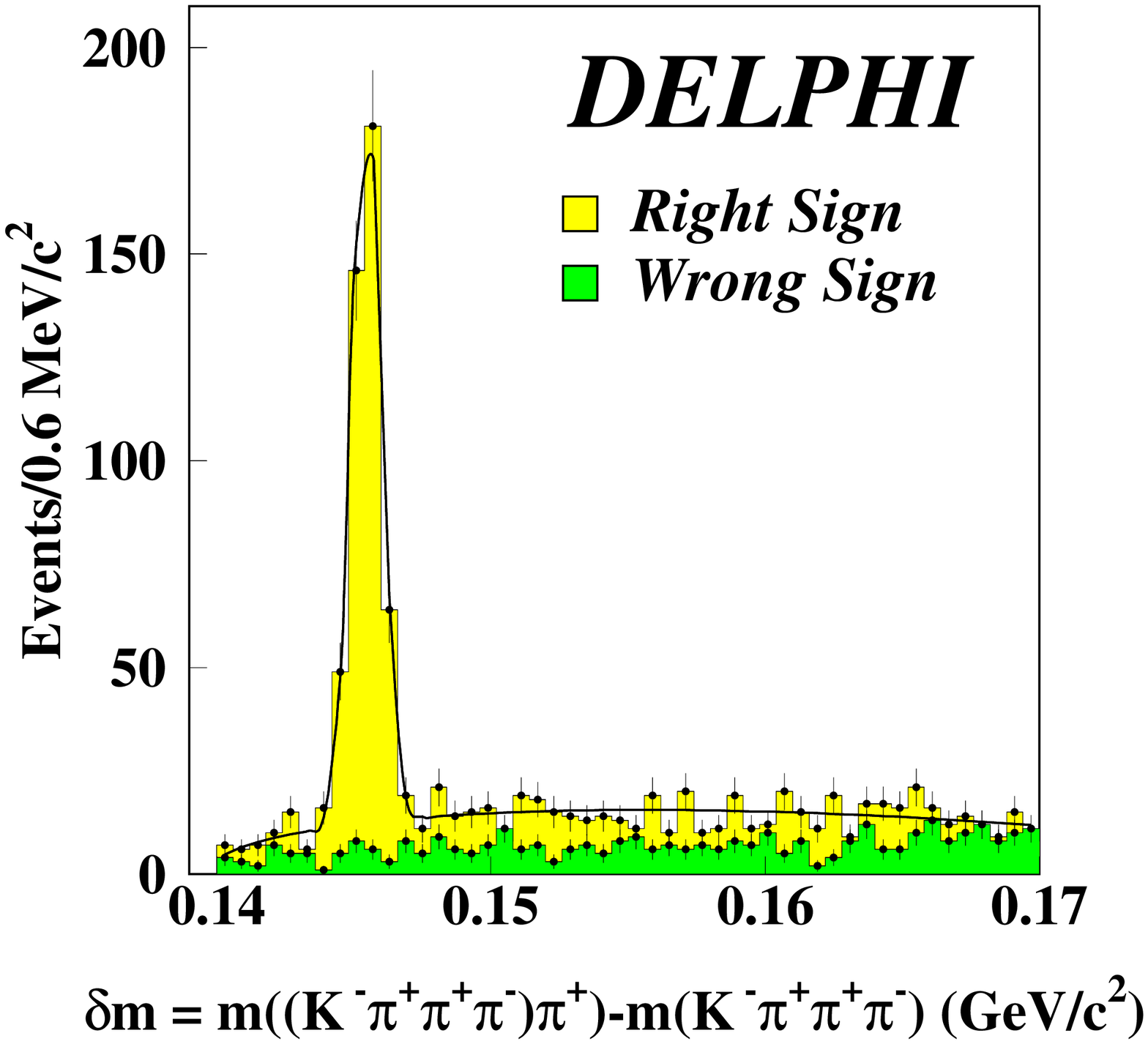,width=0.48\textwidth}}
  \end{center}
\vspace{-1.cm}
  \begin{center}
%    \mbox{\epsfig{file=masakpi02.eps,width=10cm,height=6.5cm}}
%    
%\mbox{\epsfig{file=masakpi02.eps,width=0.55\textwidth,height=0.4\textwidth}}
    \mbox{\epsfig{file=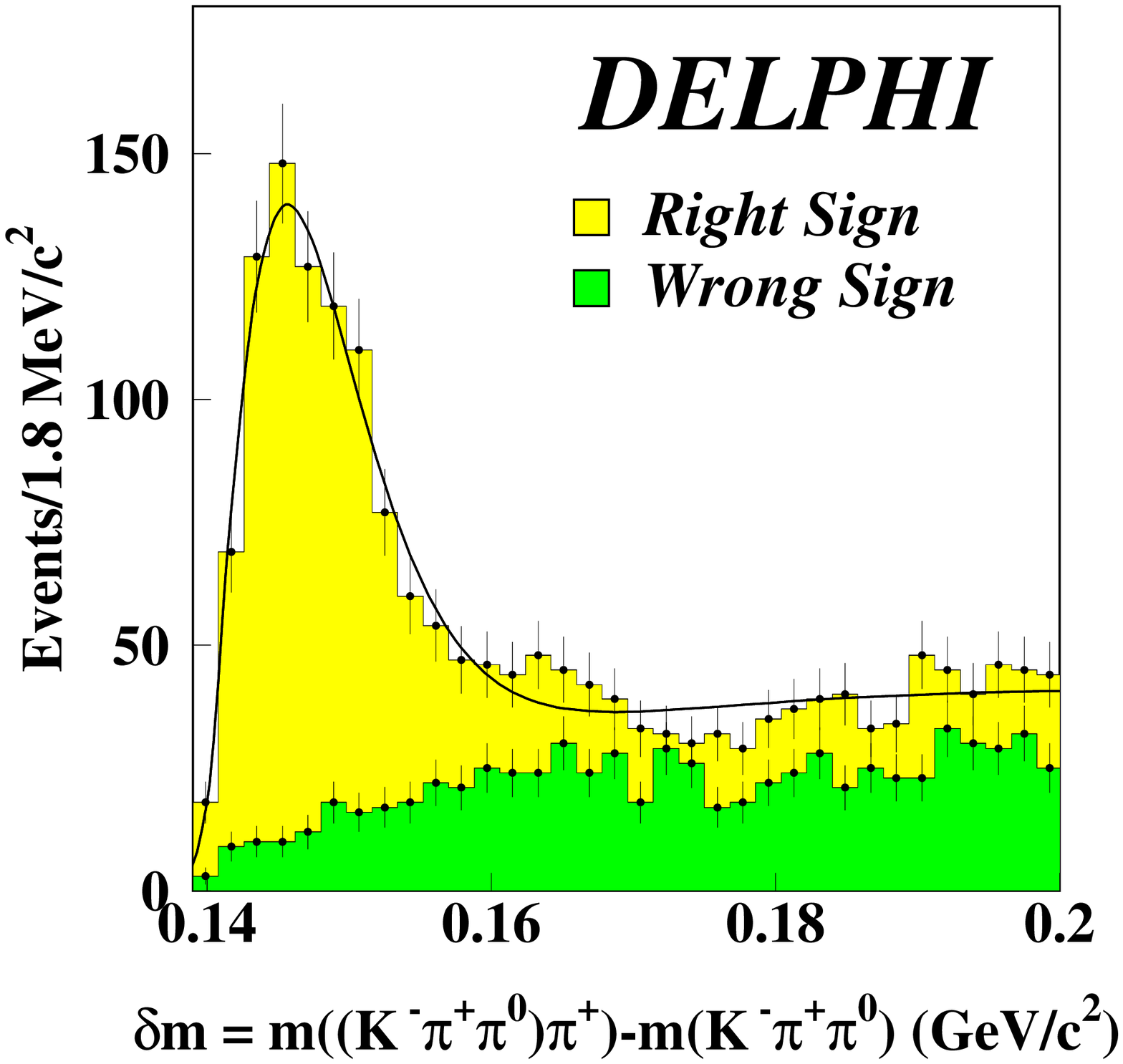,width=0.48\textwidth}}
  \end{center}
  \caption[]{\it { $\delta m=m(\Do \pi^+)-m(\Do)$ distributions for the 
  $\Do \rightarrow \Km \pi^+$ (upper),
$\Do \rightarrow \Km \pi^+  \pi^+  \pi^-$ (middle) 
and $\Do \rightarrow \Km \pi^+  (\pi^0)$ (lower) decay channels. 
Combinations with the wrong K-lepton charge correlation are superimposed as 
darker histograms.
Events registered in 92-93 and 94-95 have not been distinguished.
The curves show the fits to the right-sign distributions described in the 
text.}
   \label{fig:dmass}}
\end{figure}

\subsection{Isolation of the $\Do \rightarrow \Km \pi^+ \pi^+ \pi^-$ 
decay channel}

% SPR 28-1-03
% Change following lines to avoid duplicated sentences with respect to Kpi

%The kaon candidate corresponds to a particle of same charge as the lepton, 
%with a momentum
%larger than 1 $\GeVc$ and not identified as a pion by the standard algorithms,
%as in the previous analysed channel.
Similar selection criteria to those which were applied to isolate the
$\Do \rightarrow \Km \pi^+$ decay channel, are used. Differences
in the algorithm are related to the final state multiplicity.
Each of the three pion candidates must have a 
momentum larger than 0.5 $\GeVc$ and the total charge of the three pion system
has to be opposite to the kaon charge.
At least two, among the four charged particle track candidates for $\Do$ decay 
products
 must be 
associated to at least one VD hit in $\rphi$ and be situated in the same 
event hemisphere as 
% PR 13-11-02
the jet containing the lepton. 
% End PR 13-11-02
%The four tracks must intersect in space
%and only secondary $\Do$ vertices with a $\chi^2$ probability larger
%than 10$^{-3}$ are retained. 
As the reconstructed signal is narrower, 
$\Km \pi^+ \pi^+ \pi^-$ systems with a mass
between 1.84 and 1.90 $\GeVcd$ are selected. 
%and with a momentum larger than
%6 $\GeVc$ are selected as signal candidates. 
% SPR 9-1-03
% Sentence added
% EPR 9-1-03
% SAO 28-1-03
% Sentence added 
% EAO 28-1-03
The measured mass resolutions
are given in Table \ref{tab:resolm}.
%A $\Do$ track is formed using the parameters of the kaon and pions
%tracks fitted at their common vertex. The B decay vertex is obtained
%by intersecting the $\Do$ and the lepton trajectories. This vertex must have
%a $\chi^2$ probability larger than 10$^{-3}$ and only $\Do \ell^-$ pairs of
%total momentum larger than 10 $\GeVc$ are kept. The B decay vertex is then 
%required to be at a minimum distance from the position of the beam
%interaction point and a minimum distance is also required between the
%D and B decay points. 
% These conditions 
%depend on the number of $z$ VD hits attached to the tracks.
%If the 
%decay distance is not measured along $z$ with the VD, only cuts on decay
%distances transverse to the beam direction are applied.
Because of the higher combinatorial background, as compared with the 
$\Do \rightarrow \Km \pi^+$ decay channel, cuts 
on algebraic distances between the B and the main vertex and between
the D and the B decay vertex
are more
severe than for the previous channel. They are summarized in 
Table \ref{tab:cuts}.

%{\bf comment ambig. rejection}
The same set of four particles can give two $\Km \pi^+ \pi^+ \pi^-$
mass combinations if there is an ambiguity in the definition
of the $\Km$ and $\pi^-$ candidates. Only one combination is kept
in the analysis by 
% SPR 28-1-03 +SAO 
% Sentence below modified 
using criteria which are based on the available particle identification
information provided by the RICH and the TPC 
%unambiguous signatures 
%for these particles
 or,
if this information is missing, assuming that the $\Km$
has the larger momentum.
% EPR 28-1-03 +EAO 

The same selection criteria, as for the decay
$\Do \rightarrow \Km \pi^+$,  are applied to 
search for a $\Dstarp$ signal. 
% SPR 9-1-03
% Sentence below modified 
% EPR 9-1-03
 The global efficiencies to select signal events 
have been estimated using the simulation
(see Table \ref{tab:effic}), accounting for
all analysis steps described above, apart from the branching fractions of 
the $\Dstarp$ and 
of the $\Do$ into the selected decay channels.
%Global efficiencies to select signal events 
%(see Table \ref{tab:effic}), which include efficiencies
%of all analysis steps, apart branching fractions of the $\Dstarp$ and 
%of the $\Do$ into the considered decay channel, have been measured using 
%simulated events.

\subsection{Isolation of the $\Do \rightarrow \Km \pi^+ (\pi^0)$ 
decay channel}
\label{sec:kpi0}

The same criteria are applied, as in Section \ref{sec:kpi},
to select the $\Km$ and $\pi^+$ candidates apart from the 
cut on the $\Km \pi^+$ mass which is required now to be between
1.5 and 1.7 $\GeVcd$. This mass interval corresponds to the 
satellite peak position for the decay $\Do \rightarrow \Km \rho^+$
when the $\pi^0$ emitted from the $\rho^+$ is soft.
An estimate of the $\pi^0$ 4-vector is obtained by assuming that the
decay is of the type $\Do \rightarrow \Km \rho^+,
~\rho^+ \rightarrow  \pi^+ \pi^0$  and the $\Do$, $\rho^+$ and $\pi^0$
masses are used as constraints. 
% PR 13-11-02
In addition it has been assumed that the $\pi^0$ is contained in the 
plane defined by the $\Km$ and the $\pi^+$. When two solutions are
possible, a choice is made according to criteria which have been
defined using simulated events.
%When two solutions are possible only one is kept. 
% End PR 13-11-02

The $\Dstarp$ signal is identified by its decay to $\Do \pi^+$. Each
particle of charge opposite to the lepton candidate and emitted in the same
event hemisphere as the jet containing the lepton is considered as a
candidate for the $\pi^+$.
% SPR 9-1-03
% we have changed:''obtained'' to ``identified'' in 
% the sentence below 
% EPR 9-1-03
% identified by considering
% SPR 9-1-03
% Sentence below modified 
% EPR 9-1-03
% in turn each particle
% of charge opposite to the lepton candidate and
% emitted in the same event hemisphere as the jet containing the lepton
% to be the slow pion.
% the slow pion to be, in turn, each charged particle
% emitted in the same event hemisphere as the jet containing the lepton 
% candidate
% and of charge opposite to the lepton.
The track of this particle must form a vertex
with the $\Do$ and the charged lepton trajectories and the vertex fit
probability has to be higher than 10$^{-3}$. Signals for the cascade decay
$\Dstarp \rightarrow \Do \pi^+$ correspond to a peak in 
the distribution of the mass difference
$\delta m~=~m(\Do \pi^+)-m(\Do)$. The peak is broader than for
cases in which the $\Do$ was completely reconstructed using its
charged decay products.
% SPR 4-2-03
% Sentence below added 
% EPR 4-2-0 
Cuts on decay distances between the primary, the B and the D vertex
are given in Table \ref{tab:cuts}.

% SPR 9-1-03
% Sentence below modified 
% EPR 9-1-03
 The global efficiencies to select signal events 
have been estimated using the simulation
(see Table \ref{tab:effic}), accounting for
all analysis steps described above, apart from the branching fractions of 
the $\Dstarp$ and 
of the $\Do$ into the selected decay channels.
% The global efficiencies to select signal events 
%(see Table \ref{tab:effic}), which include efficiencies
%of all analysis steps described above, apart branching fractions of 
%the $\Dstarp$ and 
%of the $\Do$ into the considered decay channels, have been measured using 
%simulated events. 
The event selection described above does not ensure
that only $\Do$ decaying into the $\Km \pi^+ \pi^0$ channel are selected.
The simulation predicts that about 67$\%$ are of this origin and that there
are also: $\Km \ell^+ \nu_{\ell}$ (18$\%$),
$\Km \pi^+ {\rm X}$ (3$\%$) where X corresponds to neutrals,
$\Km \Kp$ (3$\%$) where the $\Kp$ is assumed to be a $\pi^+$
and 
% SPR 9-1-03
% Sentence below modified 
% EPR 9-1-03
the remaining 10$\%$ originates from various other channels. 
Apart from the 
last contribution, efficiencies have been determined for each individual 
channel using the simulation. 
Table \ref{tab:effic} shows the selection efficiency corresponding to
the weighted average for these channels.
The branching fractions measured for each of these 
channels have been used for the real data
(apart from $\Km \pi^+ {\rm X}$ which is assumed to be the same 
as in the simulation and equal to 5.6$\%$, with an error of 0.6$\%$) and a 
corresponding effective efficiency has been evaluated.
A correction factor for the remaining 10$\%$ of the events of undetermined
origin has been obtained from a fit to the simulation using the 
efficiencies of the four identified contributions (see Table
\ref{tab:effic}) and ensuring that the
% SAO 14-2-03
% Added 'simulated' to the BR.
simulated
% EAO 14-2-03
${\rm BR}(\Bdb \rightarrow \Dstarp \ell^- \overline{\nu}_{\ell})$ is
recovered. This 
%value of the (AO 18-2-03)
correction
has been used on real data with a relative uncertainty of $\pm25\%$,
corresponding to the statistical error of the fit
on simulated events.

\subsection{Selected event candidates}
\label{sec:combag}
The mass difference distributions corresponding to the variable 
$\delta m = m(\Do \pi^+)-m(\Do)$
obtained 
%for the two data taking periods and 
for the three channels, are shown in Figure \ref{fig:dmass}.

The numbers of $\Dstar$ candidates obtained by fitting
these distributions with a Gaussian
($\Do \rightarrow \Km \pi^+~{\rm and}~\Km \pi^+ \pi^+ \pi^-$) 
or a gamma distribution ($\Do \rightarrow \Km \pi^+(\pi^0)$)
for the signal, and a smooth distribution
for the combinatorial background\footnote{ The distribution selected for the 
combinatorial background
is $b_{\delta m}(\delta m)=(\delta m-m_{\pi})^{a_{0}} \left (  
\sum_{k=1}^{n}{a_{k} \delta m^{k-1}}
\right )$, with $n=2~{\rm or}~3$ and $a_{0}=0.5$.}
are given in Table \ref{tab:nbofdstar}.

%period has not been completed and it is not included in the present
% preliminary result.
%\begin{figure}[h]
%  \begin{center}
%    \mbox{\epsfig{file=~/plots_dmasa2.eps,width=14cm,height=9cm}}
%  \end{center}
%  \caption[]{\it { $\delta m~=~m(\Do \pi^+)-m(\Do)$ distributions for the 
%  $\Do \rightarrow \Km \pi^+  \pi^+  \pi^-$ channel correponding to the 
%   92-93 period of data  taking. Wrong-sign combinations are superimposed.}
%   \label{fig:massk3pi}}
%\end{figure}
%
%
%{\bf Give the plot of $\delta m$ for the K3pi channel, 
%remove the signal for 94-95.}
% SPR 9-1-03
% Presentation of Table below has been modified
% EPR 9-1-03
%\begin{table}[htb]
%  \begin{center}
%  \begin{tabular}{|c|c|}
%    \hline
% Data set & $\Dstar$ cand. \\
% \hline $\Km \pi^+$ 92-93  & 193 $\pm$ 15 \\
%        $\Km \pi^+$ 94-95  & 328 $\pm$ 16  \\
% $\Km \pi^+ \pi^+ \pi^-$ 92-93 &   144 $\pm$ 14 \\
% $\Km \pi^+ \pi^+ \pi^-$ 94-95  &  243 $\pm$ 17  \\
% $\Km \pi^+ (\pi^0)$ 92-93  & 286 $\pm$ 24 \\
% $\Km \pi^+ (\pi^0)$ 94-95  & 494 $\pm$ 27  \\
%\hline
%  \end{tabular}
%  \caption[]{\it { Number of $\Dstar-\ell$ candidate events selected
%in the two data taking periods and for the three $\Do$ decay channels.}
%  \label{tab:nbofdstar}}
%\end{center}
%\end{table}

\begin{table}[htb]
  \begin{center}
  \begin{tabular}{|c|c|c|}
    \hline
 Data set & 92-93  & 94-95 \\
 \hline 
 $\Km \pi^+$ & 193 $\pm$ 15 & 328 $\pm$ 16\\
 $\Km \pi^+ \pi^+ \pi^-$ &   144 $\pm$ 14 &  243 $\pm$ 17\\
 $\Km \pi^+ (\pi^0)$   & 286 $\pm$ 24 & 494 $\pm$ 27\\
\hline
  \end{tabular}
  \caption[]{\it { Number of $\Dstar-\ell$ candidate events selected
in the two data taking periods and for the three $\Do$ decay channels.}
  \label{tab:nbofdstar}}
\end{center}
\end{table}

%\begin{table}[htb]
%  \begin{center}
%  \begin{tabular}{|c|c|c|c|}
%    \hline
% Data set & $\Dstar$ cand. & $< m(\Dstar)-m(\Do) >$ & $\sigma[m(\Dstar)-m(\Do
%] 
%$\\
%          &                &    $ \MeVcd $         &      $ \MeVcd $       \\
% \hline $\Km \pi^+$ 92-93  & 193 $\pm$ 15 & 145.55 $\pm$ 0.05 & 
%0.63 $\pm$ 0.05  \\
%  $\Km \pi^+$ 94-95  & 328 $\pm$ 16 & 145.47 $\pm$ 0.04 & 
%0.47 $\pm$ 0.04  \\
%  $\Km \pi^+ \pi^+ \pi^-$ 92-93 &   144 $\pm$ 14 & 145.40 $\pm$ 0.07 &   
%0.63 $\pm$ 0.05  \\
%  $\Km \pi^+ \pi^+ \pi^-$ 94-95  &  243 $\pm$ 17 & 145.49 $\pm$ 0.04 & 
%0.46 $\pm$ 0.03 \\
% \hline $\Km \pi^+ (\pi^0)$ 92-93  & 286 $\pm$ 24 &  $\pm$ & 
%$\pm$   \\
% \hline $\Km \pi^+ (\pi^0)$ 94-95  & 494 $\pm$ 27 &  $\pm$ & 
%$\pm$   \\
%\hline
%  \end{tabular}
%  \caption[]{\it { Number of $\Dstar-\ell$ candidate events selected
%in the two data taking periods and for the three $\Do$ decay channels.}
%  \label{tab:nbofdstar}}
%\end{center}
%\end{table}

\subsection{$q^2$ measurement}
\label{sec:qdeux}
As explained in Section \ref{sec:principle}, to measure $\Vcb$ it is 
necessary to study the $q^2$ dependence of the differential 
semileptonic decay partial width $\dgsl/dq^2$. For signal events,
corresponding to the semileptonic decay
$\Bdb \rightarrow \Dstarp \ell^- \overline{\nu}_{\ell}$, the value
of $q^2$ has been obtained from the measurements of the $\Bdb$
and $\Dstarp$ four-momenta:
\begin{equation}
q^2=(p_{\ell}+p_{\overline{\nu}_{\ell}})^2=
(p_{\Bdb}-p_{\Dstarp})^2.
\end{equation}
The $\Dstarp$ 4-momentum is accurately measured, as all decay products
correspond to reconstructed charged particle trajectories.\footnote{ For the 
$\Do \rightarrow \Km \pi^+(\pi^0)$
decay channel the accuracy is reduced by about 10$\%$
because of the missing $\pi^0$.}
To improve on the determination of the $\Bdb$ momentum, information from
% SPR 9-1-03
% Sentence below modified ``are used'' --> ``is used''
% ``the evaluation'' --> ``an evaluation''
% EPR 9-1-03
all measured $b$-decay products is used, including an eva\-luation of the 
missing momentum in the jet containing the lepton and the positions of
the primary and of the secondary vertex, which are used as constraints to
define the direction of the $b$-hadron momentum. 
The nominal $\Bdb$ mass
is also used as a constraint in this fit. 
% SPR 9-1-03
% sentence added to explain how the missing energy is evaluated.
% EPR 9-1-03
The missing momentum in each jet has been evaluated by comparing the 
reconstructed
jet momentum with the expectation obtained by imposing energy-momentum
conservation on the whole event.
Finally, a momentum dependent correction 
is applied to the reconstructed $b$-hadron momentum so that it remains,
for simulated signal events,
centred on the generated value.

\begin{figure}

%SA0 15-1-03 
% Changed the plot and caption of this figure (resolution function). 
%EA0 15-1-03 
%    \mbox{\epsfig{file=rf022.eps,width=14cm}}
%    \mbox{\epsfig{file=rf022.eps,width=0.4\textwidth}}	
  \begin{center}
	\mbox{\epsfig{file=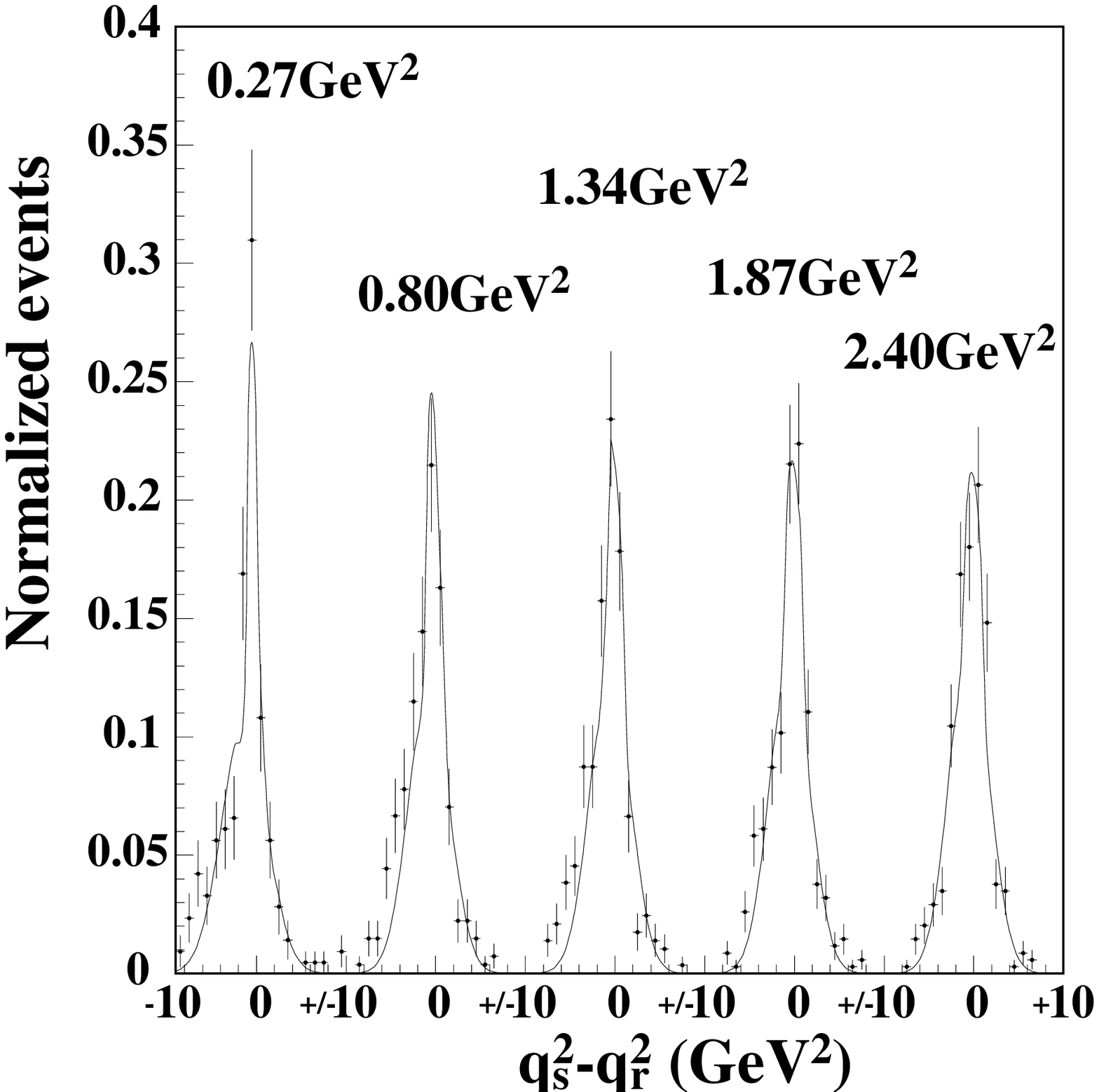,width=0.54\textwidth}
	\hspace{-1.cm}
	\epsfig{file=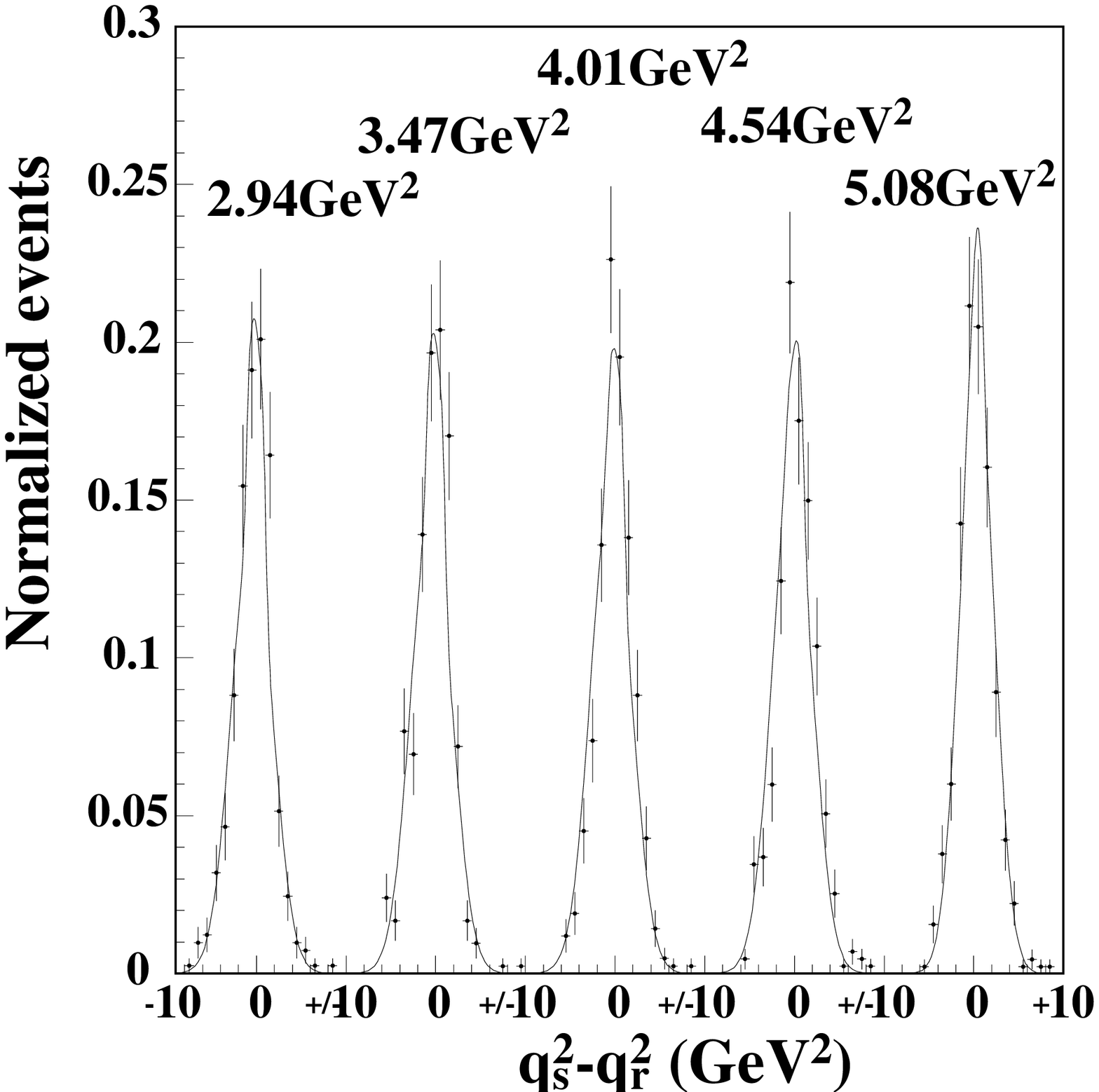,width=0.54\textwidth}}
  \end{center}
	\vspace{-1.cm}
  \begin{center}
	\mbox{\epsfig{file=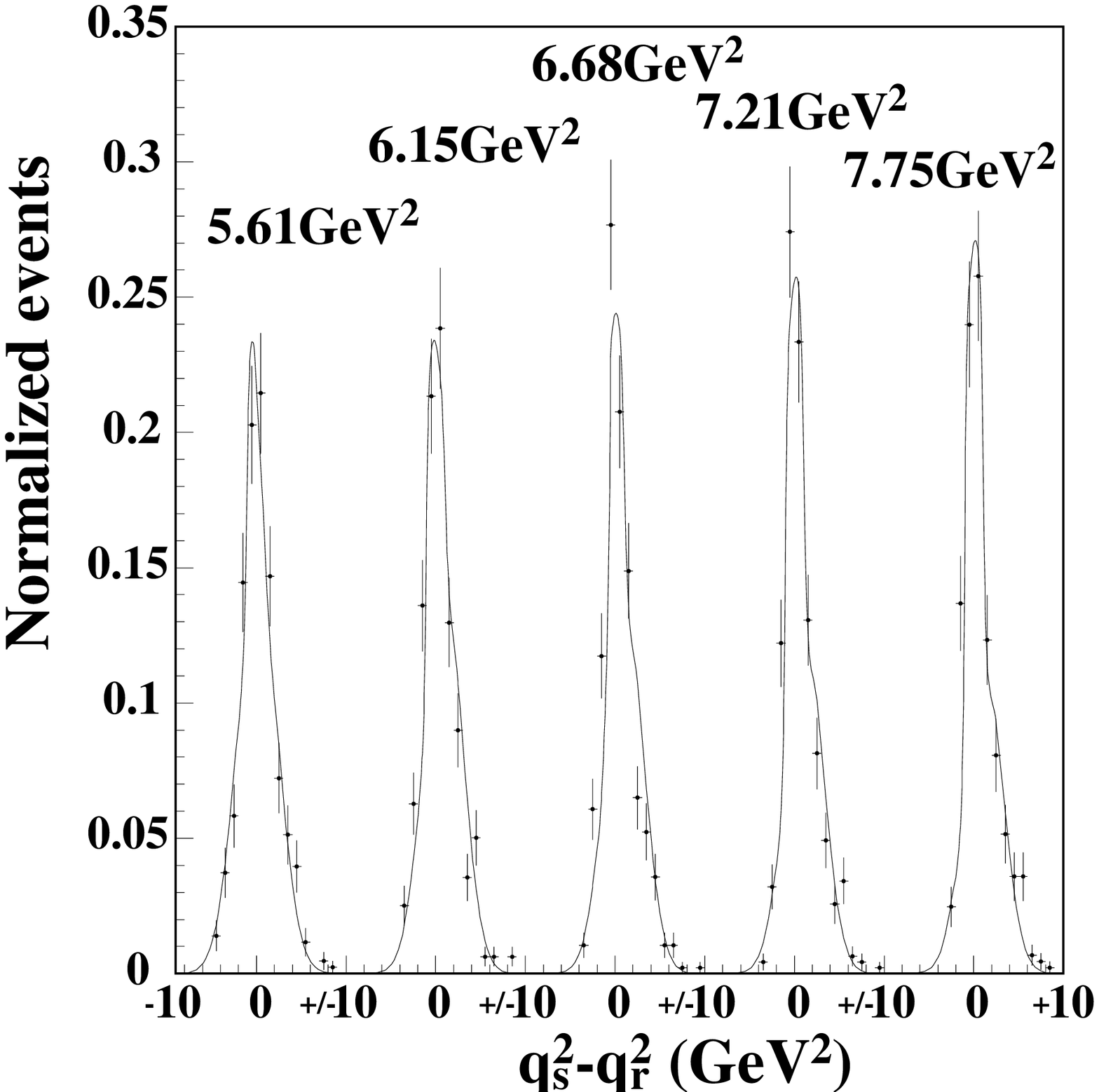,width=0.54\textwidth}
	\hspace{-1.cm}
	\epsfig{file=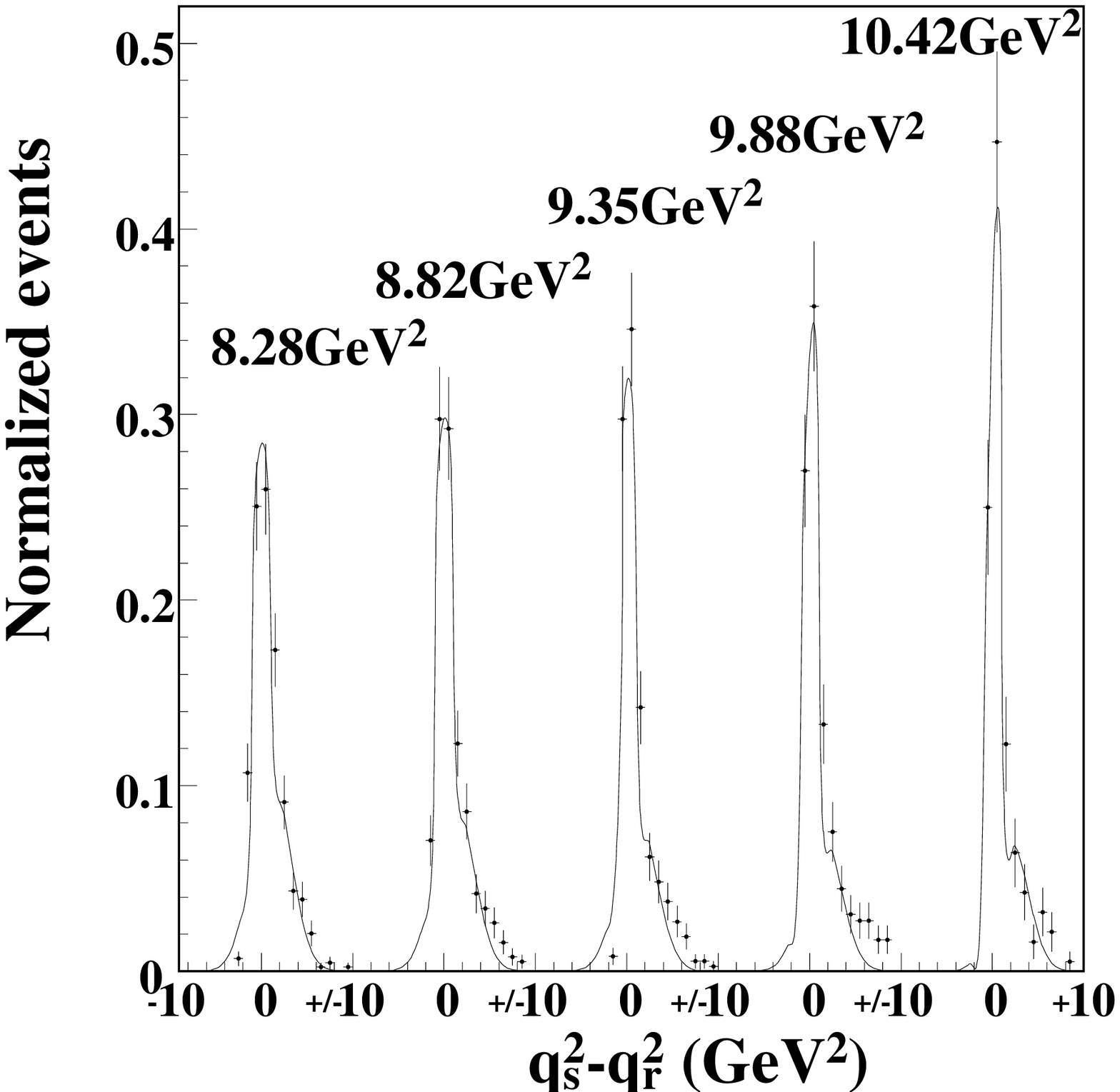,width=0.54\textwidth}}
  \end{center}
  \caption[]{\it { Fit of the ${\cal R}(q^2_s-q^2_r,q^2_s)$ slices for 
 the 94-95 data taking period, as expected from simulated events.
 The simulated $q^2_s$ central value is quoted above the corresponding slice.}
   \label{fig:res_function}}
\end{figure}

\begin{figure}
  \begin{center}
%    \mbox{\epsfig{file=comrfs2.eps,width=14cm}}
    \mbox{\epsfig{file=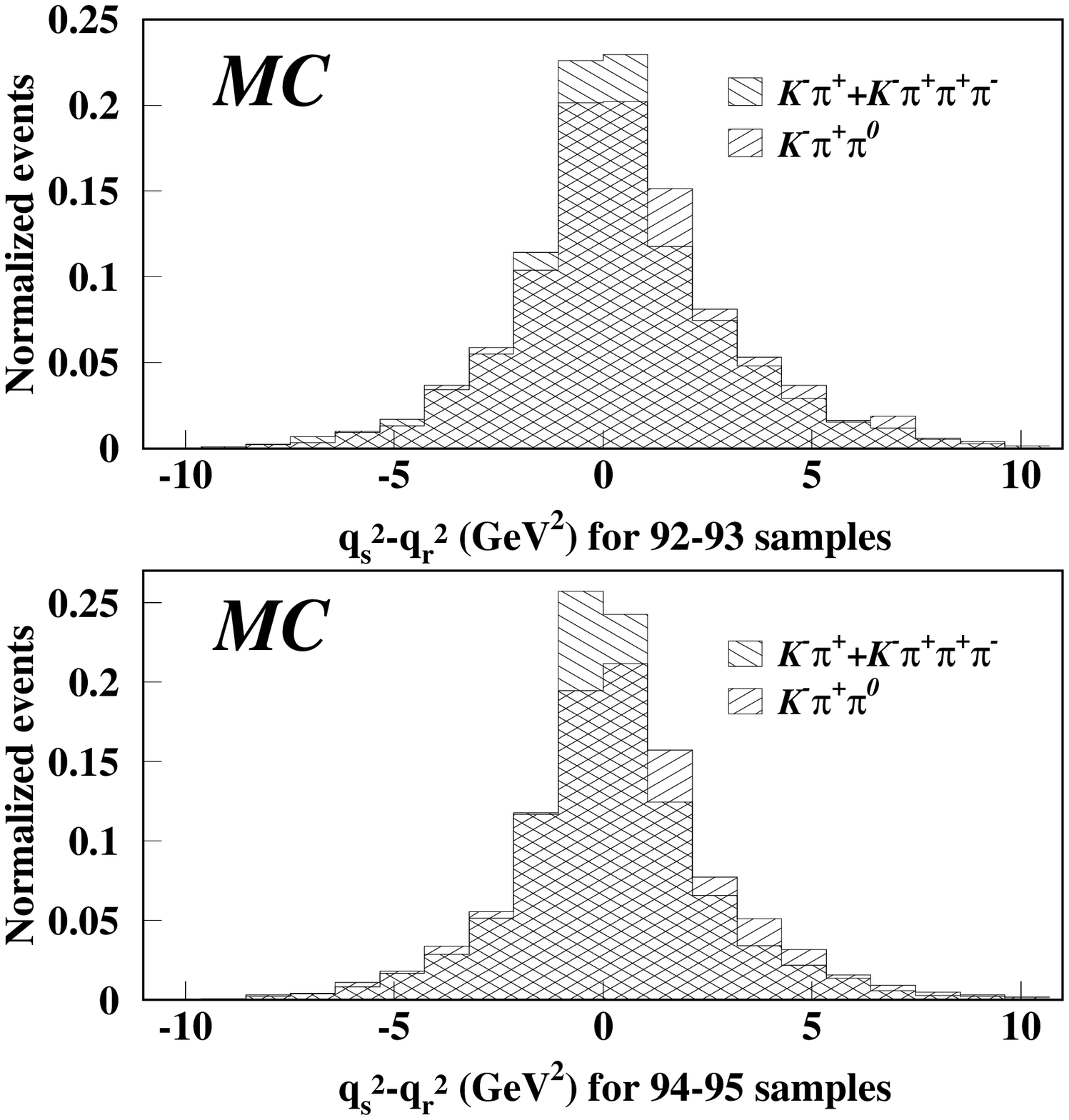,width=0.9\textwidth}}
  \end{center}
  \caption[]{\it {Comparison between the resolution functions obtained
for $\Do$ decay channels with and without a missing particle for the
92-93 and 94-95 data taking periods, as expected from simulated events.}
   \label{fig:res_kpi-kpipio}}
\end{figure}

% SPR 9-1-03
% we have changed a bit the sentence below
% EPR 9-1-03
% SAO 15-1-03
% changed the plot and place of the reference of the resolution function.
% EAO 15-1-03
 The smearing of the $q^2$ variable is studied with simulated signal events.
The function ${\cal R}(q^2_s-q^2_r,q^2_s)$ 
%is obtained which 
gives the 
distribution of the difference between the values of the reconstructed 
$q^2$, $q^2_r$, for events
generated with a given value $q^2_s$.
Twenty slices in  $q^2_s$ of the same width have been considered. Within
each slice, ${\cal R}(q^2_s-q^2_r,q^2_s)$ is parametrized as the sum
of two Gaussian distributions (see Figure \ref{fig:res_function}).
The two central positions of the Gaussians,
their standard deviations and the fraction of events corresponding to the
narrower Gaussian are parametrized with a linear dependence on $q^2_s$.
Such parametrizations are obtained independently for two sets of
ten slices. Typical values of these parametrizations correspond to
% SPR 28-1-03
%  Change unit for q^2, below
% EPR 28-1-03
$q^2$ resolutions of 0.3 and 2 $\GeV^2$ with about 50$\%$ of the events
included in the narrower Gaussian. Resolution distributions obtained
for $\Do$ reconstructed with only charged particles and for
the $\Km \pi^+ (\pi^0)$ decay channel are compared in 
Figure \ref{fig:res_kpi-kpipio}. 
%They are very similar, indicating
%that the constraints applied on the $b$-energy reconstruction largely
%reduce the effects expected from the energy reconstruction by the
%detector.

%SAO 21/5/03
% changed figures -> b&w to avoid they to be fuzzy
%EAO 21/5/03

\begin{figure}[h]
  \begin{center}
    \mbox{\epsfig{file=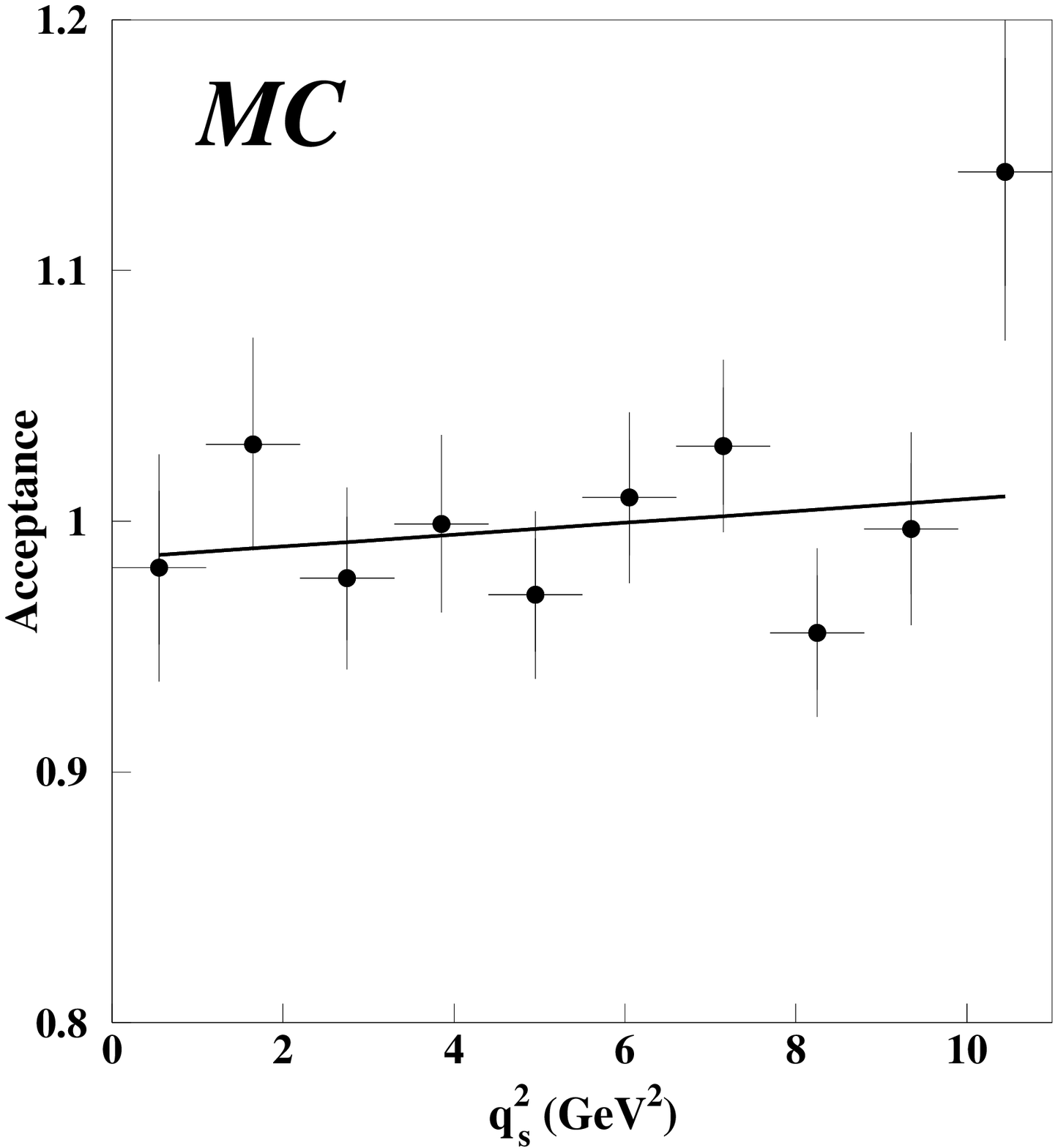,width=0.9\textwidth}}
  \end{center}
  \caption[]{\it {Stability of the acceptance 
as a function of the value of the simulated $q^2_s$.}
   \label{fig:acceptance}}
\end{figure}

The cuts applied to select the events which require a minimum momentum on 
the lepton, the $\Dstarp$ and the $\Dstarp-\ell$ system and the cut
on the minimum value for the mass of the $\Dstarp-\ell$ system can 
possibly introduce
a bias in the $q^2_s$ distribution. A $q^2_s$ dependent acceptance
correction, $\epsilon(q^2_s)$ has been evaluated by comparing the
simulated $q^2_s$ distributions for signal events
 before and after applying all analysis cuts. 
This correction has been normalized such that it does not change the
number of accepted events for which an overall efficiency has already been 
determined.
The corresponding distribution
is given in Figure \ref{fig:acceptance}. 
It is uniform and does not show evidence for
any significant bias. A linear dependence for the acceptance gives:
%SAO 27-1-03
% Modified GeV2 
%EAO 27-1-03
\begin{equation}
\epsilon(q^2_s)= (0.985 \pm 0.026)~+~ (0.0024\pm 0.0043)\times 
q^2_s,~q^2_s~{\rm in}\; \rm{GeV^2},
\end{equation} 
which is compatible with unity within quoted uncertainties.

As the cuts used in the analysis are very similar for all
data samples, the same $q^2_s$ 
dependent acceptance correction has been used for all channels and
data samples.

\section{The analysis procedure}
The purpose of this analysis is to determine the values of the parameters
% SPR 9-1-03
% sentence modified to add that we fit alo BR(b--> D** l nu)
% EPR 9-1-03
${\cal F}_{D^*}(1) \Vcb$ and $\rho_{A_1}^2$ introduced in Section 
\ref{sec:principle}, 
%and
%${\rm BR}(b \rightarrow \Dstarp {\rm X} \ell^- \overline{\nu}_{\ell})$,
using the measured $q^2_r$ distribution
of candidate events.
% SPR 28-1-03+SAO
% sentence modified again
% EPR 28-1-03+EAO
%, introduced in 
%Section  \ref{sec:principle}, 
%The first two parameters have been defined in Section \ref{sec:principle}
%and the third one corresponds to $\Dstarp$ production in the decay
%of higher mass charm states.
%using the measured $q^2_r$ distribution
%of candidate events. 
The predicted $q^2_r$ distribution for the signal is obtained
using the theoretical distribution
% SPR 9-1-03
% two lines below commented
% EPR 9-1-03
%corresponding to fixed values
%of the parameters mentioned before, 
corrected by the overall efficiency and the $q^2_s$ dependent acceptance,
%corresponding to the 
%cuts applied to the events (see Section \ref{sec:principle})
and then convoluted with the expected resolution function 
${\cal R}(q^2_s-q^2_r,q^2_s)$. The $q^2_r$ distributions for the other event
sources are taken from the simulation or from the real data
% SPR 28-1-03
% small change: for --> in the case of
% EPR 28-1-03
for the
combinatorial background.
The $q^2_r$ distributions are rather similar for the signal and 
other event categories because the procedure used to evaluate 
$q^2_r$ from the $\Bdb$ and $\Dstarp$ 4-momenta overestimates
the real $q^2$ value for background events.
% SPR 9-1-03
% Sentence added.
% EPR 9-1-03
% SPR 28-1-03
% procedures --> algorithms
% EPR 28-1-03
This is because algorithms have been defined for signal events
and thus do not include the additional hadrons emitted in
background sources.
To enhance the separation between the signal and other event sources,
three other variables have been used. As a result, the branching fraction for 
$\Dstarp$ production in the decay of higher mass charm 
states,
 ${\rm BR}(b \rightarrow \Dstarp {\rm X} \ell^- \overline{\nu}_{\ell})$, has 
also been measured. 
Before describing these quantities,
the different  event classes
% SPR 9-1-03
% in  changed to to
% EPR 9-1-03
% SPR 28-1-03+SAO 
% described --> explained 
% Changed a bit the sentence
% EPR 28-1-03+EAO 
contributing to the analysis are explained.

\subsection{The Event Sample composition}

 In addition to the signal (${\rm S}_1$), which corresponds to the decay
$\Bdb \rightarrow \Dstarp \ell^- \overline{\nu}_{\ell}$, 
there are six classes of events which contribute to the background:
\begin{itemize}
\item the combinatorial background (B) under the $\Dstarp$ peak; 
%Its 
%behaviour is studied using real data events corresponding to $\delta m$ values
%situated on the high side of the $\Dstarp$ peak 
%between 0.15 and 0.17 $\GeVcd$
%and events from wrong-sign 
%($\Dstarp-\ell^+$) combinations.

%{\bf What about a possible difference in the behaviour of right- 
%and wrong-sign
%combinatorial backg. using simulated events? What about a possible
%difference for the right-sign comb. backg. situated exactly under
%the $\Dstarp$ peak as compared with the one corresponding to higher values
%of $\delta m$.}

\item real $\Dstarp-\ell^-$ events with the $\Dstarp$ produced in the decay
of an excited charmed state (${\rm S}_2$). These events correspond 
to the decay chain
$b \rightarrow \Dstarstar \ell^- \overline{\nu}_{\ell},~
\Dstarstar \rightarrow \Dstarp {\rm X}$. 
%Their contribution is evaluated
%from the simulation using constraints from existing measurements on 
%the total and on narrow-states rates.
In the present analysis, $\Dstarstar$ includes resonant as well as 
nonresonant D$n \pi$ systems;

\item  real $\Dstarp-\ell^-$ events with the lepton originating from the decay
of another charmed hadron (${\rm S}_3$);
%or of a $\tau^-$ lepton (${\rm S}_3$). 
%Their contribution is evaluated from the simulation, using constraints
%from present rate and sample composition measurements
  
\item events in which the $\Dstarp$ is emitted during the hadronization
of a charmed quark jet in $\Zz \rightarrow c \overline{c}$ events 
(${\rm S}_4$); 
%The 
%accompanying lepton candidate is necessarily a fake.  Their contribution is 
%evaluated from the simulation.

\item $\Zz \rightarrow b \overline{b}$ events with a real $\Dstarp$ candidate
accompanied by a fake lepton of opposite sign (${\rm S}_5$);  
%Their contribution is 
%%evaluated from the simulation but corrections have been applied from
%independent studies which have compared the fake lepton rates between 
%real and simulated events.

\item  real $\Dstarp-\ell^-$ events with the lepton originating from the decay
of a $\tau^-$ lepton (${\rm S}_6$). 
%Their contribution is evaluated from the simulation, using constraints
%from present rate and sample composition measurements

\end{itemize}

\subsection{Separation of signal from background events}
\label{sec:subsepar}
There are two main classes of events which either do or do not
contain a real
$\Dstarp$.
The variable $\delta m$  ($=m(\Do \pi^+)-m(\Do)$)
allows the two classes to be separated
(see Figure \ref{fig:dmass}).
%events with a real $\Dstarp$
%signal from the combinatorial background (see Figures \ref{fig:masskpi}) and 
%\ref{fig:massk3pi}).
Variables, $d_{\pm}$, are used
to separate the different classes of events with a real $\Dstarp$.
They are obtained from a measurement of the number of charged particle
tracks (excluding the charged lepton, the pion coming from the $\Dstarp$
and the $\Do$ decay products) which are
 compatible with the $b$-decay vertex or
with the main vertex.
For the signal (${\rm S}_1$), it is expected that all other charged 
particles in the
% SPR 9-1-03
% apart for -->  apart from
% EPR 9-1-03
$b$-jet are emitted from the
beam interaction region. This will be also true for (${\rm S}_4$), 
the remaining
background from $\Zz \rightarrow c \overline{c}$ events,
and for (${\rm S}_6$). For the other
classes (${\rm S}_2$, ${\rm S}_3$ and ${\rm S}_5$) it is expected that, 
for most 
of the events, one
or more additional charged particles are produced at the $b$-vertex.

The variables $d_{\pm}$ are defined in the following way:
\begin{itemize}
\item all charged particles, other than the $\Dstarp$ decay products
and the lepton, emitted in the same event hemisphere
as the $b$-candidate,  with a momentum larger than 500 $\MeVc$, 
which form a mass with the $\Dstarp-\ell^-$ system lower than 6 $\GeVcd$
and which have values for their impact parameters to the 
$b$-decay vertex smaller than 2 and 1.5 $\sigma$ in $\rphi$ and $z$
respectively, are considered;

\item
% PR 13-11-02
selected particles, having the same $(+)$ or the opposite $(-)$
charge as the lepton are considered separately.
% End PR 13-11-02 
If there are several candidates in a class, the one with the 
largest impact
parameter to the main vertex is retained and the quantity:
\begin{equation}
x_{\pm} = \epsilon(\rphi) \times nsig(\rphi)^2 + \epsilon(z) \times nsig(z)^2
\end{equation}
is evaluated,
where $\epsilon$ and $nsig$ are, respectively,
the sign and the number of standard deviations
for the track impact parameter relative to the main vertex.
% PR 13-11-02
The sign of the impact parameter is taken to be positive (negative)
if the corresponding track trajectory intercepts
the line of the jet axis from
% SPR 9-1-03
% comma added
% EPR 9-1-03
the main vertex downstream (upstream) from that vertex.
% End PR 13-11-02
\end{itemize}
% SAO 15-1-03
% changed figures and caption
% EAO 15-1-03
% SAO 21-5-03
% changed figures again
% EAO 21-5-03
\begin{figure}
\vspace{-1.3cm}
  \begin{center}
  \epsfig{file=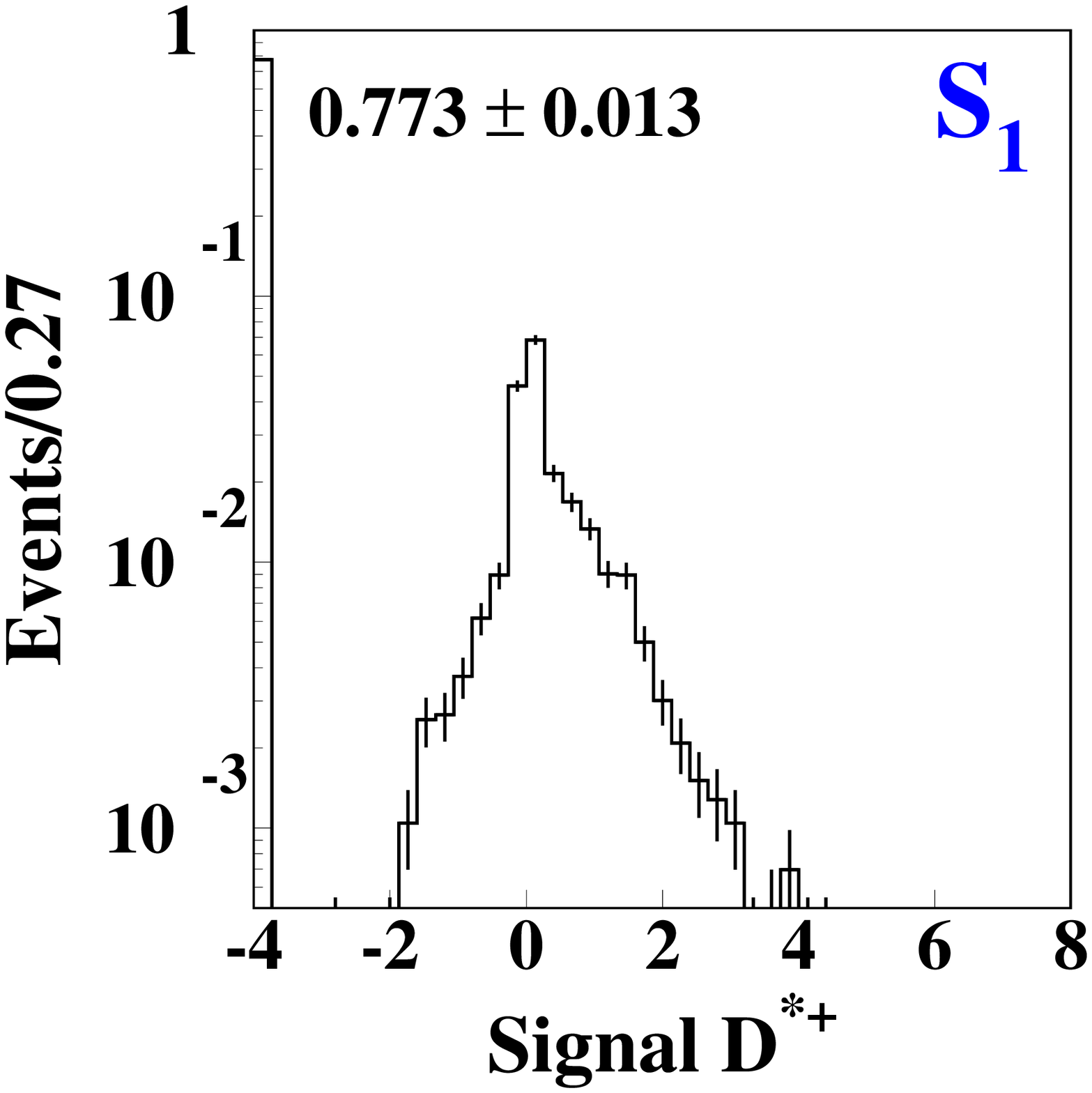,width=0.35\textwidth}
\hspace{0.4cm}
  \epsfig{file=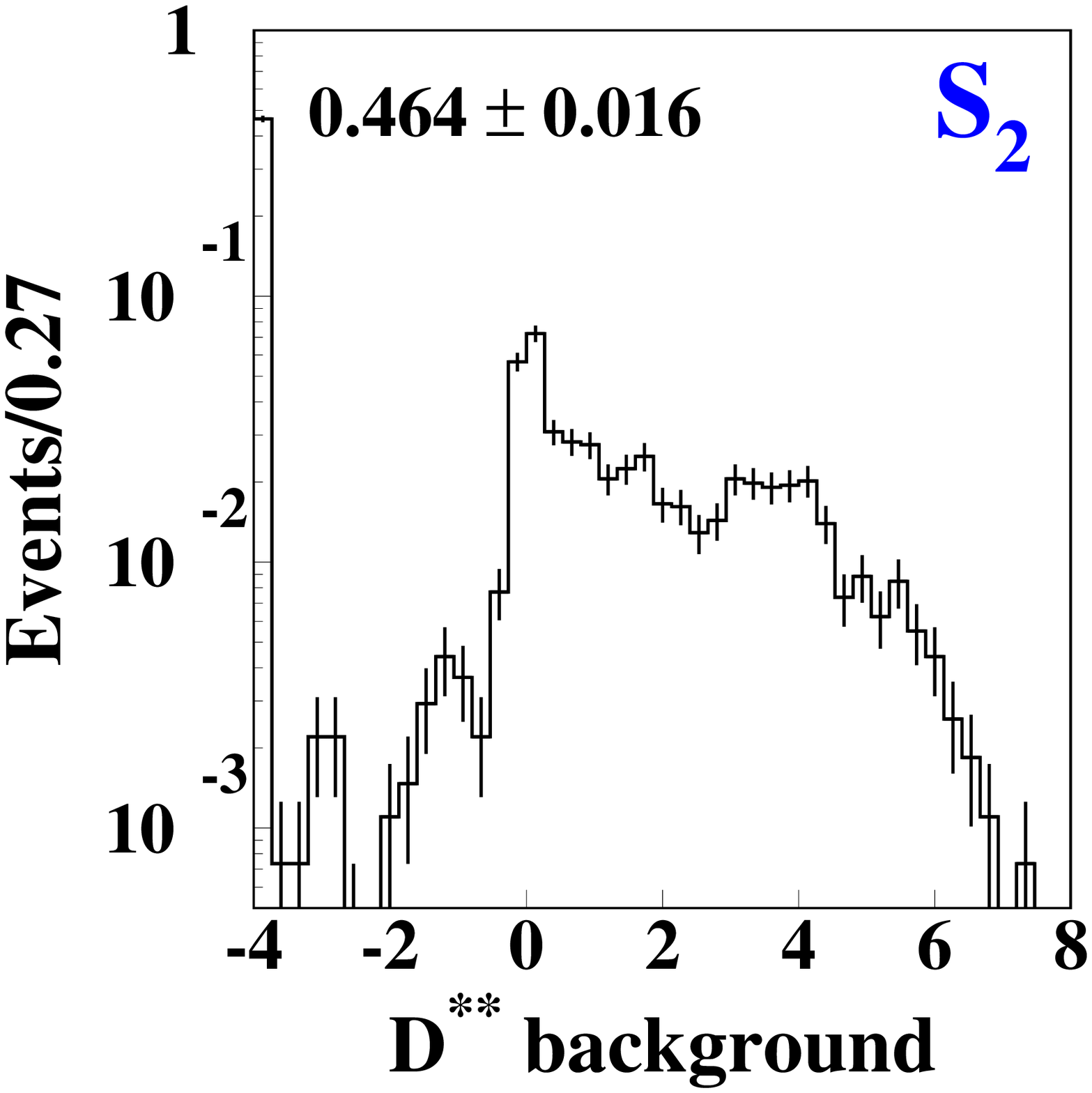,width=0.35\textwidth}
  \end{center}
\vspace{-1.2cm}
  \begin{center}
  \epsfig{file=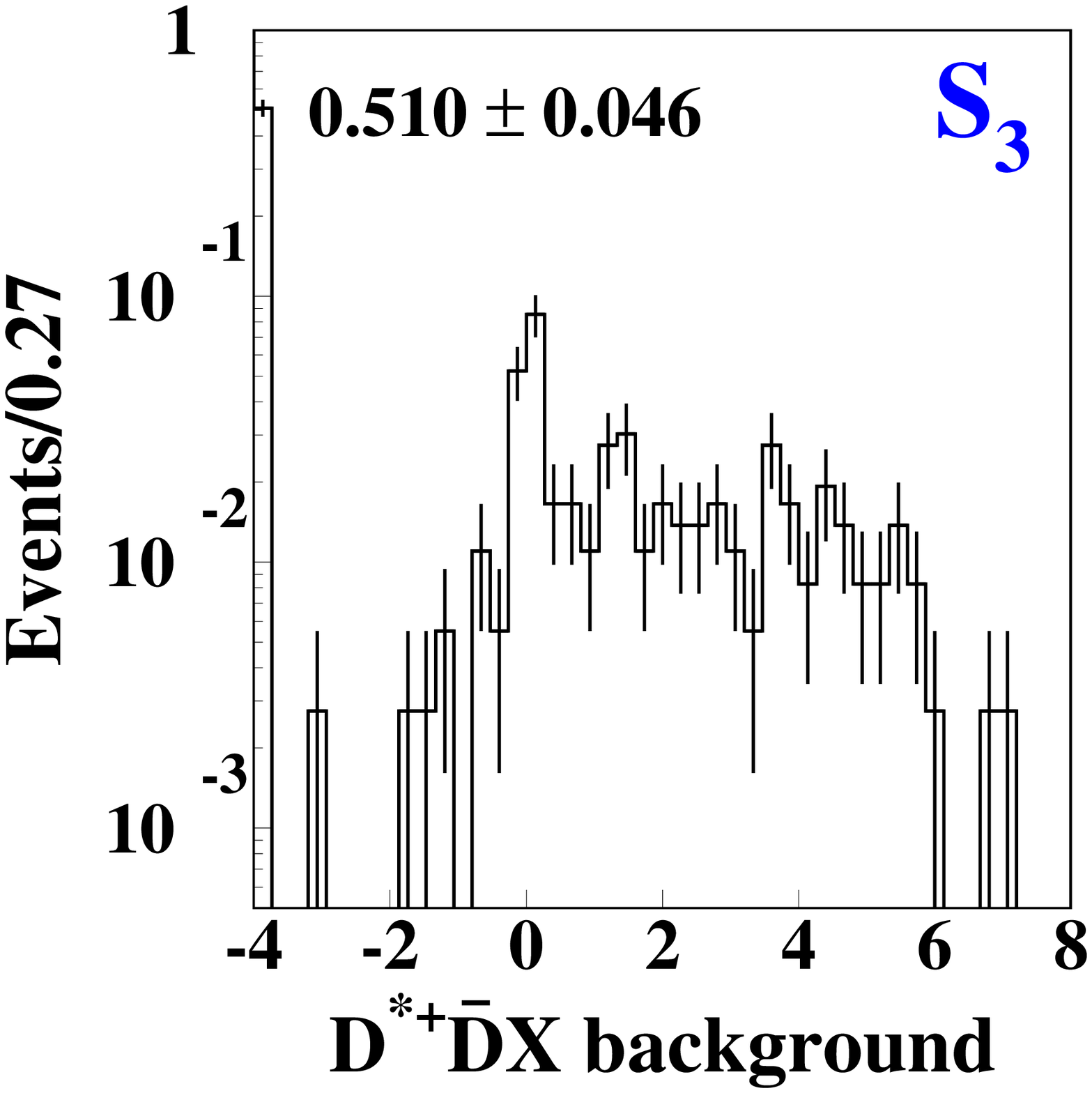,width=0.35\textwidth}
\hspace{0.4cm}
  \epsfig{file=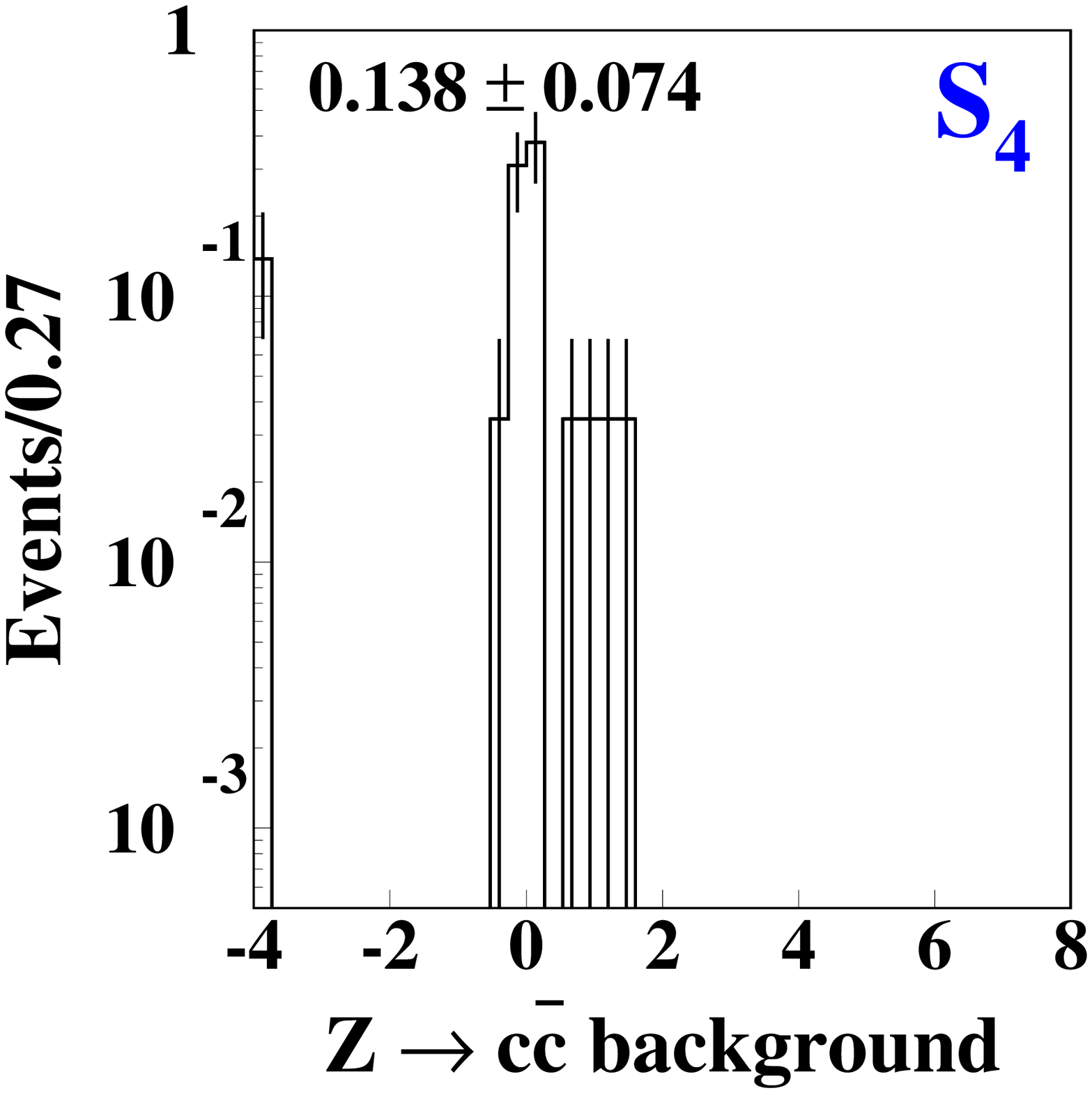,width=0.35\textwidth}
  \end{center}
\vspace{-1.2cm}
  \begin{center}
  \epsfig{file=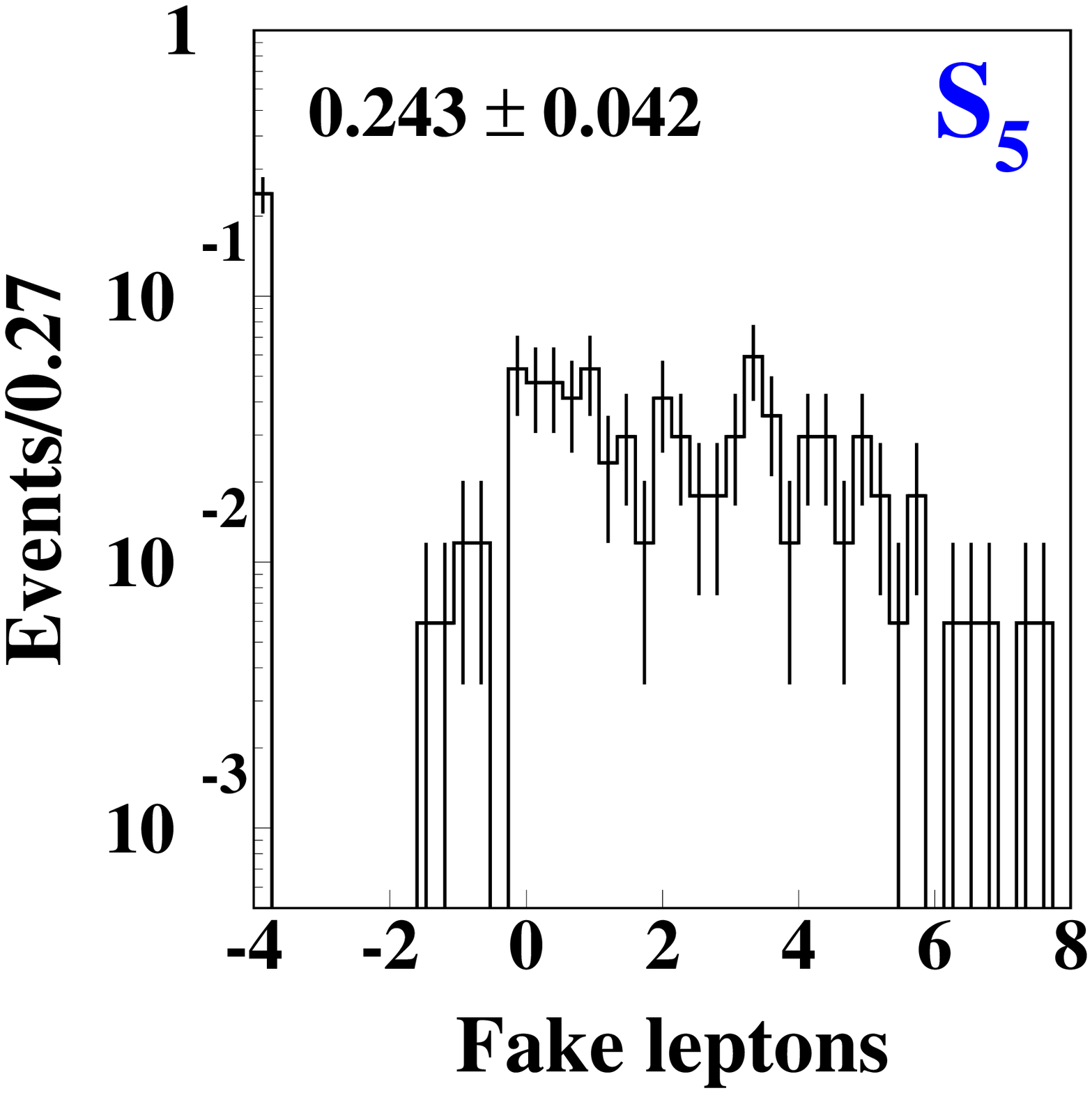,width=0.35\textwidth}
\hspace{0.4cm}
  \epsfig{file=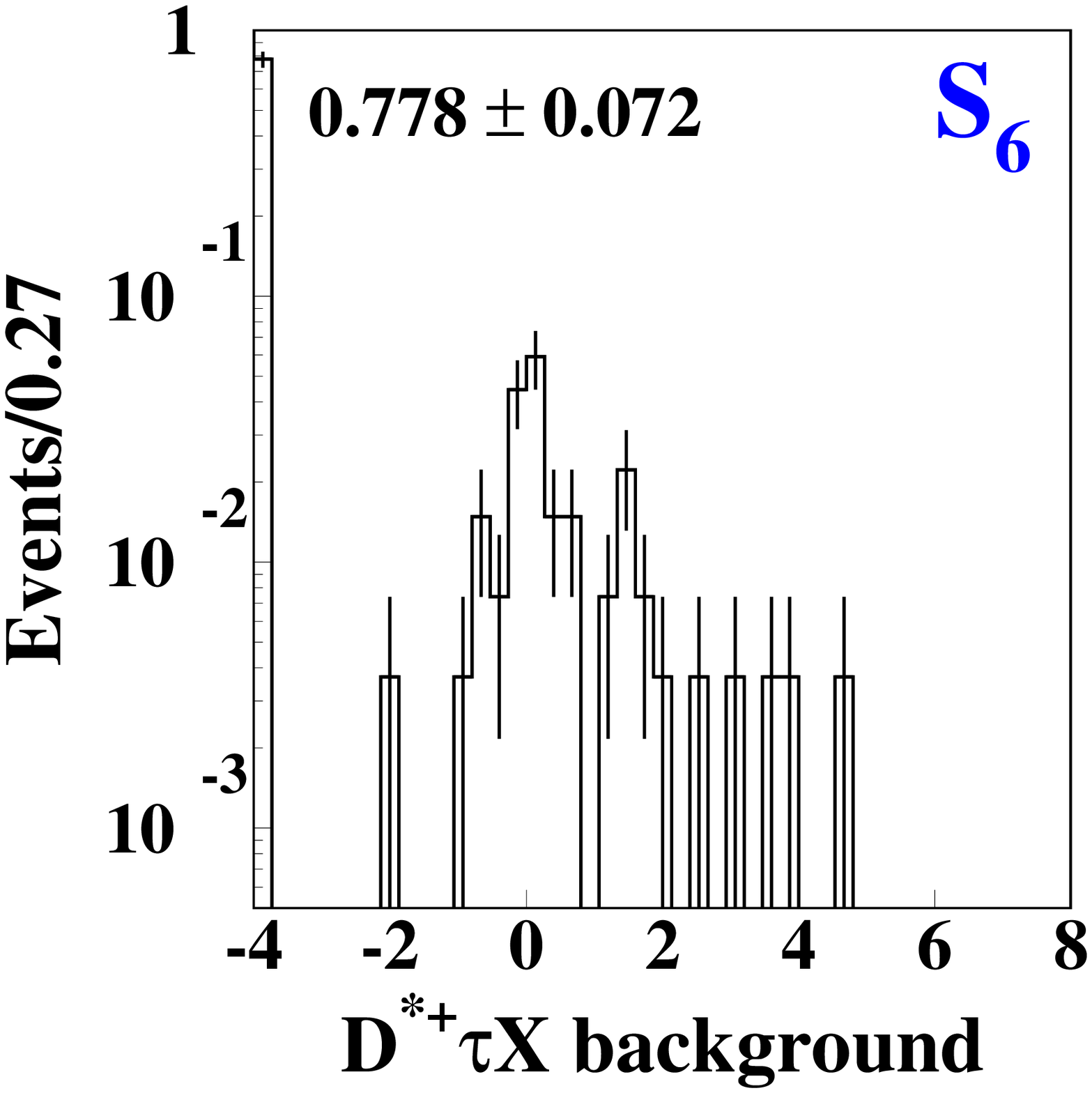,width=0.35\textwidth}
  \end{center}
\vspace{-1.2cm}
  \begin{center}
  \epsfig{file=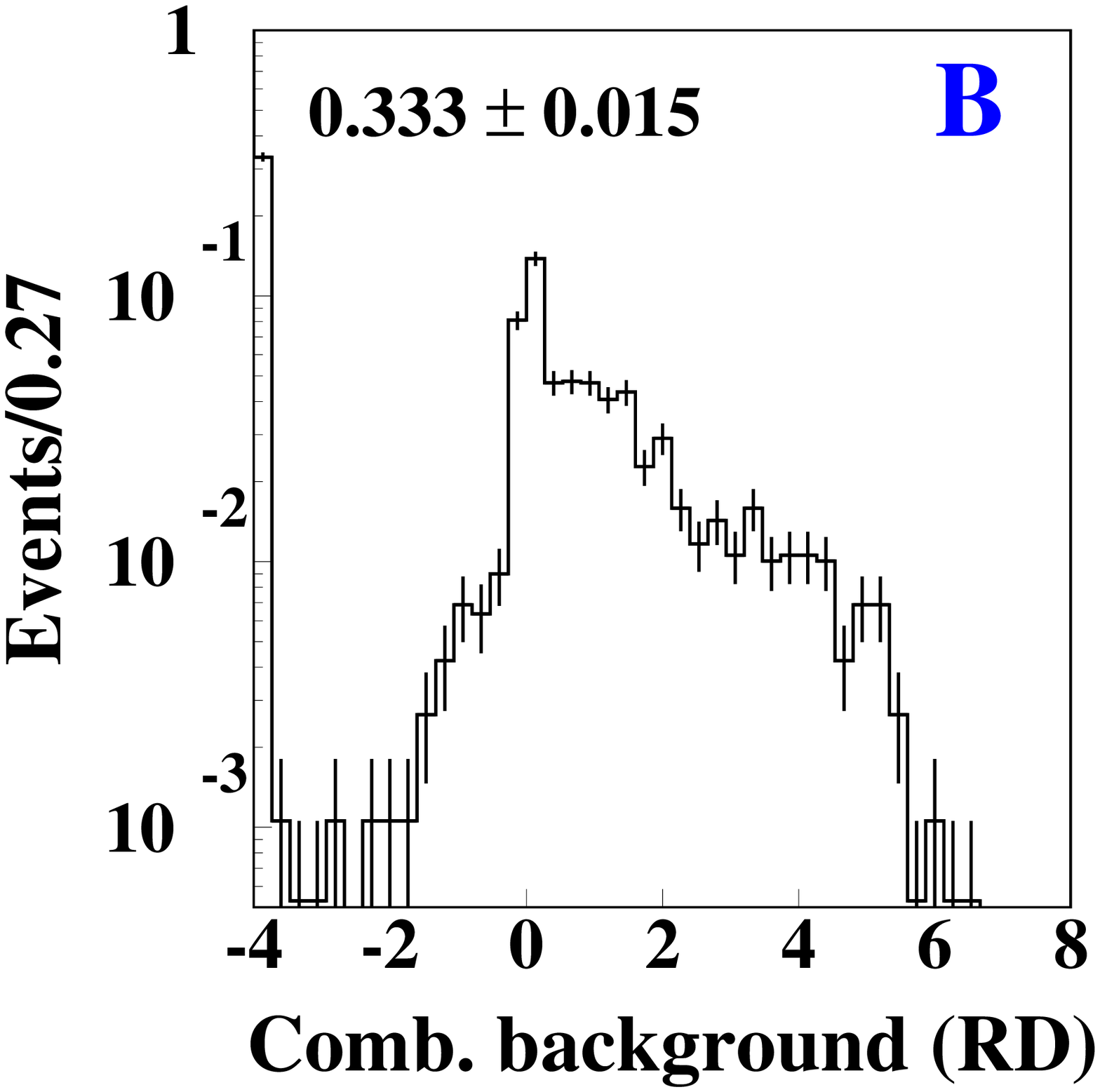,width=0.35\textwidth}
\hspace{0.4cm}
  \epsfig{file=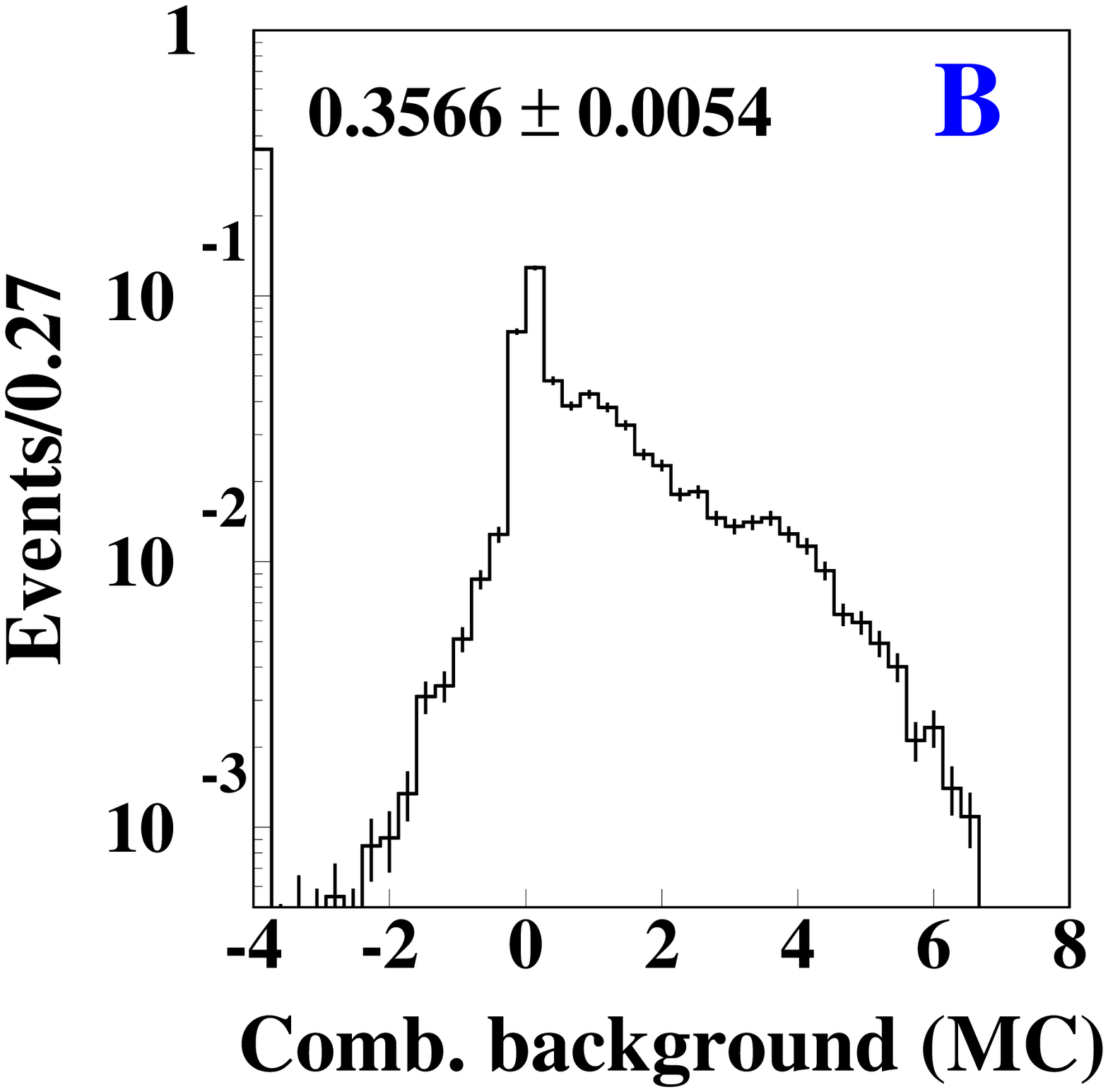,width=0.35\textwidth}

  \end{center}
  \caption[]{\it{Distributions of the $d_+$ variable for signal and 
background components. All distributions have been normalized to unity.
The content of the bin at $d=-4.$ has been inserted on each plot,
corresponding to events with no spectator track candidate. 
In the two lower plots, distributions obtained for combinatorial
background events selected in real and simulated data can be compared.
For real data the distribution for 
% SPR 4-2-03
% ``double charm'' added below and A0 18-2 'For real data' above 
% EPR 4-2-03 
double charm cascade decays ($S_3$) has been corrected as described 
in section \ref{sec:cascade}.} 
   \label{fig:d}}
\end{figure}
% SPR 9-1-03
% sentence modified
% EPR 9-1-03
% SAO 13-3-03
% Added sentence
% EA0 13-3-03
As the track impact parameters can
extend to very large values because of the 
relatively long decay time of $b$-hadrons, the variables $d_{\pm}$
are taken to be equal to the logarithm of $(1+x_{\pm}^2)$ and their
sign is taken to be the same as $x_{\pm}$. 
For events with no spectator track candidate, that is, with no additional 
tracks compatible with the $b$-decay vertex, 
a fixed value of -4. is used for $d_{\pm}$. Examples of 
distributions of the variable $d_+$ for the signal and 
for the different background components, corresponding
to all analysed channels, 
%the 94-95 period and to the $\Do \rightarrow \Km \pi^+$ decay channel
are given in Figure \ref{fig:d}.

% SPR 9-1-03
% we have changed the lines commented below
%Two quantities are also introduced:
%${\cal P}(0|0)$ and ${\cal P}(0|\neq 0)$ which are the
%probabilities for getting no-spectator candidate, respectively, when there is 
%not and when there is really such a candidate. Their values have been 
%measured using the simulation and are respectively equal to
%$75\%$ and $30\%$ with a spread of $\pm 5\%$ corresponding
%to the different years and channels. In the analysis, values extracted
%from the simulation for each sample have been used and possible differences
%between real and simulated events on these values have been evaluated.
%SAO 15-1-03
% Changed sentence just below (display -> show)  
%EAO 15-1-03
%SAO 15-1-03 
%Changed proportions according to the figure. 
%(75%-> 77% for signal, for D** it is the prop. in the tail 45% -> 54%)
%EAO 15-1-03 
% SPR 28-1-03
% Change sentence again to update numbers  
% EPR 28-1-03
% SAO 13-3-03
% Added sentence
% EA0 13-3-03
Due to track reconstruction effects, only 77$\%$
of signal events have no spectator track candidate instead of the expected 
100$\%$.
Similarly, for the $\Dstarstar$ background, 1/3 of 
events with no additional
tracks are expected by isospin at the $b$-vertex whereas 46$\%$ are observed. 
%{\bf this value has to be checked}% is measured in the simulation.
These values allow the probabilities ${\cal P}(0|0)$ and ${\cal P}(0|\neq 0)$
to be extracted for getting no spectator candidate
when, respectively, there is 
not and when there is really such a candidate at generation level.
Values for these two probabilities
are respectively equal to
$77\%$ and $30\%$ with a spread of $\pm 5\%$ corresponding
to the different years and channels. 
As it will be explained later (section \ref{sec:cascade}), 
these two quantities have been used to correct the present simulation
of 
% SPR 4-2-03
% ``double charm'' added below
% EPR 4-2-03
double charm cascade decays as it includes only the channel 
$\overline {\rm B} \rightarrow {\rm D}_s^- {\rm D} X$ whereas
the other contributions 
($\overline {\rm B}~\rightarrow~{\rm D}^- {\rm D} X~{\rm and}~
\overline\Do {\rm D} X$) have a different charged particle topology.

% EPR 9-1-03

\subsection{Fitting procedures}

Six  event  samples have been analysed separately
corresponding to different
% SPR 9-1-03
% spectrometer changed to detector below
% EPR 9-1-03
detector configurations (1992-1993 and 1994-1995) and to 
different decay channels of the $\Do$ 
($\Km \pi^+$, $\Km \pi^+ \pi^+ \pi^-$ and $\Km \pi^+ (\pi^0)$).
%\footnote{Only the 92-93 data sample has been analysed for the 
%$\Km \pi^+ \pi^+ \pi^-$ channel.}. 
%Candidate $\Dstar$ decays
%are selected by a cut on the mass difference 
%$\delta m$ which is taken to be between $0.143$ and $0.148$ $\MeVcd$.
% SPR 9-1-03
% we have changed a bit the sentence below
% EPR 9-1-03
The analysis procedure is explained for a single such sample in the following 
sections \ref{sssec:sig} to \ref{sssec:combb}.
It is applied to all samples simultaneously to 
obtain the measurements.

For each event $(i)$, four measurements have been used:
$\vec{x}_i=(q^2_r,~\delta m,~ d_+,~d_-)_i$. 
% SPR 9-1-03
% we have changed a bit the sentence below
% EPR 9-1-03
The parameters 
${\cal F}_{D^*}(1) \Vcb$, $\rho_{A_1}^2$ and the 
background from $\Dstarstar$ decays ($S_2$) are obtained by minimizing
a negative log-likelihood distribution.
% SPR 9-1-03
% we have changed a bit the sentence below
% EPR 9-1-03
Other parameters, given in the following, have to be introduced to account 
for the various 
fractions of contributing event classes and to describe their behaviour
in terms of the variables analysed.
The likelihood distribution is obtained from the product of the 
probabilities to observe $\vec{x}_i$ for each considered event. 
These probabilities can be expressed in terms of the corresponding
probabilities for each event's class and of their respective
contributions in the event samples analysed:
\begin{equation}
{\cal P}(\vec{x}_i)=\frac{B \times b(\vec{x}_i) +\sum_{j=1}^6{S_j 
\times s_j(\vec{x}_i)}}
{B+\sum_{j=1}^6{S_j}}.
\end{equation}
In this expression, $B$ and $S_j,~(j=1,...,6)$ are the numbers of events 
(fitted)
corresponding to the combinatorial background and to the different classes
of events with a real $\Dstarp$. The functions $b(\vec{x})$
and $s_j(\vec{x}),~(j=1,...,6)$ are the respective probability
distributions of 
the variable $\vec{x}$. 
Each probability distribution for the $\vec{x}$ variable is considered to
be the product of four probability distributions corresponding
to the four different variables.

These distributions can be obtained from
data ($b(\vec{x})$) or from the simulation. 
%If they are parametrized, 
%the corresponding parameters can be determined, simultaneously with the
%other quantities, during 
% SPR 9-1-03
% consists in --> involves
% EPR 9-1-03
The fitting procedure involves 
minimizing the quantity:
\begin{equation}
-\ln {\cal L}= -\sum_{i=1}^{N_{evt}}{\ln{{\cal P}(\vec{x}_i)}},
\label{eq:like}
\end{equation}
where $N_{evt}$ is the total number of events analysed.

From external measurements there are also constraints on the expected
number of events corresponding to the categories ${\rm S}_3$-${\rm S}_6$.
These constraints can be applied assuming that the corresponding 
event numbers follow Poisson distributions with fixed average values
($S_j^0$).
This is obtained by adding to Equation (\ref{eq:like}) the quantity:
% SPR 9-1-03
% we have added [] in the equation below
% EPR 9-1-03
\begin{equation}
 -\sum_{j=3}^{6}{S_j \ln{S_j^0}} + \sum_{j=3}^{6}
{\ln{\left [ \Gamma(S_j+1) \right ]}}.
\label{eq:like2}
\end{equation}

A similar expression is also added to account for the fact that
the total number of fitted events
must be compatible with the number ($N$) of selected events:
\begin{equation}
 - N_f\ln{N} + \ln{\Gamma(N_f+1)},
\label{eq:like3}
\end{equation}
in which $N_f$, the number of fitted events, is equal to:
$N_f=\sum_{i=1,6} S_i +B$.

The list of fitted parameters is given in the following for each 
component contributing to the event sample analysed.

\subsubsection{signal events}
\label{sssec:sig}

\setlength{\leftmargin}{2.2cm}
\begin{itemize}
\item [ $ s_{1,q^2_r}(q^2_r)$:] this distribution results from the convolution 
of
the theoretical expected distribution $\frac{\dgsl}{d q^2_s}$
% SPR 9-1-03
% comma replaced by parentheses
% EPR 9-1-03 
(corrected by the $q^2_s$ dependent efficiency and acceptance) with the
resolution function ${\cal R}(q^2_s-q^2_r,q^2_s)$. It depends mainly on 
$\rho_{A_1}^2$ and on the assumed $q^2_s$ dependence for the ratio
$R_1$ and $R_2$ between the different contributing form-factors. 
%This
%last contribution contributes to systematic uncertainties.

%\item [ $s_{1,m}(\delta m)$:] this is the normalized $\delta m$ distribution
%corresponding to a real $\Dstarp$ signal, a Gaussian distribution is used:
%\begin{equation}
%s_{1,m}(\delta m) = \frac{1}{\sqrt{2 \pi \sigma_{\delta_{m_0}}^2}}
%\exp{-\frac{(\delta m-\delta_{m_0})^2}{2 \sigma_{\delta_{m_0}}^2}}
%\end{equation}
%and the parameters $\delta_{m_0}$ and $\sigma_{\delta_{m_0}}$
%are fitted on real data events.

% SA0 13-3-03
% removed ~ between D0 and arrows to avoid large spaces.  
% EA0 13-3-03
\item [ $s_{1,\delta m}(\delta m)$:] 
is a Gaussian distribution
corresponding to the $\Dstarp$ signal
for the $\Do\rightarrow\Km\pi^+ ~{\rm or}~\Km \pi^+ \pi^+ \pi^-$
decay channels and a gamma distribution for 
$\Do\rightarrow\Km \pi^+ (\pi^0)$. 
The two parameters for each distribution have been obtained from a fit to data.

\item [ $s_{1,d_{\pm}}(d_{\pm})$:] these distributions are
obtained from simulated signal events. The two distributions,
for the $d_+$ and $d_-$ variables are rather similar with about
$77\%$ probability for having no spectator track candidate and
the remaining $23\%$ being concentrated around zero. 
%and
%fitted using a polynomial dependence apart for events corresponding
%to candidates with no spectator track.

\item [ $S_1$:]
the number of signal events can be expressed as:
{\small
\begin{equation}
S_1= N_H \times R_b \times 4 \times \fd \times 
{\rm BR}(\Bdb \rightarrow \Dstarp \ell^- \overline{\nu}_{\ell})
\times {\rm BR}(\Dstarp \rightarrow \Do \pi^+)
\times {\rm BR}(\Do \rightarrow {\rm X})
\times \epsilon({\rm X}).
\end{equation}
}
In this expression, $N_H$ is the number of hadronic events analysed
(Table \ref{tab:stat}),
$R_b$ is the fraction of hadronic $\Zz$ decays into $b \overline{b}$ pairs,
the factor 4 corresponds to the two hemispheres and the fact that muons and 
electrons
are used, $\fd$ is the production fraction of $\Bdb$ mesons in a $b$-quark
jet, ${\rm BR}(\Bdb \rightarrow \Dstarp \ell^- \overline{\nu}_{\ell})$
is the semileptonic branching fraction of $\Bdb$ mesons which is measured in 
this analysis,\footnote{ It is the integral of Equation \ref{eq:dgdw} 
(divided by the total $\Bdb$ width) and depends
on the two fitted quantities ${\cal F}_{D^*}(1) \Vcb$ and $\rho_{A_1}^2$.}
 the other two branching fractions correspond, respectively,
to the selected
$\Dstarp$ and $\Do$ decay channels (Table \ref{tab:external}), 
and $\epsilon({\rm X})$ are the 
efficiencies, given in Table \ref{tab:effic}, of the cuts applied in 
the analysis
to select signal events.
Note that $S_1$ is  proportional to 
$({\cal F}_{D^*}(1) \Vcb)^2$.
\end{itemize}

\subsubsection{events from $\Dstarstar$ decays}
These are events from the ${\rm S}_2$ class corresponding to the cascade decay
$b \rightarrow \Dstarstar \ell^- \overline{\nu}_{\ell},~
\Dstarstar \rightarrow \Dstarp {\rm X}$.
\begin{itemize}
\item [ $s_{2,q^2_r}(q^2_r)$:] this distribution is taken from the simulation.
Its variation for different fractions of $\Dstarstar$ states has been studied
(see Section \ref{sec:bckgm} and Figure \ref{fig:qdstar})
% SPR 9-1-03
% a few words added
% EPR 9-1-03
and accounted for as a small systematic shift and error.
%As it depends on the produced type of $\Dstarstar$ meson, 
%individual distributions corresponding to the three states expected to 
%decay into a $\Dstarp$ have been combined using actual measurements
%of the production fractions for these states.

%\item [ $s_{2,m}(\delta m)$:] This distribution is the same as
%$s_{1,m}(\delta m)$.

\item [ $s_{2,\delta m}(\delta m)$:] the same distribution 
% SPR 9-1-03
% ``is used'' is diplaced
% EPR 9-1-03
is used as for signal,
$s_{1,\delta m}(\delta m)$.

\item [ $s_{2,d_{\pm}}(d_{\pm})$:] as for the signal, 
these distributions are taken from the simulation.
It has been verified that they are not dependent on the type of
$\Dstarstar$ state which produced the $\Dstarp$. There is a marked
difference between $s_{2,d_{+}}$ and $s_{2,d_{-}}$, the latter being
rather similar to the corresponding distribution for signal events.

\item [ $S_2$:] the number of expected events 
is fitted without imposing constraints from external measurements.

%obtained using the present average of $\Dstarstar$ production
%in $b$-hadron semileptonic decays in which the $\Dstarstar$
%emits a $\Dstar$ (charged or neutral) \cite{ref:lephf}:
%\begin{equation}
%{\rm BR}(\Bdb \rightarrow \Dstarstarp \ell^- \overline{\nu}_{\ell})
%\times {\rm BR}(\Dstarstarp \rightarrow \Dstar {\rm X})=
%(1.82 \pm 0.21 \pm 0.08)\%
%\end{equation}

\end{itemize}

% SPR 4-2-03
% ``double charm'' added below
% EPR 4-2-03
\subsubsection{double charm cascade decay lepton events}
%SAO 15-1-03
% Added label for the caption of figure 5 (d distributions) 
%EAO 15-1-03
\label{sec:cascade}

These are events from the ${\rm S}_3$ class corresponding to the cascade decay
$b \rightarrow \Dstarp \overline{{\rm D}} {\rm X},~
\overline{{\rm D}} \rightarrow \ell^- \overline{\nu}_{\ell} {\rm Y}$

\begin{itemize}
\item [ $s_{3,q^2_r}(q^2_r)$:] this distribution is taken from the simulation.

%\item [ $s_{3,m}(\delta m)$:]  this distribution is the same as
%$s_{1,m}(\delta m)$.
\item [ $s_{3,\delta m}(\delta m)$:] the same distribution 
% SPR 9-1-03
% ``is used'' is diplaced
% EPR 9-1-03
is used as for signal,
$s_{1,\delta m}(\delta m)$.

\item [ $s_{3,d_{\pm}}(d_{\pm})$:] 
%as for the signal, 
%it is taken from the simulation.
%It may be noticed, from Figure \ref{fig:dvar}, that for this class of events 
%there are more tracks remaining at (or close to) the $b$-hadron decay vertex
%coming, presumably from the decay of the anti-charmed particle.
when there are spectator tracks, the distribution $s_{2,d_{+}}(d_{+})$
(with $d_+>-4$), is used. The expected fractions of events with no spectator
tracks in the $d_+$ and $d_-$ distributions have been evaluated from the
 measured contributions of
$\Dob \Dstarp$, $\Dm \Dstarp$ and $\Dsm \Dstarp$ events 
\cite{twocharma,twocharmb},
 with the $\overline{{\rm D}} \rightarrow \ell^- {\rm X}$ branching fractions
and topological decay rates for the hadronic
states X, taken from \cite{ref:PDG02}.
For $d_+$ it is expected that $(39 \pm 6)\%$ of the events have no spectator 
track and for $d_-$ this fraction is $(39 \pm 4)\%$. These numbers have to
be corrected for reconstruction effects using the variables
${\cal P}(0|0)$ and ${\cal P}(0|\neq 0)$ introduced in Section 
\ref{sec:subsepar}.

\item [ $S_3$:] the expected number of events from this source is taken from
present measurements of $b \rightarrow {\rm D \overline{D} X}$ decay rates
which correspond to:
\begin{equation}
{\rm BR}(b \rightarrow \Dstarp \ell^- X)+{\rm BR}(b \rightarrow 
\Dstarm \ell^+ X) =(0.83 \pm 0.21)\%,
\label{eq:ddrate}
\end{equation}
where the lepton originates from the $\overline{\rm D} \to Y$ 
semileptonic decay.
This value has been obtained using measurements  from 
ALEPH \cite{twocharma} and BaBar \cite{twocharmb} on exclusive 
double charm decay branching fractions of $b$-hadrons,
with a charged $\Dstar$ emitted in the final state,
and using the inclusive semileptonic decay branching fractions of
charmed particles given in \cite{ref:PDG02}.

Simulated events contain double charm decays of the type
$b \rightarrow \Dstarp \overline{{\rm D}}_s^{(*)} {\rm X}$ only,
with a corresponding
branching fraction:
${\rm BR}(b \rightarrow \Dstarp \ell^- X)=0.25 \%$. 
This rate has been rescaled to correspond to the value 
given in Equation (\ref{eq:ddrate}), assuming that the experimental
acceptance is similar
for the different contributing channels.
%{\bf Quels sont les BR utilises dans la simulation pour $b \rightarrow D_i 
%\overline{D}_j X$, comment cela se compare avec ce que l'on connait 
%actuellement ?...}

\end{itemize}

\subsubsection{$\Zz \rightarrow c \overline{c}$ events}

\begin{itemize}
\item [ $s_{4,q^2_r}(q^2_r)$:] this distribution is taken from the simulation.
%\item [ $s_{4,m}(\delta m)$:]  This distribution is the same as
%$s_{1,m}(\delta m)$.

\item [ $s_{4,\delta m}(\delta m)$:] the same distribution 
% SPR 9-1-03
% ``is used'' is displaced
% EPR 9-1-03
is used as for signal,
$s_{1,\delta m}(\delta m)$.

\item [ $s_{4,d_{\pm}}(d_{\pm})$:] as for the signal, 
it is taken from the simulation.
\item [ $S_4$:] the expected number of events from this source is taken from
the simulation after having corrected for the small difference between the 
rates for $\Dstarp$ production in $c$-jets between simulated and real events
%SAO 27-1-03
%Changed the place of the reference to avoid latex problems.
%EAO 27-1-03
\cite{ref:lephf}:
\begin{equation}
P(c \rightarrow \Dstarp)= (0.2392 \pm 0.0035)_{MC}
\leftrightarrow (0.226 \pm 0.014)_{Data}. 
\end{equation}
The remaining contamination from $ c \overline{c}$ is expected
to be very small (of the order of 1$\%$).
%{\bf Quel est le rate pour $c \rightarrow \Dstarp$ dans la simulation? }

\end{itemize}
\subsubsection{fake lepton events}
Only fake lepton events associated with a real $\Dstarp$ 
and not coming from $ c \overline{c}$ events, have to be considered
as the other contributions have been already included. 
%are already included in the sample of combinatorial
%background events.

\begin{itemize}
\item [ $s_{5,q^2_r}(q^2_r)$:] this distribution is taken from the simulation.

%\item [ $s_{5,m}(\delta m)$:]  This distribution is the same as
%$s_{1,m}(\delta m)$.
\item [ $s_{5,\delta m}(\delta m)$:] the same distribution 
% SPR 9-1-03
% ``is used'' is displaced
% EPR 9-1-03
is used as for signal,
$s_{1,\delta m}(\delta m)$.

\item [ $s_{5,d_{\pm}}(d_{\pm})$:] as for the signal, 
it is taken from the simulation.
\item [ $S_5$:] the expected number of events from this source is taken from
the simulation after
% SPR 9-1-03
% ``events samples'' --> ``event samples'' 
% EPR 9-1-03 
having applied corrections determined, using special event samples,
to account for
differences between the fake lepton rates in real and simulated data
(see Section \ref{sec:systema}).

\end{itemize}

\subsubsection{semileptonic decays with a $\tau$}
These are events from the ${\rm S}_6$ class corresponding to the cascade decay
$b~\rightarrow~\Dstarp \tau^- {\rm X},~
\tau^- \rightarrow~\ell^- \overline{\nu}_{\ell} {\rm Y}$

\begin{itemize}
\item [ $s_{6,q^2_r}(q^2_r)$:] this distribution is taken from the simulation.

\item [ $s_{6,\delta m}(\delta m)$:] the same distribution 
% SPR 9-1-03
% ``is used'' is displaced
% EPR 9-1-03
is used as for signal,
$s_{1,\delta m}(\delta m)$.

\item [ $s_{6,d_{\pm}}(d_{\pm})$:] is the same as the signal distribution
$s_{1,d_{\pm}}(d_{\pm})$

\item [ $S_6$:] the expected number of events from this source is 
obtained assuming that the production rate for $b$-hadron $\tau$ semileptonic 
decays is $0.223\pm0.004$ of the rate with a $\mu$ or $e$
\cite{ref:lig}. As for the $c \overline{c}$ background, events
from $\tau$ decays are expected to give a small contribution,
of the order of $1\%$.

%taken from
%present measurements of $b \rightarrow {\rm D \overline{D} X}$ decay rates
%which correspond to:

\end{itemize}

\subsubsection{combinatorial background events}
\label{sssec:combb}

Real data events are selected in the upper wing of the 
$\Dstarp$ mass peak between $0.15$ and $0.17 ~\GeVcd$
for $\Do \rightarrow \Km \pi^+~{\rm or}~\Km \pi^+\pi^+\pi^-$ channels, 
and in the range $0.17~-~0.22~\GeVcd$ for $\Do \rightarrow \Km \pi^+(\pi^0)$.
%all events corresponding to $\delta m<0.17 \GeV$
%are used and also wrong-sign events.

\begin{itemize}
\item [ $b_{q^2_r}(q^2_r)$:] 
this distribution is taken from real data events located in the upper 
part of the $\delta m$ distribution.

%\item [ $b_{m}(\delta m)$:]
%A smooth distribution corresponding to the following parametrization:
%\begin{equation}
%b_{m}(\delta m)=(\delta m-m_{\pi})^{a_{m,0}} \left (  
%\sum_{k=1}^{n_m}{a_{m,k} \delta m^{k-1}}
%\right )
%\end{equation}
%has been fitted on right- and wrong-sign events and the
%$n_m+1$ coefficients have been determined.
%Before using it as a probability distribution, it has to be normalized to 
%unity. The parameter $m_{\pi}=0.1396~\GeV$ is the charged pion mass.

\item [ $b_{\delta m}(\delta m)$:] the parametrization given in 
Section \ref{sec:combag} has been used.
The same mass dependence has been taken for the first four samples
while a parametrization corresponding to different values for the
coefficients has been obtained for $\Do \rightarrow \Km \pi^+(\pi^0)$ events.
Parameters of these distributions have been fitted,
outside the global likelihood fit, to the $\delta m$ 
distributions corresponding to the events  selected for the
analysis.
%the different parameters
%are fitted during the overall minimization of the negative
%log-likelihood function.

\item [ $b_{d_{\pm}}(d_{\pm})$:] 
as for the $q^2_r$ distribution, the $d_{\pm}$ distributions for combinatorial
background events are obtained from analysed events, selecting those
situated in the upper part of the $\delta m$ distribution.

\item [ $ B$:] 
%its value 
%is constrained by requiring that the total number
%of events coming from the different components is compatible with the
%total number of events analysed.
in each of the six samples, the total number of
%corresponds to the expected number of 
combinatorial
background events 
is fitted over the total $\delta m$ range.
%selected in the total $\delta m$ range.
% in the mass window selected for the signal.

\end{itemize}

\section{Measurements of ${\cal F}_{D^*}(1) \Vcb$ and $\rho_{A_1}^2$}
\setlength{\leftmargin}{1.cm}
The six event samples have been analysed in the same way. Efficiencies
and 
%parametrizations of the different 
probability distributions have been
determined independently for each sample. Common parameters corresponding
to the description of physics processes have been fitted or taken from 
external measurements. The central values and uncertainties
used for the latter are summarized in 
Table \ref{tab:external}. 

\subsection{Results on simulated events}
\label{sec:fitsim}
Signal events generated using the DELPHI simulation program
correspond to a given dynamical model, using a given modelling of the 
decay form factors. The generated $q^2_s$ distribution has been fitted
using a parametrization derived from the one 
given in Section~\ref{sec:principle}. 
As the model used in the simulation is a priori different from 
HQET
expectations, it has been necessary to add arbitrary terms
in the expression so that the fit will be reasonable over the whole
$q^2$ range.
These terms correspond to a polynomial development in powers of $(w-1)$,
starting with at least quadratic terms so that they have no effect
on the slope nor on the absolute value of the spectrum at the end-point
corresponding to $w=1$.

%SAO 7-2-03
% Removed fig 6
%EAO 7-2-03
%\begin{figure}
%  \begin{center}
%    \mbox{\epsfig{file=signal_sim2.eps,width=14cm,height=10cm}}
%    \mbox{\epsfig{file=signal_sim2.eps,width=0.9\textwidth}}
%  \end{center}
%  \caption[]{\it { Fit on pure signal simulated events.
%This spectrum corresponds to the DELPHI simulation with default parameters.
%The low content of the first bin corresponds to a phase space effect
%in semileptonic decays with a muon as the spectrum is the average between
%electron and muon contributions.}
%   \label{fig:signal}}
%\end{figure}

Using the total number of generated events to fix the normalization,
the equivalent values
for the two parameters defining the signal in the simulation:
\begin{equation}
 {\cal F}_{D^*}(1) \Vcb = 0.03552 \pm 0.00016;~\rho_{A_1}^2=1.088 \pm 0.021
\end{equation}
are obtained.
The fitted semileptonic branching fraction is equal to:
\begin{equation}
{\rm BR}(\Bdb \rightarrow \Dstarp \ell^- \overline{\nu}_{\ell})= (5.091 \pm 
0.020) \%,
\end{equation}
which agrees with the exact value of $5.103 \%$ used to generate these events.
%SAO 7-2-03
%Figure 6 removed.
%Figure \ref{fig:signal} shows the fitted $q^2_s$ distribution.
%SAO 7-2-03

The exercise is repeated on pure signal events using the
reconstructed $q^2_r$ distribution. This predicted distribution
now includes the effects of the experimental reconstruction of the 
$q^2$ variable and of the acceptance. This gives:
\begin{equation}
{\cal F}_{D^*}(1) \Vcb= 0.03549 \pm 0.00050;~\rho_{A_1}^2=1.119 \pm 0.052.
\end{equation}
The fitted semileptonic branching fraction is equal to:
\begin{equation}
{\rm BR}(\Bdb \rightarrow \Dstarp \ell^- \overline{\nu}_{\ell})= ( 5.004 \pm 
0.054 ) \%.
\end{equation}

\begin{figure}
  \begin{center}
    \mbox{\epsfig{file=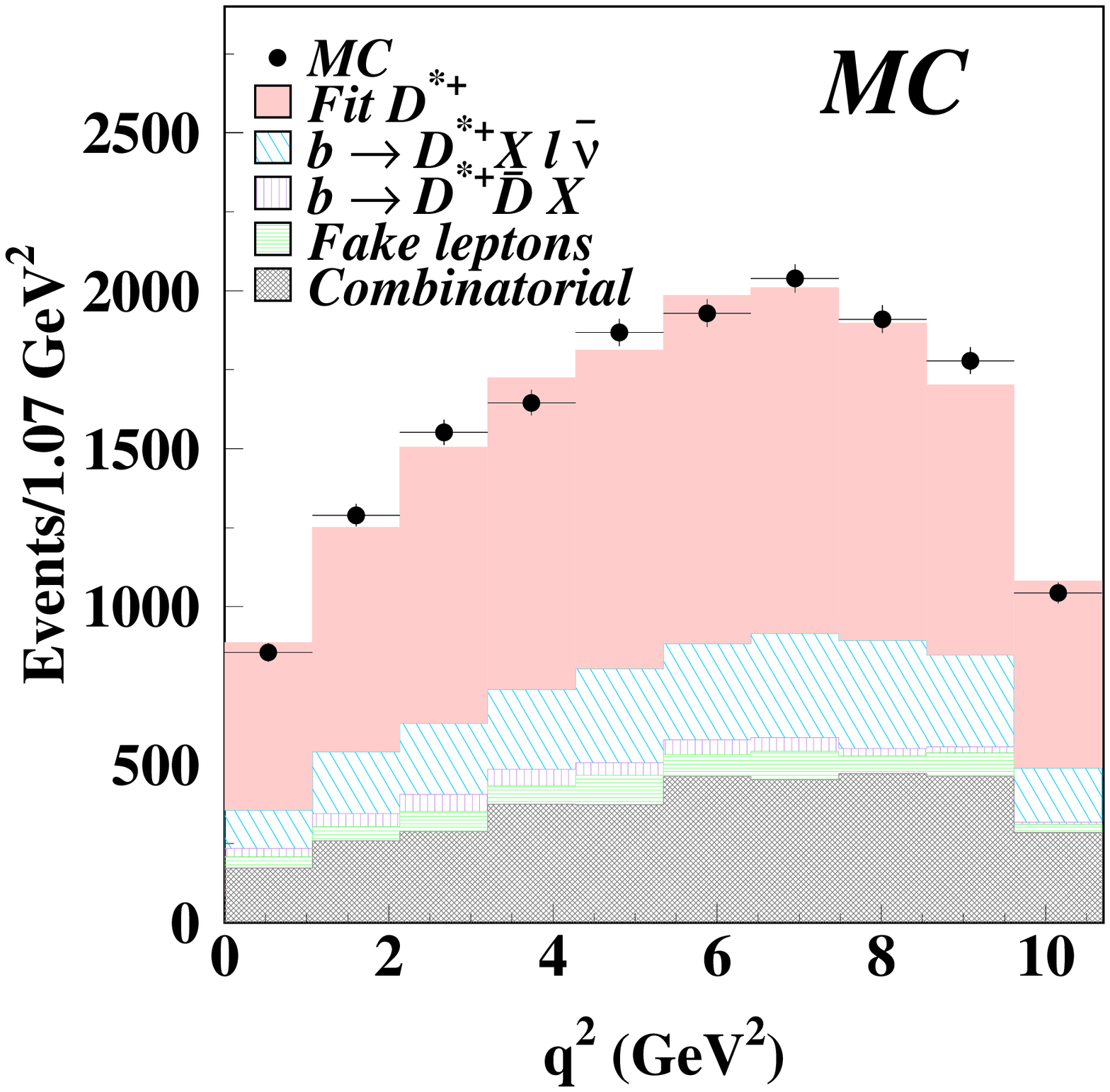,width=0.9\textwidth}}
  \end{center}
  \caption[]{\it { Fit of MC $q\overline q$ and $b\overline b$ events. 
The three analysed $\Do$ decay channels and the two data taking periods
have been combined. Only events selected within the $\delta m$
mass interval corresponding to the $\Dstarp$ signal are displayed. 
The small contribution of $\Zz \rightarrow c \overline{c}$ 
decays and leptons originating from $\tau^-$ events has been 
included in the fake lepton component.}
   \label{fig:mc}}
\end{figure}

Finally, using the sample of $\Zz \rightarrow q \overline{q}$ 
and $b \overline{b}$ simulated events,
the signal parameters are determined, including the different background 
components giving (see Table \ref{tab:resultsMC}):
\begin{equation}
{\cal F}_{D^*}(1) \Vcb= 0.03579 \pm 0.00063;~\rho_{A_1}^2=1.122 \pm 0.061.
\end{equation}
The fitted semileptonic branching fraction is equal to:
\begin{equation}
{\rm BR}(\Bdb \rightarrow \Dstarp \ell^- \overline{\nu}_{\ell})= ( 5.081 \pm 
0.065 ) \%,
\end{equation}
demonstrating that the fitting procedure gives the expected values
for the signal parameters correctly.
The $q^2$ distribution for MC events selected 
within the $\delta m=~[0.144,~0.147]~\GeVcd$ interval for the $\Km \pi^+$ and 
$\Km \pi^+ \pi^+ \pi^-$ channels and within the $\delta 
m=~[0.14,~0.17]~\GeVcd$ interval for $\Km \pi^+(\pi^0)$ is shown in  Figure 
\ref{fig:mc} with the 
contributions from the fitted components. 

% SPR 9-1-03
% Horizontal lines added in Table below
% EPR 9-1-03
\begin{table}[htb]
\begin{center}
  \begin{tabular}{|c|c|c|c|}
    \hline
 Data set &${\cal F}_{D^*}(1) \Vcb$  & $\rho_{A_1}^2 $& ${\rm BR}(\Bdb 
\rightarrow \Dstarp \ell^- \overline{\nu}_{\ell})~(\%)$\\
    \hline
  $\Km \pi^+$ 92-93  & 0.0375 $\pm$ 0.0020 &  1.27 $\pm$ 0.17 &  
5.16 $\pm$ 0.21  \\
  $\Km \pi^+$ 94-95  & 0.0356 $\pm$ 0.0013 &  1.16 $\pm$ 0.13 & 
 4.94 $\pm$ 0.14 \\
\hline
  $\Km \pi^+ \pi^+ \pi^-$ 92-93  & 0.0356 $\pm$ 0.0020  
& 1.03 $\pm$ 0.21 &  5.28 $\pm$ 0.23   \\
  $\Km \pi^+ \pi^+ \pi^-$ 94-95  & 0.0363 $\pm$ 0.0014  
& 1.13 $\pm$ 0.13  &  5.20 $\pm$ 0.15 \\
\hline
  $\Km \pi^+(\pi^0)$ 92-93  & 0.0355 $\pm$ 0.0018 &  1.14 $\pm$ 0.17 &  4.95 
$\pm$ 0.19   \\
  $\Km \pi^+(\pi^0)$ 94-95  & 0.0351 $\pm$ 0.0013 &  1.05 $\pm$ 0.13 &  5.06 
$\pm$ 0.14   \\
\hline
  Total sample & 0.03579 $\pm$ 0.00063 & 1.122 $\pm$ 0.061 &  
5.081 $\pm$ 0.065 \\
\hline
  \end{tabular}
  \caption[]{\it {Fitted values of the parameters in $\Zz \rightarrow q 
\overline{q}~and~b \overline{b}$ simulated events. 
% SPR 9-1-03
% Sentence below modified 
% EPR 9-1-03
The quoted uncertainties are statistical.}
%Quoted uncertainties are only of statistical origin.}
  \label{tab:resultsMC}}
\end{center}
\end{table}

\subsection{Results from data}

To analyse real data events, additional corrections have been applied
to account for remaining differences between real and simulated events.
Central values and uncertainties on these corrections are explained
in the following when evaluating systematic uncertainties attached
% SPR 9-1-03
% THE added below
% EPR 9-1-03
to the present measurements.

% SPR 9-1-03
% sentence below changed a bit to avoid mis interpretation on how
% the final result is obtained.
% EPR 9-1-03
The results obtained on the six data samples and using the total
statistics are 
given in Table \ref{tab:resultsRD}.

% SPR 9-1-03
% Horizontal lines added in Table below
% EPR 9-1-03
\begin{table}[htb]
\begin{center}
  \begin{tabular}{|c|c|c|c|}
    \hline
 Data set &${\cal F}_{D^*}(1) \Vcb$  & $\rho_{A_1}^2 $& ${\rm BR}(\Bdb 
\rightarrow \Dstarp \ell^- \overline{\nu}_{\ell})~(\%)$\\
    \hline
%%% SAO 13/12/03 Changed table 7 according to the referee suggestion 
%%% Added a sentence in the caption. 
%%  $\Km \pi^+$ 92-93  & 0.0385  $\pm$ 0.0060 & 1.11 $\pm$ 0.50 & 6.38 
%% $\pm$ 0.86 \\
%%  $\Km \pi^+$ 94-95  & 0.0336  $\pm$ 0.0044 & 0.69 $\pm$ 0.47 & 6.03
%%  $\pm$ 0.61 \\
%%\hline
%%  $\Km \pi^+ \pi^+ \pi^-$ 92-93  & 0.0438 $\pm$ 0.0056 & 1.47 $\pm$ 0.40 & 
%%6.76 $\pm$ 0.77\\
%%  $\Km \pi^+ \pi^+ \pi^-$ 94-95  & 0.0356 $\pm$ 0.0043 & 1.17 $\pm$ 0.39 & 
%%5.27 $\pm$ 0.53\\
%%\hline
%%  $\Km \pi^+(\pi^0)$ 92-93  & 0.0423  $\pm$ 0.0042 & 1.38 $\pm$ 0.32 & 6.64  
%%$\pm$ 0.58 \\
%%  $\Km \pi^+(\pi^0)$ 94-95  & 0.0369  $\pm$ 0.0037 & 1.41 $\pm$ 0.28 & 4.97  
%%$\pm$ 0.44 \\
%%
%%%EAO 13/12/03
  $K^- \pi^+$ 92-93  & 0.0394  $\pm$ 0.0055 & 1.15 $\pm$ 0.48 & 6.55 $\pm$ 0.77 \\
  $K^- \pi^+$ 94-95  & 0.0340  $\pm$ 0.0041 & 0.71 $\pm$ 0.45 & 6.11 $\pm$ 0.55 \\
\hline
  $K^- \pi^+ \pi^+ \pi^-$ 92-93  & 0.0410 $\pm$ 0.0058 & 1.43 $\pm$ 0.46 & 6.06 $\pm$ 0.77\\
  $K^- \pi^+ \pi^+ \pi^-$ 94-95  & 0.0342 $\pm$ 0.0042 & 1.12 $\pm$ 0.41 & 5.01 $\pm$ 0.51\\
\hline
  $K^- \pi^+(\pi^0)$ 92-93  & 0.0407  $\pm$ 0.0043 & 1.36 $\pm$ 0.35 & 6.22 $\pm$ 0.57 \\
  $K^- \pi^+(\pi^0)$ 94-95  & 0.0404  $\pm$ 0.0031 & 1.48 $\pm$ 0.24 & 5.70 $\pm$ 0.40 \\
%\hline
%  electron alone  &     &     &    \\
%  muon alone  &     &     &    \\
\hline
  Total sample  & $0.0381 \pm 0.0018$ & $1.23 \pm 0.15$ & $5.83 \pm 0.22$\\
\hline
  \end{tabular}
  \caption[]{\it{Fitted values of the parameters on real data events.
The quoted uncertainties are statistical. 
As 92-93 event samples have a reduced sensitivity to the ${\rm D}^{**}$
background (S2), fitted values quoted in this Table, when corresponding to
individual event samples, have been obtained using a fixed value for the
branching fraction ${\rm BR}(b \rightarrow \Dstarp {\rm X} \ell^- \overline{\nu}_{\ell})$
(=0.67~$\%$, see Equation \ref{eq:dsstrate}). Results quoted for the 
total statistics have been obtained letting free this quantity 
to vary in the fit.}
% SPR 9-1-03
% Sentence below modified 
% EPR 9-1-03}
%Quoted uncertainties are only of statistical origin.
  \label{tab:resultsRD}}
\end{center}
\end{table}

 \begin{figure}
  \begin{center}
    \mbox{\epsfig{file=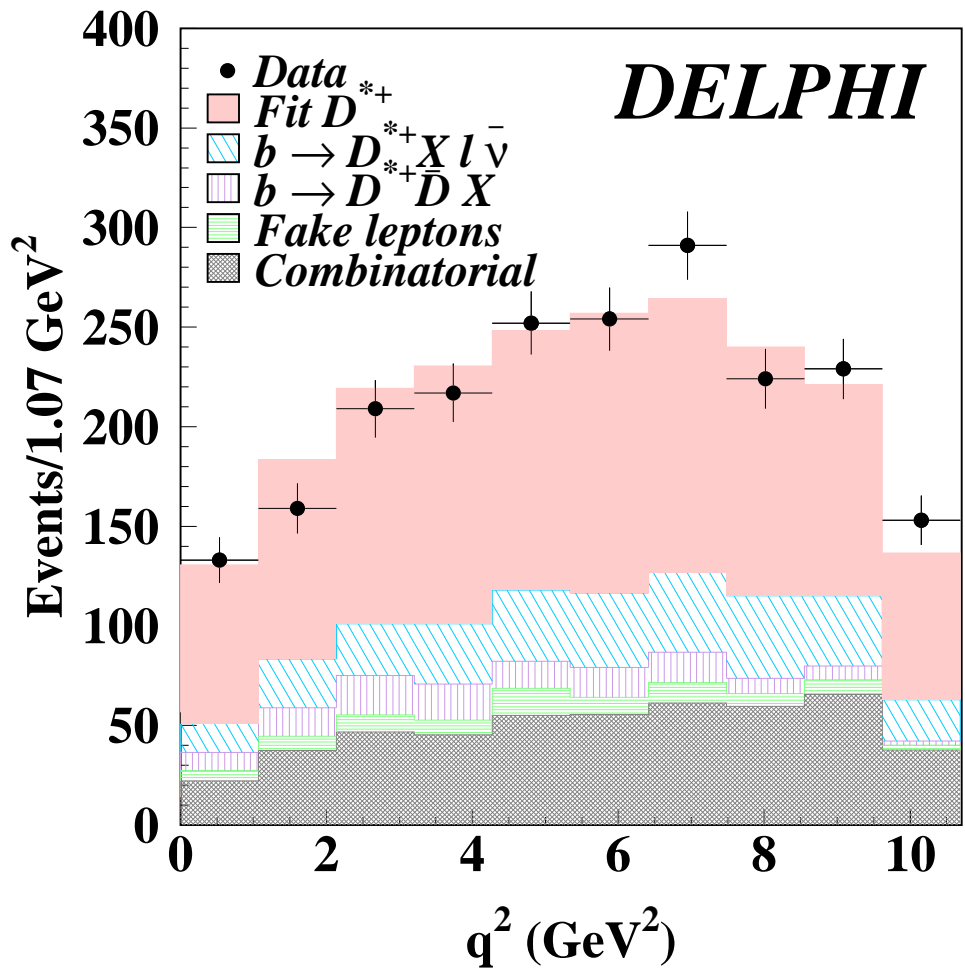,width=0.9\textwidth}}
  \end{center}
  \caption[]{\it { Fit on real data events. All periods are combined.
Only events selected within the $\delta m$
mass interval corresponding to the $\Dstarp$ signal are displayed.
The small contribution of $\Zz \rightarrow c \overline{c}$ 
decays and leptons originating from $\tau^-$ events has been 
included in the fake lepton component.}
   \label{fig:data}}
\end{figure}
%Correcting this result for the bias observed when the 
%same fitting procedure is applied to simulated events
%(see Section \ref{sec:fitsim}), 
%The following values have thus been obtained:

% SPR 9-1-03
% sentence below changed a bit ``average'' --> ``obtained''
% EPR 9-1-03
The values obtained are:
\begin{equation}
{\cal F}_{D^*}(1) \Vcb= 0.0381 \pm 0.0018 ;~\rho_{A_1}^2= 1.23 \pm 0.15,
\end{equation}
which correspond to a  branching fraction equal to:
\begin{equation}
{\rm BR}(\Bdb \rightarrow \Dstarp \ell^- \overline{\nu}_{\ell})= 
( 5.83 \pm 0.22 ) \%.
\end{equation}

The correlation coefficient $\rho({\cal F}_{D^*}(1)\Vcb,\rho_{A_1}^2) $ 
is equal to 0.894.
%SAO 14-2-03
% Removed sentence with the number of real D*
%SAO 14-2-03
%>From Table \ref{tab:nbofdstar}, 1688 $\pm$ 48 events are selected within
%the $\delta m$ mass interval
%corresponding to a real $\Dstarp$ signal. 
Fitted fractions of the different components
%(including the combinatorial background which does not 
%correspond to real $\Dstarp$ events)
are given in Table \ref{tab:fractions}.

\begin{figure}[th!]
\begin{center}
\begin{tabular}{cc}
\mbox{\epsfxsize8.cm \epsfysize8.cm
\epsffile{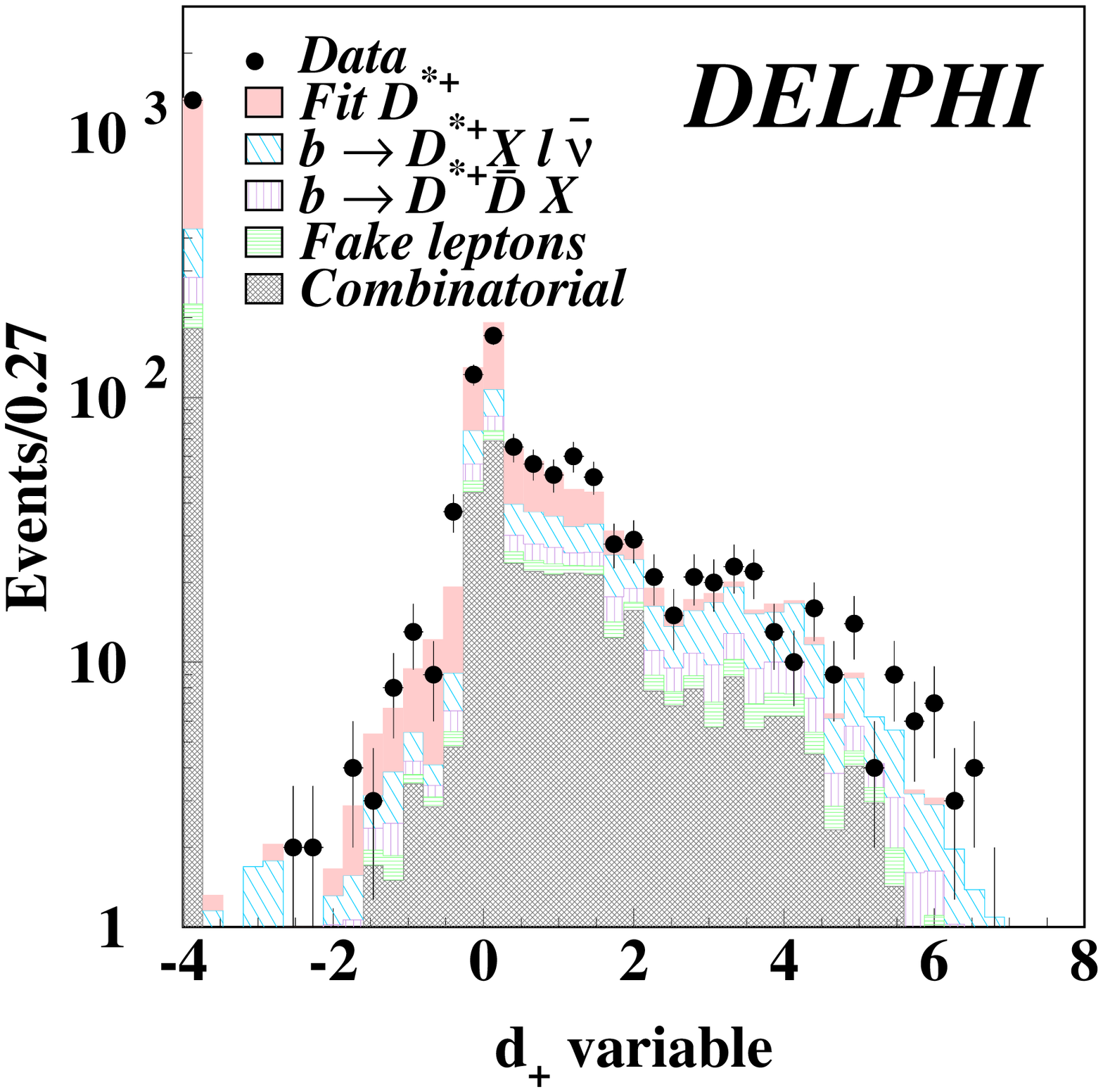}} & 
\mbox{\epsfxsize8.cm \epsfysize8.cm\epsffile{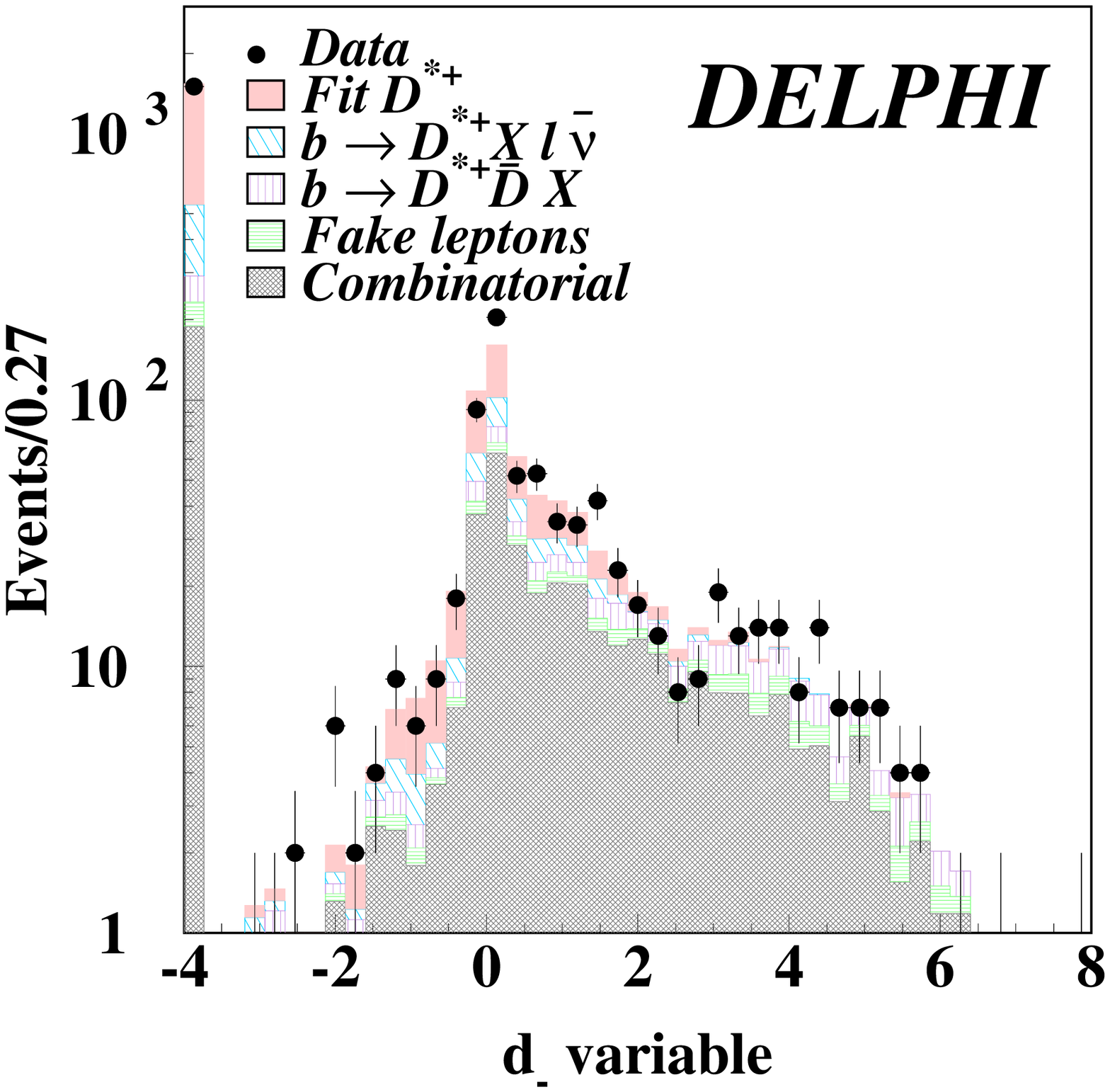}} \\
\end{tabular}
\caption{{\it Distributions for the $d_{\pm}$ variables, for events selected 
within the
$\delta m$ interval of the $\Dstarp$ signal and corresponding 
contributions from the fitted
components. The small contribution of $\Zz \rightarrow c \overline{c}$ 
decays and leptons originating from $\tau^-$ events has been 
included in the fake lepton component.
The measured (fitted) number of events in the first bin, at 
-4., are 
1324 $\pm$ 36 (1358) and 1498 $\pm$ 39 (1540), for the $d_+$ and $d_-$ 
distributions respectively.}
\label{fig:dvarsel}}
\end{center}
\end{figure}

Distributions of the $q^2$ and $d_{\pm}$ variables for events selected 
within the $\delta m=~[0.144,~0.147]~\GeVcd$ interval for the $\Km \pi^+$ and 
$\Km \pi^+ \pi^+ \pi^-$ channels and within the $\delta 
m=~[0.14,~0.17]~\GeVcd$ interval for $\Km \pi^+(\pi^0)$ are shown in Figures 
\ref{fig:data} and \ref{fig:dvarsel} with the 
contributions from the fitted components.

{\normalsize
\begin{table}[htb]
\begin{center}
\begin{tabular}{|p{1.5cm}|p{1.5cm}|p{1.5cm}|p{1.5cm}|p{1.5cm}|p{1.4cm}|
p{2.2cm}|}
% \begin{tabular}{|c|c|c|c|c|c|c|}
    \hline
\centering Signal & \centering $\Dstarstar$ &  \centering Cascade &  
\centering 
Charm &  \centering Fake lept. &  \centering $\tau$ &  Comb. Backg.\\
(${\rm S}_1$) & (${\rm S}_2$) & (${\rm S}_3$) & (${\rm S}_4$) & (${\rm S}_5$) 
& (${\rm S}_6$) & B \\
    \hline
\hline
$1196\pm35$ & $319\pm38$ & $129\pm11$ & $12\pm3$ & $67\pm8$ & $16\pm4$ & 
$523\pm23$ \\
 & $26.7 \pm 3.2$ & $10.8 \pm 0.9$ & $ 1.0\pm 0.3$ & $ 5.6 \pm 0.7$ & $ 1.3\pm 
0.3$ &$ 43.7 \pm 1.9$  \\
\hline
  \end{tabular}
  \caption[]{\it {Number of events
and fitted fractions 
(in $\%$ of signal events) attributed to the different 
components of the analysed sample of 
 events selected within the $\delta m$ mass interval 
corresponding to the $\Dstarp$ signal.}
  \label{tab:fractions}}
\end{center}
\end{table}}

\subsection{Evaluation of systematic uncertainties}
\label{sec:systema}
Values for the parameters taken from external measurements and hypotheses
used in the present analysis have been varied within their corresponding 
range of 
uncertainty. The results are summarized in Table \ref{tab:systvcb}.

% SPR 28-1-03
% Syst. from MC statistics added, syst from b-frag added
% EPR 28-1-03
% EAO 28-1-03
% Signs added
% SAO 28-1-03
\begin{table}[htb]
{\small
\begin{center}
  \begin{tabular}{|c|c|c|c|c|}
    \hline
parameter  & central value  & rel. err. on & rel. err. on& rel. err. on  \\
or hypothesis & and uncert. & ${\cal F}_{D^*}(1) \Vcb(\%)$&$\rho_{A_1}^2(\%)$
& BR$(\%)$\\
\hline
{\bf External parameters} & & & & \\ 
\hline
 Rates and BR &Table \ref{tab:external} and \ref{tab:sys_external} 
 &$\mp$2.5 & 
$\pm$0.4 &$\mp$5.3 \\
 $\Km \pi^+ X$ rates & $1.100 \pm 0.025$ &$\mp$0.5 & 0.0 &$\mp$1.1 \\ 
 $b$-hadron frag. &see text &$\mp$0.8  &$\mp$3.0  & 0.0 \\
 $\tau(\Bdb)$  & $(1.542 \pm 0.016)\ps$ & $\mp$0.5 & 0.0 & $\mp$1.0\\ 
\hline
{\bf Detector performance} & & & & \\
\hline
 Tracking efficiency & see section \ref{sec:sysdetector} & 
 $\mp$ 1.1 & 0.0  & $\mp$ 2.2  \\
 Lepton identification &$\pm 1.5\%~(e),~\pm 2.0\%~(\mu)$ &$\mp$0.5 &$\pm$0.1 & 
$\mp$1.0\\
 Fake Lepton rates & Table \ref{tab:fakel} &0.0 & $\pm$0.1 & 0.0\\
 $q^2$ resolution & see text & $\pm$2.3 & $\pm$6.2 &$\mp$0.4\\
% $b$-tagging capability & & & & \\
 $q^2$ acceptance & see section \ref{sec:channel}& $\pm$0.5 & $\pm$1.5 & 
$\mp$0.2 \\
 Control of $d_{\pm}$ dist.& see text &$\pm$3.0  &$\pm$1.1  &$\pm$5.5  \\
 Selection efficiency & Table \ref{tab:effic} &$\mp$0.6 &$\pm$0.1 &$\mp$1.2 \\
 MC statistics & see section \ref{sec:fitsim} &$\pm$1.8 &$\pm$5.9 &$\pm$1.3 \\
\hline
{\bf Signal modelling} & & & & \\
\hline
 $R_1(w)$ and $R_2(w)$ & see text &$\pm$1.0& $\pm$22.8& 0.0 \\
\hline
{\bf Backg. modelling} & & & & \\
\hline
 $\Dstarstar$ states & see text & $\pm$2.2 & $\pm$5.3 & $\pm$0.6  \\
% $b \rightarrow {\rm D \overline{D} X}$  & $0.0090 \pm 0.0025 $& 0.4 & 3.2\\
 Double charm cascade decay rate  & $0.0083 \pm 0.0021 $& $\pm$0.4 & $\pm$1.7& 
$\mp$0.4\\
% Cascade decay fractions  &  &  &  & \\
 $\overline B^0_d \rightarrow \Dstarp \tau^- \overline \nu_{\tau} X$ &$0.0127 
\pm 0.0021$& $\mp$0.4& $\mp$0.4& $\mp$0.6\\  
 P($c \rightarrow \Dstarp {\rm X}$) & $0.226 \pm 0.014$  & 0.0 & $\pm$0.1&
 0.0\\
\hline
{\bf Total systematics} &  & $\pm$5.8 & $\pm$25.2 &$\pm$8.4\\
\hline
  \end{tabular}
  \caption[]{\it { Systematic uncertainties given as relative values
expressed in $\%$. The total systematics are obtained by summing the
components in quadrature.}
  \label{tab:systvcb}}
\end{center}
}
\end{table}

\subsubsection{Uncertainties related to external parameters}
\begin{table}[htb]
\begin{center}
  \begin{tabular}{|c|c|}
    \hline
parameter  & central value \\
or hypothesis & and uncert.\\
\hline
 R$_b$ & $0.21664 \pm 0.00068$ \\
 P($b \rightarrow \Bdb$)  & $0.388 \pm 0.013$\\
 BR$(\Dstarp \rightarrow \Do \pi^+)$ & $0.677 \pm 0.005$\\
 BR$(\Do \rightarrow \Km \pi^+)$ & $0.0380 \pm 0.0009$ \\
 BR$(\Do \rightarrow \Km \pi^+ \pi^+ \pi^-)$ & $0.0746 \pm 0.0031$\\
 BR$(\Do \rightarrow \Km \pi^+ \pi^0)$ & $0.131 \pm 0.009$\\
 BR$(\Do \rightarrow \Km \ell^+ \nu_\ell)$ & $0.0686\pm 0.0030$\\
 BR$(\Do \rightarrow \Km K^+)$ & $0.00412 \pm 0.00014$\\
\hline
  \end{tabular}
  \caption[]{\it {Values for the external parameters used in the analysis.
The quoted value for BR$(\Do \to \Km \ell^+ \nu_{\ell})$ corresponds
to the sum of the branching fractions for the electron and muon final states.}
  \label{tab:external}}
\end{center}
\end{table}

\begin{itemize}
\item Values for D and $\Dstar$ branching fractions into the analysed 
final states, the R$_b$ value and the $b$-hadron lifetime have been 
taken from \cite{ref:PDG02}.
%taken from the last report of the LEPEWWG 
%\cite{ref:lepewwg} whereas values for $b$-hadron
%lifetimes and production rates correspond to averaged values
%obtained by the LEPHF group \cite{ref:lephf}. 
A summary of the values used in the present analysis is given in 
Table \ref{tab:external}. They have been varied within 
the corresponding range of uncertainty and the systematic errors 
induced in ${\cal F}_{D^*}(1)\Vcb(\%)$, 
$\rho_{A_1}^2(\%)$ and the ${\rm BR}(\Bdb \rightarrow \Dstarp 
\ell^- \overline{\nu}_{\ell})$ are given in Table \ref{tab:sys_external}.

\item Global efficiencies to select $\Do \to \Km \pi^+ (\pi^0)$ events 
have been estimated using the simulation as described in section 
\ref{sec:kpi0}. Measured branching fractions of several $\Do \to \Km \pi^+
X$ decay channels have been used for real data and a correction factor has
been applied to take into account events from undetermined origin. 

% SPR 9-1-03
% Version number for JETSET added AO 18-2-03 and REF
% EPR 9-1-03
\item Simulated events have been generated using the JETSET 7.3 program 
with the parton shower option \cite{ref:luc}. The non-perturbative 
part of the fragmentation of $b$-quark jets is taken to be a Peterson 
distribution which depends
on a single parameter, $\epsilon_b$:
\begin{equation}
D(z)=\frac{N}{z\left[1 -\frac{1}{z} -\frac{\epsilon_b}{1-z}  \right ]^2} .
\end{equation}
In this expression, $N$ is a normalization factor and 
$z=\frac{E_B+p_{L,B}}{E_b+p_{L,b}}$ with $B$ and $b$ indicating the
B hadron and the $b$-quark. The average fraction of the beam energy taken
by weakly decaying $b$-hadrons has been evaluated in \cite{ref:lephf} to be
$<X_E>=\frac{<E_B>}{E_{beam}}=0.702 \pm 0.008$. Simulated events,
generated with the parameter $\epsilon_b=0.002326$,
correspond to $<X_E>=0.7035$. 
% SPR 9-1-03
% Text changed below to include the new parametrization of the b-frag.
% EPR 9-1-03
The effects of a variation of the average
value $<X_E>$ and of the shape of the fragmentation distribution, 
on the results of the analysis have been studied by weighting
events generated with a known value of the variable $z$
so that they correspond to recent measurements \cite{ref:delphibfrag}. 
% SPR 28-1-03
% Comment added on the fact that XE increases by 2%
% EPR 28-1-03
In addition to a change in the slope of the B momentum distribution,
$<X_E>$ increases by 2$\%$.

%To reproduce
%the central value and the uncertainties measured for $<X_E>$, the parameter
%$\epsilon_b$ has to be varied in the range: $0.00247^{+0.00080}_{-0.00067}$.
% SPR 28-1-03
% Sentences below updated

%These variations on the $b$-quark fragmentation function induce two effects:
%\begin{itemize}
%\item a variation of the acceptance for signal events amounting to 
%$(-0.5 \pm 1.2)\%$ in relative value,

%\item a variation of the acceptance for background events originating
%from $b$-hadron decays,
%%{\bf to be evaluated}

%\item a variation of the resolution function on the $q^2$ variable. This 
%variation comes from the fact that a correction has been applied to the 
%fitted 
%$b$-hadron energy such that it remains centred on the simulated value
%independently of this value (see Section \ref{sec:qdeux}). 
%Resolution functions have been determined 
%on simulated weighted events to measure this effect which is found to be 
%negligible (of the order of $0.1\%$).

%\end{itemize}
A new parametrization of the resolution function 
${\cal R}(q^2_s-q^2_r,q^2_s)$ described in section \ref{sec:qdeux} 
has been determined on simulated weighted events and the analysis has 
been redone giving relative variations of $-0.8\%$ and $-3\%$,
respectively, on ${\cal F}_{D^*}(1) \Vcb $ and $\rho_{A_1}^2$.

% EPR 28-1-03
\item The $b$-hadron lifetime used in the simulation is equal to $1.6~ps$
and is independent of the type of produced $b$-hadrons. Events have been 
weighted so that their lifetime agrees with present measurements.
%No significant variation on the acceptance for signal events has
% been observed.
%It may be noted that the reconstructed lifetime measured on 94 simulated
%events ($(1.627\pm 0.038)~ps$) agrees with the input value.
%The quoted uncertainty for the MC statistics corresponds only to 
%$\Zz \rightarrow q \overline{q}$ events.
Efficiencies given in Table \ref{tab:effic} have been 
determined using weighted 
events and uncertainties related to the present accuracy on the $\Bdb$
lifetime measurement can be neglected.

\begin{table}[htb]
\begin{center}
  \begin{tabular}{|c|c|c|c|}
    \hline
parameter  & rel. err. on & rel. err. on & rel. err. on  \\
or hypothesis & ${\cal F}_{D^*}(1)\Vcb(\%)$ & $\rho_{A_1}^2(\%)$ & BR$(\%)$\\
\hline
 R$_b$ & $\mp$0.15  & $\pm$0.01  & $\mp$0.32\\
 P($b \rightarrow \Bdb$)   & $\mp$1.69  & $\pm$0.02 & $\mp$3.39\\ 
 BR$(\Dstarp \rightarrow \Do \pi^+)$ & $\mp$0.37 &$\pm$0.03 & $\mp$0.76\\
 BR$(\Do \rightarrow \Km \pi^+)$  &$\mp$0.31  & $\pm$0.01 &$\mp$0.63\\ 
 BR$(\Do \rightarrow \Km \pi^+ \pi^+ \pi^-)$ & $\mp$0.68 & 0.0 
 & $\mp$1.35\\ 
 BR$(\Do \rightarrow \Km \pi^+ (\pi^0))$   &$\mp$1.70  & $\pm$0.35 
 & $\mp$3.64\\ 
\hline
 Total  & $\mp$2.5& $\pm$0.4& $\mp$5.3 \\ 
\hline
  \end{tabular}
  \caption[]{\it {Systematic uncertainties from external parameters as 
 relative values expressed in $\%$. The uncertainty 
 related to the BR$(\Do \rightarrow \Km \pi^+ (\pi^0))$ 
 includes the contribution of the} $\Do \rightarrow \Km \pi^+ \pi^0$,
 $\Do \rightarrow \Km \ell^+ \nu_\ell$ \it{and} 
 $\Do \rightarrow \Km K^+$ \it{ branching fractions. 
 Central values and errors of these parameters are given 
in Table \ref{tab:external}.}
  \label{tab:sys_external}}
\end{center}
\end{table}

\end{itemize}

\subsubsection{Uncertainties from the detector performance}
\label{sec:sysdetector}
\begin{itemize}
\item Differences between simulated and real data events on the tracking 
efficiency have been studied in \cite{ref:tracking} and correspond
to $\pm0.3\%$ for each charged particle. For the soft pion 
coming from the $\Dstarp$ an uncertainty of $\pm1\%$ has been assumed. 

\item Differences  between simulated and real data events on lepton 
identification have been measured using dedicated samples of real data events
\cite{ref:leptons} and the real data to simulation ratios are equal to 
($88.5 \pm 1.5)_{92-93}\%$~and~($94.0 \pm 1.5)_{94-95}\%$ 
for electrons. For muons, the real data to simulation ratios are equal to 
$96\%$ for the two periods with a $\pm 2\%$ uncertainty.

\item Differences between fake lepton rates in which the 
lepton is a misidentified hadron, have also been measured
using dedicated samples of real data events and compared with the simulation
\cite{ref:leptons}
to obtain correction factors which are summarized in Table~\ref{tab:fakel}.
\begin{table}[htb]
\begin{center}
  \begin{tabular}{|c|c|c|}
    \hline
 Data set & electron & muon \\
    \hline
  92-93  & $0.69 \pm 0.03$    & $1.44 \pm 0.03$    \\
  94-95  & $0.77 \pm 0.03$    & $1.61 \pm 0.03$    \\
\hline
  \end{tabular}
  \caption[]{\it {Correction factors to apply to simulated events
in which the candidate lepton is a misidentified hadron.}
  \label{tab:fakel}}
\end{center}
\end{table}

\item Resolution of the $q^2$ variable.

%{\bf To be written}
A resolution function, common to all three $\Do$ decay channels has been used.
This function is determined independently for the 92-93 and 94-95 data 
samples. To quantify the importance of controlling the experimental resolution
on $q^2$ the widths of the fitted Gaussians have been increased by 
5$\%$. 
% SPR 9-1-03
% Sentence added
% EPR 9-1-03
This value is two times larger than observed differences between the averaged
missing energy measured in jets for real and simulated events. It 
corresponds to the increase in smearing of the resolution 
function when including events with a missing $\pi^0$.
%SAO 27-1-03
% Added sign to systematics resolution function.
%EAO 27-1-03
The induced variations in 
${\cal F}_{D^*}(1) \Vcb$ and
$\rho_{A_1}^2$ are $+0.3\%$ and $+1.2\%$ respectively. 

The uncertainty on the parametrization of the resolution 
distributions has been evaluated by varying the number of fitted
groups of slices in $q^2$  on which a linear variation of the parameters
of the two Gaussian distributions were evaluated. Results
obtained with two groups of 10 slices and with five groups
of four slices have been compared. This corresponds to relative variations on 
${\cal F}_{D^*}(1) \Vcb$ and
$\rho_{A_1}^2$ of $+2\%$ and $+5\%$ respectively. 

Results obtained when
including or excluding $\Km \pi^+ (\pi^0)$ events,
which have a poorer resolution, in the determination
of the resolution function have been also compared.
This corresponds to relative variations on 
${\cal F}_{D^*}(1) \Vcb$ and
$\rho_{A_1}^2$ of $\pm 1.0\%$ and $\pm 3.5\%$ respectively.
 
Measured differences obtained from these comparisons have
been summed in quadrature.
%the fitted parameters
%within their quoted uncertainties. The corresponding relative variations 
%induced on fitted parameters correspond to the values given
%in Table \ref{tab:systvcb}. 
%%The other half is our present estimate
%%for effects induced by using a different parametrization of the
%%resolution function.

The value of $q^2$ is obtained from the measurements of the B and $\Dstarp$
4-momenta (see Section \ref{sec:qdeux}). The B momentum is obtained from a 
constrained fit, imposing the B meson mass, which includes information
from primary and secondary vertex positions and from the energy 
% SPR 9-1-03
% AND added below
% EPR 9-1-03
and momentum of 
the particles belonging to the jet that provide an estimate of the B momentum
and direction. Uncertainties on the polar and azimuthal angles giving the 
B direction, and on the magnitude of the B momentum, which were determined 
from the measurement of the $\Dstarp$, charged lepton and missing-jet momenta,
have been varied by $\pm 30\%$ and new resolution distributions 
for $q^2$ have been 
obtained.
Corresponding variations on fitted values for 
${\cal F}_{D^*}(1) \Vcb$ and $\rho_{A_1}^2$ are found to be negligible.

\item Control of the $d_{\pm}$ distributions.
Distributions of the $d_{\pm}$ variables obtained for events selected 
for values of $\delta m$ higher than the $\Dstarp$ signal, in real
and simulated events, have been compared (see the bottom two
distributions in Figure \ref{fig:d}). The probabilities
for having no spectator track differ by $ (2.5\pm1.0)\% $ between data 
and the simulation. To account for this difference the corresponding 
probabilities for  no spectator track have been varied by $\pm 3\%$,
simultaneously for signal and background components with a
real $\Dstarp$. Such a variation does not apply for events
from the combinatorial background as the shape of the corresponding
distributions has been taken from real events.

%SAO 18-2-03
%Added control shape d+-
%EAO 18-2-03

The effect of a different shape of the $d_{\pm}$ distributions 
has also been evaluated for double charm cascade decays $(S_3)$.
A flat distribution has been considered for $d_\pm >2$ to account 
for the different topologies of $\Dob \Dstarp$, $\Dm \Dstarp$ and 
$\Dsm \Dstarp$. The effect of this variation has been found 
to be negligible.
 
\item The effect of a possible difference between the tuning of 
the $b$-tagging 
% SPR 9-1-03
% reference added below
% EPR 9-1-03
%\cite{ref:btag}
\cite{ref:delphi}
between real and simulated data events has been neglected
because loose criteria have been used in this analysis.

\end{itemize}

\subsubsection{Uncertainties on signal modelling}
These uncertainties correspond to the use of the $w$ dependent
ratios $R_1(w)$ and $R_2(w)$ defined in Equation (\ref{eq:ratios}).
Values for these quantities, using different models, have been obtained
by the CLEO collaboration \cite{ref:cleoform}: $R_1=1.18 \pm 0.30 \pm 0.12 $, 
$R_2=0.71 \pm 0.22 \pm 0.07$ with a correlation $\rho(R_1,R_2)=-0.82$ between 
the uncertainties on these two measurements. Relative variations induced in
the fitted parameters  
${\cal F}_{D^*}(1)\Vcb$, $\rho_{A_1}^2$ and ${\rm BR}(\Bdb \rightarrow 
\Dstarp \ell^- \overline{\nu}_{\ell})$ have been obtained varying
the values of $R_1$ and $R_2$ within their corresponding range of uncertainty.
They are given in Table \ref{tab:sysr1r2}.

\begin{table}[htb]
\begin{center}
  \begin{tabular}{|c|c|c|c|}
    \hline
  & $\frac{\Delta{\cal F}_{D^*}(1)\Vcb}{{\cal F}_{D^*}(1)\Vcb}\;(\%)$ & 
$\frac{\Delta\rho_{A_1}^2}{\rho_{A_1}^2}\;(\%)$ & 
$\frac{\Delta{\rm BR}(\Bdb \rightarrow \Dstarp \ell^- \overline{\nu}_{\ell})}
{{\rm BR}(\Bdb \rightarrow \Dstarp \ell^- \overline{\nu}_{\ell})}\;(\%)$\\
    \hline
$\Delta R_1=\pm 1\sigma$ & $\mp$ 1.3  & $\pm$ 0.3  &  0.0 \\
$\Delta R_2=\pm 1\sigma$ & $\mp$ 1.8  & $\mp$ 22.5  & 0.0 \\
\hline
  \end{tabular}
  \caption[]{\it {Relative variations of the fitted parameters due to the 
 $R_1$ and $R_2$ measurements. The 1$\sigma$ variation corresponds to the sum
 in quadrature of the statistical and systematic uncertainties.}
  \label{tab:sysr1r2}}
\end{center}
\end{table}

% SPR 28-1-03
% sentence added below
% EPR 28-1-03
As observed already in previous analyses, the uncertainty on $R_2$ 
dominates the systematic uncertainty on $\rho_{A_1}^2$.

\subsubsection{Uncertainties on background  modelling}
\label{sec:bckgm}
\begin{itemize}
\item The fraction of $\Dstarp$ mesons originating from decays of $\Dstarstar$
mesons depends on the total production rate of these states and on their
relative fractions.

Combining present measurements, the production rate of $\Dstarp$ mesons
originating from $\Dstarstar$ decays and 
accompanied by an opposite sign lepton is \cite{ref:lephf}:
\begin{equation}
{\rm BR}(b \rightarrow \Dstarp {\rm X} \ell^- \overline{\nu}_{\ell})
= (0.8 \pm 0.1)\%.
\label{eq:dsstarrate}
\end{equation}

This information is not included in the fit as $\Dstarp$ events produced
in $\Dstarstar$ decays are directly fitted,
% SPR 28-1-03
% added below the fact that the D** rate is fitted simultaneously with the rest
% EPR 28-1-03
simultaneously with ${\cal F}_{D^*}(1)\Vcb$ and $\rho_{A_1}^2$,
 to the data giving:
\begin{equation}
{\rm BR}(b \rightarrow \Dstarp {\rm X} \ell^- \overline{\nu}_{\ell})
= (0.67\pm 0.08\pm0.10)\%,
\label{eq:dsstrate}
\end{equation}
which is compatible with the expectation given in 
Equation (\ref{eq:dsstarrate}).
% The quoted systematic has been evaluated 
%by considering the same sources of errors as they are listed in 
%Table \ref{tab:systvcb}.
%\item 
%SAO 22-1-03  + SPR 28-1-03
The statistical error correlation coefficients 
%between the fitted 
%${\rm BR}(b \rightarrow \Dstarp {\rm X} \ell^- \overline{\nu}_{\ell})$ and 
%the parameters fitted previously 
are: $\rho({\cal F}_{D^*}(1)\Vcb,{\rm BR}(b \rightarrow \Dstarp {\rm X} \ell^- 
\overline{\nu}_{\ell}))= -0.171$
and $\rho(\rho_{A_1}^2,{\rm BR}(b \rightarrow \Dstarp {\rm X} \ell^- 
\overline{\nu}_{\ell})) = 
-0.061$. 
%EAO 22-1-03
% SPR 28-1-03
% Sentence below added
% EPR 28-1-03
The quoted systematic, in Equation (\ref{eq:dsstrate}), has been evaluated 
by considering the same sources of errors as are listed in 
Table \ref{tab:systvcb}.

To evaluate the effect of the uncertainty in the sample composition 
of produced $\Dstarstar$ states, the model of \cite{ref:leibo} has been used.
Parameters entering into this model have been varied so that the 
corresponding production rates of the narrow states remain within the 
$\pm 1 \sigma$ measured ranges
defined in Equations (\ref{eq:darate}, \ref{eq:dbrate}):

\begin{eqnarray}
{\rm BR}(\overline{{\rm B}} \rightarrow 
{\rm D}_1 \ell^-\overline{\nu_{\ell}} ) 
&=& (0.63 \pm 0.10) \% \label{eq:darate}\\
{\rm BR}(\overline{{\rm B}} \rightarrow 
{\rm D}_2^* \ell^-\overline{\nu_{\ell}} ) 
&=& (0.23 \pm 0.08)\%~{\rm or}~<0.4\% ~{\rm at~the~95\%~CL} 
\label{eq:dbrate} \\
{\rm R}^{**} = \frac{{\rm BR}(\overline{{\rm B}} \rightarrow {\rm D}^*_2 
\ell^-\overline{\nu_{\ell}})}
{{\rm BR}(\overline{{\rm B}} \rightarrow 
{\rm D}_1 \ell^-\overline{\nu_{\ell}})} 
&=& 0.37 \pm 0.14 ~{\rm or}~<0.6 ~{\rm at~the~95\%~CL}, \label{eq:Rss}
\end{eqnarray}
where ${\rm R}^{**}$ is the ratio between the production rates 
of ${\rm D}^*_2$ and
${\rm D}_1$ in $b$-meson semileptonic decays.

A dedicated simulation program has been written to generate the
decay distributions of the different $\Dstarstar$ states. Correlations between
the lepton and hadron momenta induced by the decay dynamics are included.
The $w$ dependence of the different form factors has been parametrized
according to the model given in \cite{ref:leibo}. It has been assumed
that, in addition to narrow states whose production fractions
are given in Equations (\ref{eq:darate}-\ref{eq:Rss}), broad $\Dstar \pi$
final states, emitted in a relative S wave, are produced.

 The two sets of model parameters giving the two most displaced central 
values for the $q^2$ distribution are used to evaluate the systematic 
uncertainty coming from the sample composition of $\Dstarstar$ 
states\footnote{ Values of the parameters (see \cite{ref:leibo})
corresponding to Model 1 are:
$\tau^{\prime}=-0.2,~\tau(1)=0.5, \hat{\tau}_1=-0.375~{\rm and}~
\hat{\tau}_2=0.375$. The corresponding values for Model 2 are:
$\tau^{\prime}=-2.0,~\tau(1)=0.83, \hat{\tau}_1=0.~{\rm and}~
\hat{\tau}_2=0.75$.}. 
These two distributions
are shown in Figure \ref{fig:qdstar} and the fitted values
obtained with these two models are given in Table \ref{tab:vcbrhomodel}.
%As the standard DELPHI simulation used up to now in this analysis was not
% basedon a really justified dynamical model for $\Dstarstar$ production,
\begin{table}[htb]
\begin{center}
  \begin{tabular}{|c|c|c|}
    \hline
  & Model 1 & Model 2 \\
    \hline
${\cal F}_{D^*}(1) \Vcb(\%)$    & $0.0400 \pm 0.0017$  
  & $0.0383 \pm 0.0018$   \\
$\rho_{A_1}^2$   & $1.38 \pm 0.13$    & $1.25 \pm 0.15$    \\
${\rm BR}(\Bdb \rightarrow \Dstarp \ell^- \overline{\nu}_{\ell})(\%)$
&$5.92\pm0.21$ &$5.84\pm0.22$ \\
\hline
  \end{tabular}
  \caption[]{\it {Fitted values corresponding to the two models describing
$\Dstarstar$ production.}
  \label{tab:vcbrhomodel}}
\end{center}
\end{table}

The average of these two results is used
to determine the central values for ${\cal F}_{D^*}(1) \Vcb$ and
$\rho_{A_1}^2$ and half of their difference is taken as systematic
uncertainty; this gives:

\bc
${\cal F}_{D^*}(1)\Vcb= 0.0392 \pm 0.0018 ;~\rho_{A_1}^2= 1.32 \pm 
0.15$.
\ec
\begin{figure}[th!]
\begin{center}
\begin{tabular}{cc}
\mbox{\epsfxsize8.cm \epsfysize8.cm
\epsffile{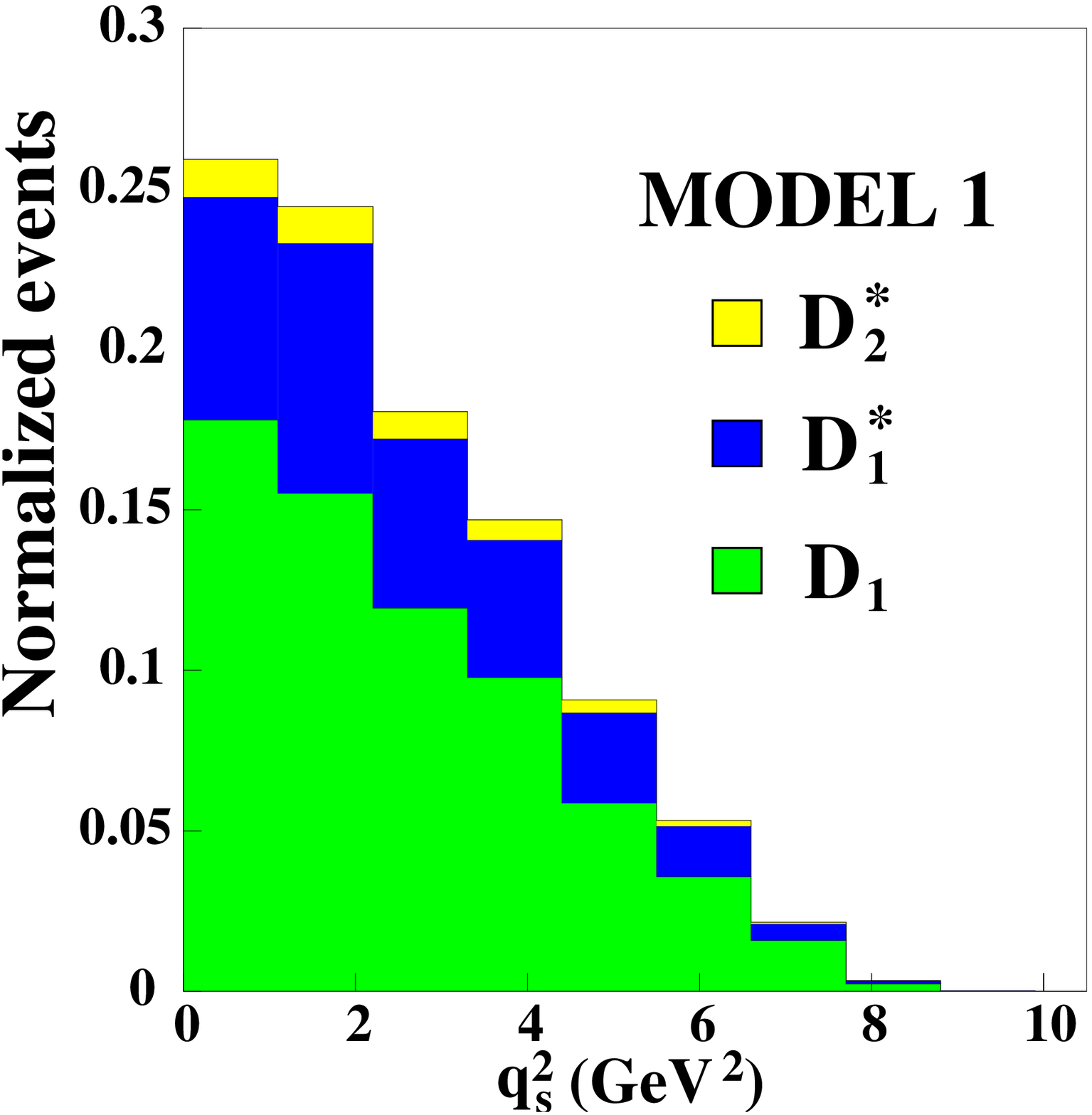}} & 
\mbox{\epsfxsize8.cm \epsfysize8.cm\epsffile{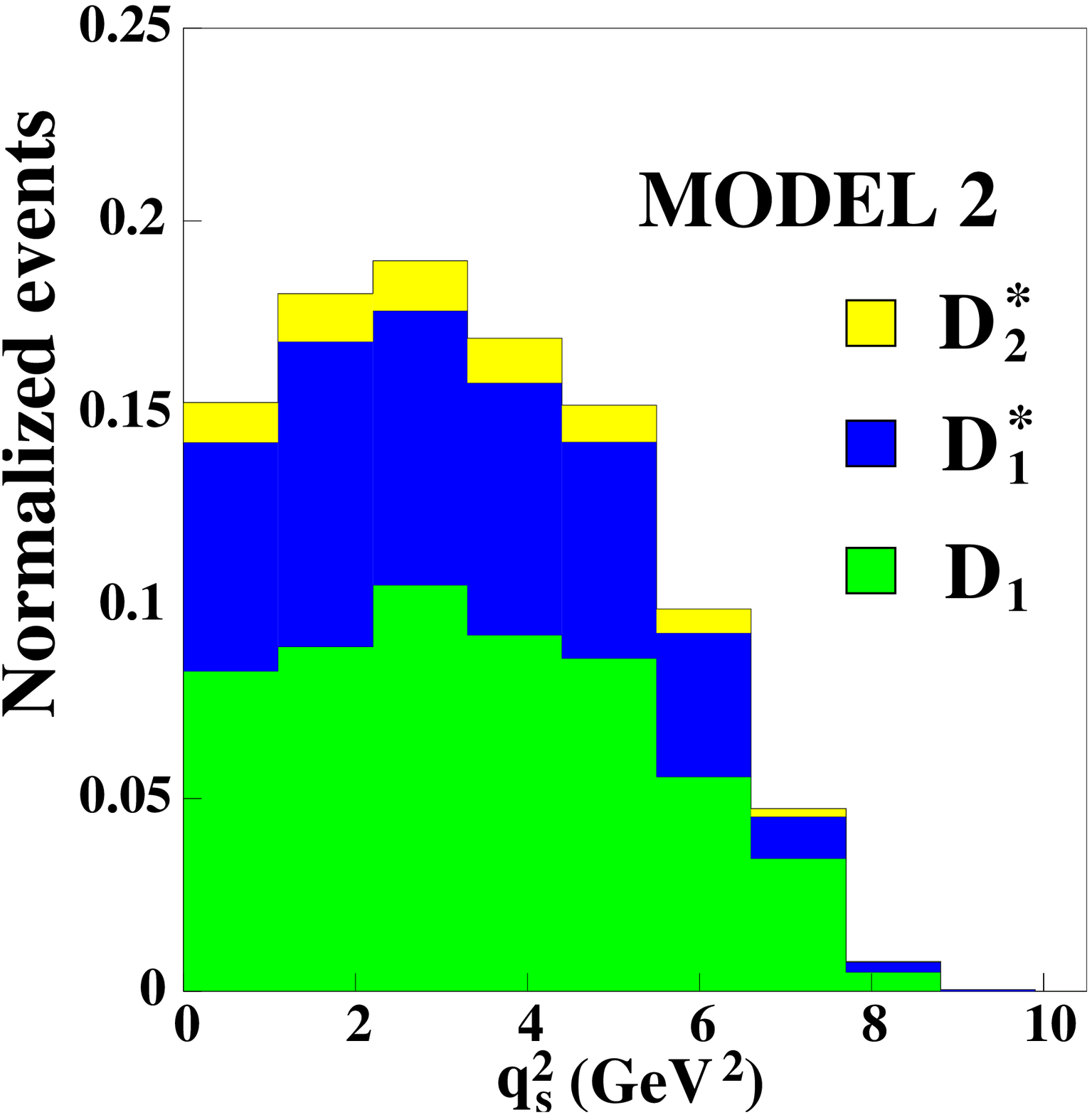}} \\
\end{tabular}
\caption{{\it $q^2$ distributions, normalized to unity,
obtained using the two sets of
parameters of the model \cite{ref:leibo}, which correspond
to the largest variation in the central value of these distributions
and which give production rates for narrow $\Dstarstar$ states that are
compatible with present measurements. The precise definition 
for models 1 and 2 is given in section \ref{sec:bckgm}.
%SAO 28-1-03
%Added sentence about models 1 and 2
%EAO 28-1-03
The three components given in each
histogram correspond, from top to bottom, to narrow 2$^+$, broad
1$^+$ and narrow 1$^+$ states.}
%It can be noted that fitted values obtained using the
%modelling of $\Dstarstar$ production, as given by the default DELPHI 
%simulation, correspond to those obtained with the parameters of Model 2.
\label{fig:qdstar}}
\end{center}
\end{figure}

\item
% SPR 4-2-03
% ``double charm'' added below
% EPR 4-2-03
The rate for the double charm cascade decay background, evaluated from 
simulated events,
has been rescaled to agree with present measurements
(see Equation (\ref{eq:ddrate})).

\item
The small components of tau and charm backgrounds have been evaluated using 
present measurements.

\item
The modelling uncertainty  of the combinatorial component corresponding
to events situated under the $\Dstarp$ peak has a negligible contribution.

\end{itemize}

% EAO 13-4-03
% Added this section by Patrick suggestion

\section{Combined result}
The present measurement of ${\cal F}_{D^*}(1)\Vcb$ and $\rho_{A_1}^2$ has been 
combined with the previous DELPHI result \cite{ref:DELPHI_vcb} 
obtained with a more inclusive analysis in which the $\Dstarp$ was 
reconstructed using the charged pion and tracks attached 
to a secondary vertex, accounting for the $\Do$ decay products. The values 
obtained in the previous DELPHI analysis were:
\bc
 ${\cal F}_{D^*}(1)\Vcb = 0.0355 \pm 0.0014 ^{\>+\> 0.0023}_{\>-\> 0.0024};~
  \rho_{A_1}^2 =  1.34 \pm 0.14 ^{\>+\> 0.24}_{\>-\> 0.22}$
 and
 ${\rm BR}(\Bdb \rightarrow \Dstarp \ell^- \overline{\nu}_{\ell})=(4.70 \pm 
0.13 ^{\>+\> 0.36}_{\>-\> 0.31})\%$.
\ec
Modifying the central values and uncertainties of the parameters entering in
this analysis so that they correspond to the 
values taken for the present measurement yields the results:
\bc
 ${\cal F}_{D^*}(1)\Vcb = 0.0372 \pm 0.0014 \pm 0.0025;~
  \rho_{A_1}^2 =  1.51 \pm 0.14 \pm 0.37$ and
 ${\rm BR}(\Bdb \rightarrow \Dstarp \ell^- \overline{\nu}_{\ell})=(5.20 \pm 
0.14 \pm 0.42)\%$.
\ec

The average with the present measurement
has been obtained using the method adopted by the LEP Vcb
working group \cite{ref:lephf}.
Common sources of systematic uncertainties between the two analyses have 
also been identified and properly treated in the averaging procedure.
The statistical correlation between the two measurements has been 
evaluated to be 8$\%$ (it was evaluated to be 5$\%$ in 
\cite{ref:firstvcb} where a similar combination was done).

Averaging the two measurements gives the following results:
\bc
   ${\cal F}_{D^*}(1)\Vcb = 0.0377 \pm 0.0011 \pm 0.0019;~
    \rho_{A_1}^2 =  1.39 \pm 0.10 \pm 0.33$ and
 ${\rm BR}(\Bdb \rightarrow \Dstarp \ell^- \overline{\nu}_{\ell})=(5.39 \pm 
0.11 \pm 0.34)\%$.
\ec
Using ${\cal F}_{D^*}(1)= 0.91 \pm 0.04$ gives:

\bc
 $\Vcb = 0.0414 \pm 0.0012 \pm 0.0021 \pm 0.0018$,
\ec
where the last uncertainty corresponds to the systematic error from 
theory.

% EAO 13-4-03

\section{Conclusions}
Measurements of ${\cal F}_{D^*}(1) \Vcb$, $\rho_{A_1}^2$
and of ${\rm BR}(\Bdb \rightarrow \Dstarp \ell^- \overline{\nu}_{\ell})$ 
have been obtained using exclusively reconstructed
$\Dstarp$ decays by the DELPHI Collaboration. Variables have been defined
which allow different decay mechanisms producing $\Dstarp$ mesons in the 
final state to be separated. 
% SPR 28-1-03
% Sentence below commented
% EPR 28-1-03
% SAO 28-1-03
% Changed systematics 
% EAO 28-1-03
% SAO 5-2-03
% Changed systematics again (to symmetrize them) 
% EAO 5-2-03
% SAO 18-2-03
% Added sentence and Vcb value 
% EAO 18-2-03
%In this way the rate of
%$\Dstarstar$ mesons, decaying into a $\Dstarp$, has been also measured.

The following values have been obtained:

\bc
${\cal F}_{D^*}(1)\Vcb= 0.0392 \pm 0.0018 \pm 0.0023;~\rho_{A_1}^2= 1.32 \pm 
0.15\pm 0.33$,
\ec
which correspond to a branching fraction: 
\bc
${\rm BR}(\Bdb \rightarrow \Dstarp \ell^- \overline{\nu}_{\ell})
=(5.90 \pm 0.22 \pm 0.50)\%$.
\ec

%Using ${\cal F}_{D^*}(1)=0.91 \pm 0.04$, gives: 
%\bc
%$\Vcb= 0.0418 \pm 0.0020 \pm 0.0023 \pm 0.0017$
%\ec
%where the last uncertainty corresponds to the systematic error from the 
%theory.

These values are in agreement with previous measurements obtained
by the ALEPH~\cite{ref:ALEPH_vcb}, DELPHI \cite{ref:DELPHI_vcb} and 
OPAL \cite{ref:OPAL_vcb} collaborations.

The $b$-quark semileptonic branching fraction into a $\Dstarp$ emitted
from higher mass charmed excited states has also been measured to be:
\bc
${\rm BR}(b \rightarrow \Dstarp {\rm X} \ell^- \overline{\nu}_{\ell})
= (0.67\pm 0.08\pm0.10)\%$.
\ec

% SAO 13-4-03
% Added DELPHI average
% EAO 13-4-03

Combining the present measurements with the previous analysis from DELPHI
\cite{ref:DELPHI_vcb} which was done
using a more inclusive approach, yields:
\bc
   ${\cal F}_{D^*}(1)\Vcb= 0.0377 \pm 0.0011 \pm 0.0019;~
   \rho_{A_1}^2 = 1.39 \pm 0.10 \pm 0.33$ and  
  ${\rm BR}(\Bdb \rightarrow \Dstarp \ell^- \overline{\nu}_{\ell})=(5.39 \pm 
0.11 \pm 0.34)\%$.
\ec
Using ${\cal F}_{D^*}(1)= 0.91 \pm 0.04$ this gives:
\bc
 $\Vcb = 0.0414 \pm 0.0012 \pm 0.0021 \pm 0.0018$,
\ec
where the last uncertainty corresponds to the systematic error from
theory.

%         Modified on 04-06-1999 by dimartino
%-------------------------------------------------------------------
\subsection*{Acknowledgements}
\vskip 3 mm
 We are greatly indebted to our technical 
collaborators, to the members of the CERN-SL Division for the excellent 
performance of the LEP collider, and to the funding agencies for their
support in building and operating the DELPHI detector.\\
We acknowledge in particular the support of \\
Austrian Federal Ministry of Education, Science and Culture,
GZ 616.364/2-III/2a/98, \\
FNRS--FWO, Flanders Institute to encourage scientific and technological 
research in the industry (IWT), Federal Office for Scientific, Technical
and Cultural affairs (OSTC), Belgium,  \\
FINEP, CNPq, CAPES, FUJB and FAPERJ, Brazil, \\
Czech Ministry of Industry and Trade, GA CR 202/99/1362,\\
Commission of the European Communities (DG XII), \\
Direction des Sciences de la Mati$\grave{\mbox{\rm e}}$re, CEA, France, \\
Bundesministerium f$\ddot{\mbox{\rm u}}$r Bildung, Wissenschaft, Forschung 
und Technologie, Germany,\\
General Secretariat for Research and Technology, Greece, \\
National Science Foundation (NWO) and Foundation for Research on Matter (FOM),
The Netherlands, \\
Norwegian Research Council,  \\
State Committee for Scientific Research, Poland, SPUB-M/CERN/PO3/DZ296/2000,
SPUB-M/CERN/PO3/DZ297/2000 and 2P03B 104 19 and 2P03B 69 23(2002-2004)\\
JNICT--Junta Nacional de Investiga\c{c}\~{a}o Cient\'{\i}fica 
e Tecnol$\acute{\mbox{\rm o}}$gica, Portugal, \\
Vedecka grantova agentura MS SR, Slovakia, Nr. 95/5195/134, \\
Ministry of Science and Technology of the Republic of Slovenia, \\
CICYT, Spain, AEN99-0950 and AEN99-0761,  \\
The Swedish Natural Science Research Council,      \\
Particle Physics and Astronomy Research Council, UK, \\
Department of Energy, USA, DE-FG02-01ER41155, \\
EEC RTN contract HPRN-CT-00292-2002. \\
%=========================================================================%

%SAO 3-2-2003
%=========================================================================%

%SAO 18-2-2003
% removed repeated references 
%EAO 18-2-2003

\end{document}